\documentclass[%
 reprint,
unsortedaddress,
 amsmath,amssymb,
 aps,
]{revtex4-1}

\usepackage{graphicx}% Include figure files
\usepackage{dcolumn}% Align table columns on decimal point
\usepackage{bm}% bold math
\usepackage{soul}
\usepackage{url}
\usepackage{hyperref}

\usepackage{multirow} % for the complex table
\usepackage{siunitx} % for the complex table
\usepackage{float} % for the complex table

\begin{document}

\preprint{PoP (tentative)}

\title{ {\it DustNET}: enabling machine learning and AI models of dusty plasmas}
%\title{Mini-traps for ultracold neutrons}

%Manuscript Title:\\with Forced Linebreak}% Force line breaks with \\

\author{Zhehui Wang}
\email{Correspondence: zwang@lanl.gov (ZW)}
\affiliation{Los Alamos National Laboratory, Los Alamos, NM 87545, USA}%

\author {Justin C. Burton}
\affiliation{Department of Physics, Emory University, 400 Dowman Dr., Atlanta, GA 30322, USA}

\author{Niklas Dormagen$^{1,2}$}
%\affiliation{}
\address{$^1$Institute of Physics, Justus Liebig University, D 35392 Giessen, Germany}
\address{$^2$NanoP, THM University of Applied Sciences, D 35390 Giessen, Germany}
%\ead{niklas.dormagen@nanop.thm.de}

\author{Cheng-Ran Du$^{3,4}$}
\address{$^3$College of Physics, Donghua University, Shanghai 201620, China}
\address{$^4$Member of Magnetic Confinement Fusion Research Centre, Ministry of Education, Shanghai 201620, China}

 \author{Yan Feng}%$^{1}$$^\ast$}
 % \affiliation{%$^{1}$ 
 \affiliation{Institute of Plasma Physics and Technology, Jiangsu Key Laboratory of Frontier Material Physics and Devices, School of Physical Science and Technology, Soochow University, Suzhou 215006, China} %\\
   % $^\aSpace Gravityst$The author to whom correspondence may be addressed: fengyan@suda.edu.cn}

\author{John E. Foster}
\affiliation{Nuclear Engineering and Radiological Sciences, University of Michigan, Ann Arbor, MI 48109, USA}

\author{Susan S. Glenn}
\affiliation{Los Alamos National Laboratory, Los Alamos, NM 87545, USA}%

\author{Max Klein$^{1,2}$}
\address{$^1$ Institute of Physics, Justus Liebig University, D 35392 Giessen, Germany}
\address{$^2$ NanoP, THM University of Applied Sciences, D 35390 Giessen, Germany}

\author{Christina A. Knapek}
\affiliation{Institute of Physics, University of Greifswald, 17487 Greifswald, Germany}

\author{Lorin Matthews}
%\affiliation{}
\affiliation{Center for Astrophysics, Space Physics, and Engineering Research, Baylor University, Waco, TX 76798-7316, USA}

\author{Andr\'e Melzer}
\affiliation{Institute of Physics, University of Greifswald, 17487 Greifswald, Germany}

\author{Edward Thomas}
\affiliation{Physics Department, Auburn University, Auburn, AL 36849, USA}

\author{Chuji Wang}
\affiliation{Mississippi State University, Starkville, MS 39759, USA}

\author{Jalaan Avritte}
\affiliation{Physics Department, Auburn University, Auburn, AL 36849, USA}

\author{Shan Chang}
\affiliation{School of Information and Intelligence Sciences, Donghua University, Shanghai 201620, China}

\author{Neeraj Chaubey}
\email{Current address: Applied Materials, Inc.}
\affiliation{Department of Physics and Astronomy, University of California, Los Angeles, California 90095-7099, USA}

\author{Pubuduni Ekanayaka}
\email{Current address: Desert Research Institute (DRI), Division of Atmospheric Sciences, Reno, NV 89512, USA.}
\affiliation{Mississippi State University, Starkville, MS 39759, USA}

\author{John A. Goree}
\affiliation{Department of Physics and Astronomy, University of Iowa, Iowa City, Iowa 52242, USA}

\author{Truell Hyde}
%\affiliation{}
\affiliation{Center for Astrophysics, Space Physics, and Engineering Research, Baylor University, Waco, TX 76798-7316, USA}

\author{Chen Liang}
 \affiliation{Institute of Plasma Physics and Technology, Jiangsu Key Laboratory of Frontier Material Physics and Devices, School of Physical Science and Technology, Soochow University, Suzhou 215006, China} 
 
 \author{Zhuang Liu}
 \affiliation{Institute of Plasma Physics and Technology, Jiangsu Key Laboratory of Frontier Material Physics and Devices, School of Physical Science and Technology, Soochow University, Suzhou 215006, China}

\author{Zhuang Ma}
 \affiliation{Institute of Plasma Physics and Technology, Jiangsu Key Laboratory of Frontier Material Physics and Devices, School of Physical Science and Technology, Soochow University, Suzhou 215006, China}

\author{Ilya Nemenman}
\affiliation{Department of Physics, Department of Biology, Initiative in Theory and Modeling of Living Systems, Emory University, 400 Dowman Dr., Atlanta, GA 30322, USA}
%\affiliation{Emory University, 1510 Clifton Rd., Atlanta, GA, 30322, USA  }
%\affiliation{Emory University, 400 Dowman Dr., Atlanta, GA 30322, USA  }

\author{Elon Price}
\affiliation{Physics Department, Auburn University, Auburn, AL 36849, USA}

\author{A. S. Schmitz}
\address{ Institute of Physics, Justus Liebig University, D 35392 Giessen, Germany}

\author{M. Schwarz}
\address{NanoP, THM University of Applied Sciences, D 35390 Giessen, Germany}

\author{Saikat C. Thakur}
\affiliation{Physics Department, Auburn University, Auburn, AL 36849, USA}

\author{M. H. Thoma}
\address{ Institute of Physics, Justus Liebig University, D 35392 Giessen, Germany}

\author{Hubertus M. Thomas}
\affiliation{Institut für Frontier Materials auf der Erde und im Weltraum, Deutsches Zentrum für Luft- und Raumfahrt, 51147 Köln (Cologne), Germany}

\author{L. Wimmer}
\address{ Institute of Physics, Justus Liebig University, D 35392 Giessen, Germany}

\author{Wei Yang$^{3,4}$}
%\email{Member of Magnetic Confinement Fusion Research Centre, Ministry of Education, Shanghai 201620, China}
%\affiliation{College of Physics, Donghua University, 201620 Shanghai, China}
\address{$^3$College of Physics, Donghua University, Shanghai 201620, China}
\address{$^4$Member of Magnetic Confinement Fusion Research Centre, Ministry of Education, Shanghai 201620, China}

\author{Zimu Yang}
\affiliation{Nuclear Engineering and Radiological Sciences, University of Michigan, Ann Arbor, MI 48109, USA}

\author{Xiaoman Zhang}
 \affiliation{Institute of Plasma Physics and Technology, Jiangsu Key Laboratory of Frontier Material Physics and Devices, School of Physical Science and Technology, Soochow University, Suzhou 215006, China}

\date{\today}% It is always \today, today,

\begin{abstract}

Dusty plasmas are ubiquitous throughout the universe, ranging from laboratory ionized matters, industrial plasmas, and nuclear fusion devices to the Earth's atmosphere and above, cometary tails,  lunar surface, planetary rings, interstellar and intergalactic clouds. Despite decades of research, many aspects of dusty plasmas remain poorly understood within a unified framework, {\it i.e.}  `{\it the standard model for dusty plasmas}'. Numerous theoretical and numerical models have been developed to describe diverse dusty-plasma phenomena, such as particle charging, transport, waves, and self-organized structures. Fully predictive models, spanning the wide range of spatial and temporal scales present in both laboratory and natural systems, remain nevertheless elusive. In standard plasma models, physical quantities, including the densities, momenta, and energies of electrons, ions, neutrals, and dust grains, evolve according to coupled differential equations. In practice, however, these equations are rarely solvable analytically, and numerical solutions are often limited by computational cost, numerical errors, and incomplete knowledge of boundary, initial conditions, or transport coefficients.

Recent advances in machine learning (ML), particularly deep neural networks (DNNs), provide new opportunities to augment traditional physics-based modeling. In this work we review ML and artificial intelligence (AI) approaches, collectively termed \emph{bottom-up} data-driven methods, for dusty plasma research. Central to this effort is the community-driven development of \emph{Dust Neural nEtworks Technology} (DustNET), a curated dataset initiative inspired by ImageNet that is designed to support the training, benchmarking, and evaluation of ML models for dusty plasmas. By integrating experimental measurements, numerical simulations, and AI-generated synthetic data, DustNET aims to enable improved predictive modeling, uncertainty quantification, multi-scale and multi-physics applications.

DustNET-driven models may also be deployed in real-time experiments or applications, where considerations such as edge computing, computational efficiency, power consumption, and memory constraints become important for autonomous dusty-plasma  diagnostics and control. Emerging multi-modal data integration, knowledge synthesis and generation using AI foundation models and AI agents, when combined with DustNET datasets, offer a new pathway toward a more holistic understanding of dusty plasmas and their applications in laboratory, industry, space, and astrophysical settings.

\end{abstract}

\pacs{Valid PACS appear here}% PACS, the Physics and Astronomy
\maketitle

\clearpage

\begin{widetext}
\tableofcontents
\end{widetext}

\clearpage

%\vspace{10in}

\section{Introduction \label{sec:1}} 

Excluding the elusive dark matter and dark energy~\cite{Sah:2005}, dusty plasmas, also called complex plasmas, are ubiquitous in the observable universe and make the universe dusty~\cite{MeRo:1994,BPFG:2004,GaSe:2014,FoMo:2009}. Dusty plasmas are multi-phase systems that contain condensed matter (in the form of dust grains, which are usually micro-meter size objects, but could be smaller or larger), and plasma at minimum, and usually other common phases of matter, such as liquid or gas. Smaller objects with sizes in the range of 1-10 nm are called very small grains or VSGs~\cite{MeRo:1994}.  Besides laboratory experiments and industrial applications such as semiconductor  materials processing for microelectronics~\cite{Bouc:1999} and nuclear fusion, dusty plasmas are frequently encountered in the Earth's atmosphere (lightning, ionosphere, high-energy density events including nuclear explosion), the solar system (the Moon surface and above, cometary tails, Saturn's rings)~\cite{merlino2021dusty}, near other stars (planet formation), galaxies, and the intergalactic media. 

Dusty plasmas provide fertile ground to study
and test multiscale (spatial and temporal), multi-phase plasma physics and plasma-material interactions, leveraging the fact that individual dust grains and their motion may be measured directly using imaging and other sensing methods~\cite{WXKW:2020}. Dust absorption and emission of light in the form of interstellar reddening, which occur when dust particles absorb photons and re-emit photons at a longer wavelength, effectively place dust masks over stars and galaxies, and blur our vision of the universe and its evolution since the earliest time~\cite{BPFG:2004,KSKB:2023}. 

Complex plasmas can also serve as model systems for studying fundamental interactions at smaller length scales that are difficult to observe directly. The relatively large dust particles (compared with atoms, molecules and other nanoparticles) and slow dynamic evolution enable direct observation and analysis of individual particle motions, a feature that is not easily achievable for atoms or molecules. Complex plasmas have been used to study processes such as phase transitions, wave propagation, and collective behavior at fine details. However, these insights can only be obtained if the dust particle positions and trajectories are measured with high precision.

The first motivation of this paper is to summarize and highlight machine learning (ML) and artificial intelligence (AI) approaches, collectively called ‘data-driven’ methods, for modeling and other applications in dusty
plasmas. ML and AI use neural networks (NNs), and more often deep neural networks (DNNs), as universal function approximators to
augment and accelerate research, and address the 
deficiency of traditional scientific methods~\cite{WFD:2023, SFJ:2023}.  Examples include NN-augmented computations and simulations, offline and online automated processing of large datasets, data interpretation,
active experiment controls, generating and testing hypotheses.

 In the domain of dusty plasmas, traditional models have been developed to understand rich phenomena, such as dust charging, dust mobility, dust destruction and growth, dust and plasma waves, dust and plasma structures, {\it etc.} yet a fully predictive integrated model unifying different spatial and temporal scales does not exist for either laboratory or natural dusty plasmas. Computational cost associated traditional {\it standard physics models} for dusty plasmas increases rapidly with the model spatial and temporal resolution, when physical quantities such as the densities, momenta, energies, temperatures of the electrons, ions, atoms and dust are described as functions of position and time. The spatial and temporal evolutions of the quantities are governed by a set of differential equations. On many occasions, the functions for the physical quantities may not be known analytically or explicitly. The accuracy of numerical solutions to the differential equations is often limited by computational errors and incomplete knowledge about boundary and initial conditions, and transport coefficients.

Central to  the data-driven framework is to establish Dust Neural nEtworks Technology (DustNET), which, besides NN algorithms, also includes a curated dataset initiative modeled after ImageNet~\cite{ImageNet,ImageNet2}, designed
to support the training and evaluation of DNNs for dusty plasmas. ImageNet, an open image database with more than 14 million labelled images, has been instrumental in advancing image classification, computer vision, deep learning, and AI in 2010s, a period which established the computational, algorithmic foundations of DNNs for ML and AI.  Similar to ImageNet, images of dusty plasmas are available from almost every laboratory dusty plasma experiment, as described below. The existing experimental datasets, some of which have already been analyzed by neural networks, may form the initial base datasets of DustNET. 

 It is laborious and time consuming to manually label millions of dusty plasma images, even if the resources allow. Scalable data augmentation methods, as a part of {\it dataset development strategy}, have emerged to aid NN model training and reduce reliance on large-scale labeled datasets by domain experts, such as dusty plasma physicists. Traditional image augmentation techniques involve geometric and other deterministic  transformations of labeled data, such as translation, rotation, flipping, cropping, noise injection, filtering, zooming, {\it etc.} to expand the training dataset while preserving label consistency~\cite{KS:2019}. These methods have shown to be effective in natural image domains but often fail to capture more subtle quantitative variations in scientific images, including medical images~\cite{CMV:2021}. 
 
 To overcome these limitations, learned augmentation techniques have gained traction. For instance, Generative Adversarial Networks (GANs)~\cite{goodfellow2020generative} and Variational Autoencoders (VAEs)~\cite{KiWe:2013} can generate realistic synthetic data by learning the underlying distribution of the input data. These generative models are especially useful in domains like medical imaging or materials science, where acquiring labeled samples is costly. Recent advances in diffusion models have significantly raised the bar for generative quality and diversity. These models learn to reverse a noise process and generate high-fidelity samples conditioned on a prompt or class label~\cite{HoJA:2020,RDNC:2022}. Diffusion models such as Stable Diffusion, and transformer models such as DALL$\cdot$E (see more below) are now widely used for image synthesis and have demonstrated strong performance in data-limited regimes. In scientific contexts, diffusion and transformer models are increasingly explored to generate physically plausible synthetic data to augment training sets~\cite{PTDC:2022}. 
  
 Another class of data augmentation approaches incorporates domain knowledge, including physics, geometry, topology, and statistical principles, to supplement and compensate for limited labelled data availability. Physics knowledge is now combined with machine learning architecture in many different forms, for example, Physics-informed NN (PiNN, PINN)~\cite{RaPK:2019}, Physics-guided NN (PgNN)~\cite{FPFR:2023}, physics-encoded neural networks (PeNNs), neural operators (NOs)~\cite{KLZL:2023}, and other physics-imbedding methods.  Known physical laws or symmetry constraints are directly integrated into model training or data generation. Methods like PINNs and simulation-informed data synthesis allow researchers to create synthetic datasets grounded in first principles. This ensures that augmented data respects the underlying behavior of physical systems, which is crucial for applications in materials science, fluid dynamics, and other domains where model interpretability and physical realism are essential. It has also been suggested that the exceptional simplicity of physics-based functions
hinges on properties such as symmetry, locality, compositionality and polynomial log-probability,
and these properties can be translate into exceptionally simple neural networks approximating images of natural phenomena~\cite{LiTR:2017}. 
 
 Recent years have seen rapid progress in generative artificial intelligence (GenAI) technology. Transformer-based architectures have revolutionized representation learning across modalities~\cite{VSPU:2023}. GenAI models, such as DALL$\cdot$E (from openAI), can generate realistic or imaginative images based on natural language descriptions, commonly known as `prompt engineering'. Other types of GenAI  can create new content like text, images, audio, and video, based on the patterns and relationships it learns from vast amounts of training data.  GenAI capability stems from significant advances in both model architectures such as diffusion models and transformers, and training strategies that involve billions or more model parameters. Some examples of using generative AI methods to generate images from text, will be given in examples below. In the near term, GenAI methods may be used for scientific visualization, education, and conceptual exploration. In the longer term, the GenAI methods may be used to augment simulation and experimental data, and also to validate NN models.

Many of the latest ML and AI methods, including NN and data augmentations, have not yet being adopted by dusty plasma research. The breadth of the interdisciplinary field of applied AI for dusty plasma research requires a community approach to handle the growingly complex and large body of dusty plasma data and scientific knowledge. In parallel to the explosive growth of scientific data from experiments, simulations and observations, the body of scientific knowledge about dusty plasma has also grown large enough that few can claim to be an expert for all of them. This already happened in many sub-field of science including physics. For example, condensed matter physics~\cite{Cohe:2006}. Here the knowledge consists of experimental and observation data, meta data associated with the experimental and observational, models to interpret the data and for predictions. Besides applications in experimental data processing, data interpretation, predictions (inferences) and uncertainty quantification, DustNET-enabled deep neural networks may also be used for real-time experimental dusty plasma controls and optimization~\cite{WLTC:2024}, and searching for new physics beyond the Standard Model of Physics, including the new physics beyond the `Standard Model of Dusty Plasma Physics'.

Another motivation of this work is to apply dusty plasma physics to advance AI models, including the `world models'~\cite{HaSc:2018}. From fundamental science point of view, dusty plasmas play a very unique role as an interdisciplinary field that marries the principles of physics, chemistry, materials science, and astronomy. One latest trend in AI and machine learning is multi-modal data and knowledge integration towards AI foundational models (FoMos). An AI FoMo for dusty plasmas, combined with DustNET, provides a transformative approach towards new holistic understanding and applications of dusty plasmas in laboratory and in nature.

The rest of the paper is organized as follows. Sec.~\ref{sec:pheno} provides an overview of dusty plasma phenomena at different scales, how these dynamic processes give rise to different classification problems that can be, if not already, addressed by ML/AI methods. Sec.~\ref{sec:model} reviews standard physics-based models and their limitations, and how ML/AI has been used and can be further expanded. Sec.~\ref{sec:dataset} highlights existing dusty plasma datasets, including synthetic data generation and text data. In Sec.~\ref{Sec:dFuion}, we present more details in DustNET initiative, including algorithms for data fusion, and Dusty-plasma Unified Sensing, Testing and Multi-modal Analysis Platform (DUST-MAP), a possible multi-modal foundation model for dusty plasmas. Some applications of DustNET and DUST-MAP, including addressing open problems in dusty plasmas, are discussed in Sec.~\ref{sec:Apps}. A summary is given in Sec.~\ref{sec:Summ}. %discusses future directions and conclud. {\it Observations, theory, interpretation, UQ}

\section{Diverse dusty plasma phenomena on different scales \& Classification \label{sec:pheno}}

%{\it What classification problems have been solved in dusty plasmas using machine learning?}

\begin{figure*}[!htb]
    \centering
   \includegraphics[width=1.85\columnwidth]{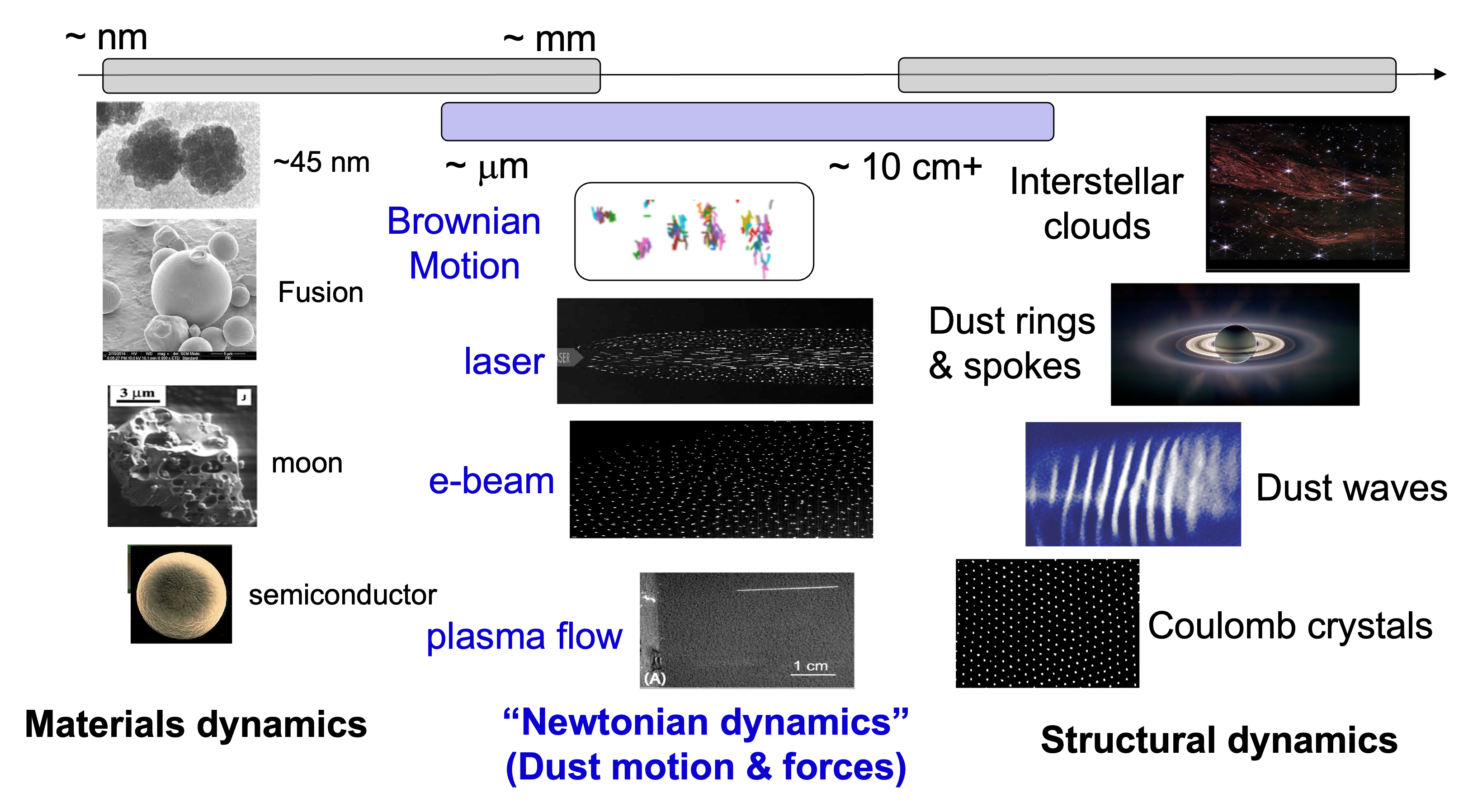}
    \caption{ Examples of dusty plasma dynamics on different temporal and spatial scales. On the smallest length scale ($L \le r_d$), materials dynamics are important. %`Materials Genomics' for dust?-- the composition and structural of laboratory, space, interstellar and inter-galatic dust. 
The dust scale ($L \sim r_d$), based on direct measurements, can be used to examine dust motion using Newton's laws. External forces using lasers, electron beams (e-beam), and plasma flows have been reported. The largest scale ($L >  r_d$) shows rich structures and their evolution, ranging from Coulomb crystals, dust waves to planetary rings, interstellar and intergalactic dust clouds. The thumbnail inserts are from the web searches or scientific literature, such as~\cite{BoBo:1994,Wint:2004,TSWS:2021,merlino2021dusty,GLLW:2022}.} %45 nm: L. Boufendi, A. Bouchoule, Plasma Source Sci. Technol. 3, 262 (1994).}
    \label{fig:pheno1}
\end{figure*}

Dusty plasmas occur broadly in nature with little to no human interference, and can also be readily produced in both highly controlled laboratory environment or industrial settings~\cite{BBB:2023}. These multi-phase multi-scale systems exhibit a wide range of intriguing phenomena, often recorded as images produced by dust-scattered visible and infrared light, as shown in Fig.~\ref{fig:pheno1}. Individual dust particles from dusty plasmas may also be collected and inspected by microscopy down to atomic resolutions. 

A natural classification scheme of dusty plasmas is based on the local plasma and dust properties, such as electron (ion, dust) temperature, electron (ion, dust) density, magnetic fields, gravity, ion and dust species,  power source(s) for plasma generation, {\it etc.} We may divide dusty plasmas into two classes: dusty plasmas in laboratory and industry, and dusty plasmas in nature, based on their significant differences in image field of view and resolution in comparison with a typical dust size ($r_d$). Laboratory and industrial dusty plasmas are usually less than 1 m and individual dust can often be clearly resolved in an image. Natural dusty plasmas can easily exceed 10s of km on the Earth, and much larger in space, interstellar and inter-galactic medium. It is not yet possible to resolve individual dust particles when the field of view is $ > 10^6$ $r_d$ in natural settings (technologies to do so do exist, but they can be expensive to deploy for now). A dusty plasma becomes more complex when the dust size and shape distributions need to be taken into account. 

Laboratory and industrial dusty plasma phenomena may still be divided into finer classes according to length scales, since different models or theories are used for data interpretation. A fully integrated dusty plasma model that captures the quantum physics at atomic scale, and fluid physics at macroscale is yet to be developed and can be computationally costly using traditional methods, which partly motivate accelerated integrated model using ML and AI. On the microscopic or sub-dust length scale ($ \sim <$ 0.1 r$_d$), dominant force, time scale is driven by electrons, modulated by the ionic ambient. Dust charging, chemical bonding often engage quantum physics and plasma chemistry. %Dust motion is significantly affected by ion and neutral dynamics (motion). 
On the intermediate or mesoscale scale, dust diffusion, dust hopping, dust growth, dust destruction are driven by Newtonian dynamics of individual dust grains. %On the macroscopic or laboratory scale, . {\it Coulomb crystals} of dust have been observed in laboratory dusty plasmas. The particles typically arrange themselves in regular 2D and 3D geometric patterns, such as hexagonal or cubic lattices, similar to crystals of ions and atoms. Phenomena that are deviated from Coulomb crystals of dust include dust Coulomb liquid, melting, other linear perturbation phenomena, dust acoustic waves, dust lattice waves, void formation ( in microgravity), dust filamentation, nonlinear phenomena such as shock waves. 
On the macroscopic scale, the collective dust motion can be described by kinetic or fluid equations. Dusty plasmas have found to self-organize into highly ordered structures known as Coulomb crystals under certain conditions, where dust grains arrange into regular 2D or 3D lattices (e.g., hexagonal or cubic patterns). These structures are formed when inter-particle potential energy dominates over thermal energy and neighboring dust motion is strongly coupled. In contrast, when coupling weakens, the system can transition into dusty liquids or exhibit melting, dust acoustic waves, shock fronts, or void formation, especially under microgravity. Laboratory dusty plasmas are thus popular platforms for studying both individual dust particles and collective phenomena, such as waves, structure formation, and phase transitions, offering insights into fundamental plasma physics, industrial applications, and naturally occurring space and astrophysical processes. 

Naturally occurring dusty plasmas on Earth result from interactions between dust particles, originating from both natural sources and human activities, such as rocket launches, and atmospheric or ionospheric plasmas. Examples include noctilucent clouds (also known as polar mesospheric clouds), auroral plasmas that produce the northern and southern lights, and lightning storms. The noctilucent clouds, with a size of 10 to 100 km that scatter sunlight, has an estimated electron density ($n_e$) of 10$^3$ - 10$^5$ cm$^{-3}$, a dust density $n_d \sim$ 10$^{-3}$ - 10$^{-1}$ cm$^{-3}$ and the average dust diameter about 0.3 \textmu m. Temperatures below 150 K or -120 $^o$C are required to form noctilucent clouds. On the planetary and space scale, dusty plasmas,  such as comet tails, Saturn's rings and spokes~\cite{HHHM:2004, merlino2021dusty}, can now be observed at close encounters or even in-situ through lunar and planetary missions~\cite{wahlund2009detection}.  On the galactic scale, cosmic dusty plasmas are now routinely observed by using the Earth-bound and space telescopes~\cite{MeRo:1994}. Lately, the James Webb Space Telescope has revealed intricate details of interstellar dust and gas clouds, allowing astronomers to map the three-dimensional structures of the interstellar medium for the first time~\cite{NASA:ISM}. 

As imaging and other datasets in dusty plasma experiments and observation continue to expand, automated image analysis has become increasingly necessary.A variety of neural network architectures have been developed for these purposes. Convolutional neural networks (CNNs) and more recent attention-based architectures such as Vision Transformers (ViTs)~\cite{DBK:2021} are widely used for image classification and recognition tasks. Autoencoders provide efficient frameworks for unsupervised feature extraction and dimensionality reduction, which can reveal latent structures in high-dimensional image datasets. Object detection and segmentation networks are particularly relevant for dusty plasma research because they enable the identification and tracking of individual dust grains in time-resolved imaging data.

Generative models, including generative adversarial networks (GANs) and variational autoencoders (VAEs), can be used for image synthesis and data augmentation, helping to mitigate limited training data. In addition, self-attention–based architectures such as ViTs are effective at capturing long-range spatial correlations in images. Hybrid models combining visual encoders with sequence models (RNNs or transformers) further enable higher-level scene interpretation and temporal analysis in image sequences. More discussions are given in Sec.~\ref{Sec:dFuion}.

Here we propose an alternative classification scheme for dusty-plasma phenomena based on image data at spatial scales relative to the dust grain size $r_d$: \emph{material dynamics}, \emph{particle motion (Newtonian dynamics)}, and \emph{collective structural dynamics}.

\subsection{Materials dynamics}

The materials dynamics of dust–plasma interactions encompass diverse multi-physics processes, such as thermodynamics, plasma physics, materials science, chemistry, and atomic physics, that govern the behavior, transformation, and evolution of individual dust grains both on the dust surface and underneath. Besides the materials and surface properties of the dust grains, these interactions are highly sensitive to plasma parameters such as electron and ion temperature, plasma density, ion mass and charge, plasma chemistry, and plasma sheath or boundary conditions. Nucleation, dust growth, evaporation, and surface reactions can occur, resulting in sophisticated evolution of dust materials.

The nucleation process in a plasma is an elementary step in dust formation, often initiated by plasma-mediated chemical reactions or phase transitions~\cite{Girs:2024}. Clusters of atoms or molecules aggregate and surpass a critical size to become stable, leading to the formation of nanometer-scale seeds. These seeds grow through accretion or coagulation into micrometer-sized or larger grains, particularly in astrophysical plasmas or plasma-enhanced chemical vapor deposition environments~\cite{Bouc:1999}.

In-situ formation of dust particles in plasma processing tools,  rather than dust originating from external sources, was first recognized as a significant contamination source in the semiconductor industry in the late 1980s~\cite{SHJH:1991}. These particles, often composed of materials like silicon dioxide (SiO$_2$), would become negatively charged and levitate within the positive column of the plasma due to force balance. When the plasma was turned off, the particles would settle onto the wafer surface, causing wafer defects and electronics malfunctions. Understanding and controlling dust formation in real-time are still vital for advancing semiconductor manufacturing.  As electronic device feature size continue to shrink (now less than 10 nm for the state-of-the-art), maintaining adequate processing environments becomes increasingly critical to ensure high wafer yields and reliable semiconductor devices. Ongoing research focuses on real-time monitoring and regulation of particle formation and behavior within plasmas, aiming to develop more effective mitigation techniques, such as inductively coupled plasmas, modification of plasma chemistry, in-situ dust cleaning and reduction. Recently, human-AI collaboration for improving semiconductor process development has shown promising results~\cite{KOLT:2023}. %Dust materials dynamics In industrial setting, plasma processing, including high-power lasers, electron and ion beams, semiconductor and other materials, 

%Laboratory observations of dust materials dynamics. Dedicated experiment on dust growth and destruction in plasmas, dust growth in microgravity.

The hottest laboratory dusty plasmas occur at the edge of thermonuclear fusion devices.  %Here is an overview of dusty plasma phenomena across these scales: Dust glows (UFOs) in fusion plasmas, nanoparticle and dust growth in plasma reactors. In high-temperature plasmas, such as those found in fusion devices
In tokamaks such as JET,  ITER, spherical tokamaks such as NSTX, MAST, stellerators such as W7-X, dust grains on or near the wall can experience intense ion bombardment, leading to surface heating, thermal desorption, and sputtering. When hot ions, such as deuterium, tritium, strike dust grains at energies ranging from tens to hundreds of electron volts, they can displace surface atoms through momentum transfer, leading to erosion and, under prolonged exposure, complete evaporation of the dust particle~\cite{Wint:2004,RBK:2022}. Understanding of these processes are critical in designing plasma-facing components, where dust generation, migration, and re-deposition can affect plasma stability, contamination, and even safety due to tritium retention or explosion risks in ITER-like environments~\cite{Wint:2004}.

  %[ The Physics of Dusty Plasmas by P. K. Shukla and A. A. Mamun.]

`Dust fossils' preserved in terrestrial  sediments, lunar regolith, meteorites, and interplanetary or interstellar samples offer an extraordinary window into the ancient material dynamics that shaped both the early Solar System, our local galaxy the Milky Way, and potentially further away from the Earth in the universe. These grains function as natural time capsules, recording information about cosmic background, high-energy transients such as supernovas, elemental formation processes, and environmental conditions spanning billions of years. Approximately 100–300 tonnes of cosmic dust
enter Earth’s atmosphere daily~\cite{Plan:2012}. For example, pre-solar grains with isotopic anomalies (e.g., in H, N) indicates origin from the earliest stages of solar
system formation~\cite{Mess:2000}. Analysis of micrometeorites from the Earth's
stratosphere, Antarctic snow, deep-sea sediments, and Greenland ice cores has revealed amorphous silicates (the most abundant presolar mineral), silicon carbide(SiC)~\cite{HGKT:2020}, and materials~\cite{HLKB:2022}. 

%Remote dust observation and analysis in the interstellar and inter-galactic environment.

The datasets for materials dynamics come from a diverse set of instruments. In-situ real-time measurement tools include high-speed imaging based on laser light scattering (Mie or Rayleigh scattering), laser-induced fluorescence, quartz crystal microbalances, optical emission spectroscopy, Langmuir probes. Ex-situ measurement tools include scanning electron microscopy (SEM), mass spectrometry, secondary ion mass spectrometry (SIMS)~\cite{Mess:2000}, and surface analysis techniques (e.g., X-Ray Photoelectron Spectroscopy, Energy Dispersive Spectroscopy) are used post-experiment to analyze the changes in dust composition, structure, and surface chemistry.

%\clearpage

\subsection{Newtonian dynamics}

Newtonian dynamics described rich phenomena of individual dust motion under different settings, and is probably the most studied phenomena in laboratory~\cite{Tsyt:1997}. In addition to passive observations, such as dust particles in positive columns of gas
discharges~\cite{EmBr:1970}, finer controls of dust motion is now possible using lasers or `optical tweezers' for dust manipulation.

When a particle is levitated in its sheath equilibrium position in a dusty plasma by balancing the electrostatic force \( F_e \), gravitational force \( F_g \), ion drag force \( F_{id} \), neutral drag force \( F_{nd} \), and thermophoretic force, it can be set in motion by various methods. Observations of subsequent motions can lead to measurement of the basic plasma parameters, such as the particle charge \( Q_d \), charge variations, \( E \), \( V \), inter-particle coupling, sheath structure, etc. For example, using the driven damped oscillation model of a particle levitated in the plasma sheath, the particle charge \( Q_d \) was estimated\cite{ref8,ref9,ref10}; the plasma crystals melt-up was studied by a laser sheet interacting with an array of particles levitated in the plasma sheath \cite{ref11}. However, due to the constraint of gravity, which restricts dust particle experiments to the plasma sheath where charged particles are levitated \cite{ref8,ref9,ref10,ref11, ref12}. Unless the micro-gravity experiments, a ground-based laboratory experiment would require additional force to eliminate the constraint of gravity, enabling dust particles to be levitated beyond the plasma sheath region. Thanks to the optical trapping (OT) technology that has been demonstrated in recent experiments to be able to trap and transport single dust particles in any location of the whole plasma volume \cite{ref13, ref14, ref15}.

The concept of OT and detailed calculations of optical trapping forces, e.g., radiation pressure and photophoretic forces, can be seen elsewhere \cite{ref16, ref17}. But their orders of magnitudes can be illustrated, for example, using a carbon nanofoam particle of 2 µm in diameter and under the illumination of a laser beam at 10 W/mm\(^2\). The thermal conductivity is 0.7 Wm/K at room temperature. The density of bulk carbon nanotubes is 140 kg/m\(^3\). The estimated radiation pressure force and photophoretic force are $\sim 3 \times 10^{-15} \, \text{N}$ and $\sim 4 \times 10^{-11} \, \text{N}$, respectively. The gravitational force is $\sim  1 \times 10^{-16} \, \text{N}$. If the particle’s initial and settling speeds are 1 m/s and $\sim$ 10 mm/s in air, respectively, the drag force is $\sim 1 \times 10^{-13} \,\text {N}$. Depending on particle materials, e.g., light absorbing, non-absorbing, or between, one of the two types of optical forces can be more dominant or even several orders of magnitude larger than the other. For a charged particle in a dusty plasma, OT forces can be an effective component of the forces that levitate the particle or drive the particle’s motion in the plasma.

Different from the optical configuration used in the initial demonstration of particle levitation, in which a single tightly focused laser beam (termed as optical tweezers) was used \cite{ref18}, in our studies, two counter-propagating hollow beams are used to form a universal optical trap (UOT) \cite{ref19, ref20}. The UOT is capable of trapping single particles with diverse properties, ranging from transparent to absorbing, from inorganic to biological, from solids to droplets, and from spherical to irregularly shaped particles, spanning size from sub-micron to approximately 50 µm in various media including air, solution, reactive gases, and plasmas. Moreover, the UOT offers high trapping rigidity and trapping efficiency, attributes particularly crucial for plasma trapping. Here, we highlight our latest results involving the trapping and transporting of single particles in plasmas using a UOT and high-precision measurements of the electric field using a single trapped particle in an RF plasma.

\begin{figure}[H]
\centering
\includegraphics[width=0.5 \textwidth]{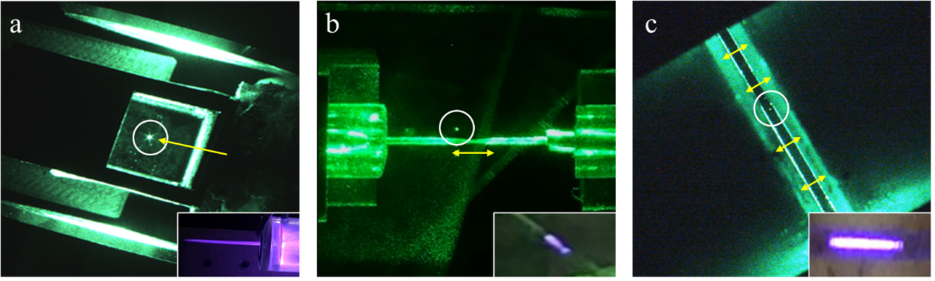} 
\caption{Optical trapping of single particles in atmospheric plasmas, (a) atmospheric argon plasma jet, (b) DC discharge in air, (c) helium DBD in air. The insets show corresponding plasma images.}
\label{fig:plasma}
\end{figure}

Figure \ref{fig:plasma} shows images of a single particle trapped in three different plasmas. Fig. \ref{fig:plasma}(a) shows a single carbon nanotube particle of 10 -20 µm in size trapped downstream of an atmospheric pressure argon plasma jet. Fig. \ref{fig:plasma}(b) illustrates a single carbon nanotube particle trapped near an atmospheric DC discharge. Fig. \ref{fig:plasma}(c) depicts single carbon nanotube particles trapped within a dielectric barrier discharge. Each trapped particle is circled, and an image of the plasma is shown in the inset. Trapping particles in atmospheric plasmas such as a plasma jet, DC discharge, or DBD is extremely challenging due to open-air turbulence and interactions with plasma gas flow (Figs.\ref{fig:plasma}(a) and \ref{fig:plasma}(c)), which pose significant hurdles in stabilizing trapped particles. Even with the compact size of these plasmas, microscope objectives are impractical for forming a UOT due to their short working distance. In these plasma trapping experiments (and others detailed below), all UOTs are created using optical lenses with a low numerical aperture. These experiments demonstrate the feasibility of trapping particles in atmospheric plasmas. Further development efforts are encouraged to improve particle trapping in atmospheric plasmas.

\begin{figure}[H]
\centering
\includegraphics[width=0.5 \textwidth]{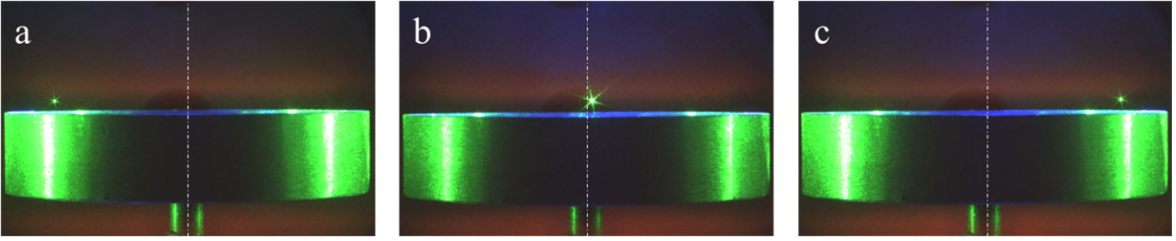} 
\caption{Side views of optical trapping of single particles in an RF argon dusty plasma in various locations, (a) the left edge of the electrode, (b) the center of the electrode, and (c) the right edge of the electrode.  The trapped particle is a 6.2 \textmu m glass sphere.}
\label{fig:trap}
\end{figure}

Figure \ref{fig:trap} shows images of a single particle trapped in an RF dusty plasma generated by a 13.56 MHz RF plasma generator (RF VII, RF-3-XIII) and confined in a metal chamber (30 cm x 30 cm x 20 cm)\cite{ref15}. The input RF power ranged from 1.0 to 10.0 W, operating at gas pressure from 2.7 Pa to 27 Pa. Fig. \ref{fig:trap} (a, b, and c) demonstrates that the particle can be trapped in various locations inside the plasma: at the left edge of the electrode (a), the center of the electrode (b), and the right edge of the electrode (c), viewed from a side perspective perpendicular to both the gravitational field and the optical axis of the UOT. Similarly, particles can be trapped at different heights above the electrode surface. The trapped particle in Fig. \ref{fig:trap} is a 6.2 µm glass sphere. The RF plasma electrode is illuminated by scattered light from the 532-nm trapping laser beam, while a blue illuminating beam is used for imaging the trapped particle in the plasma. In experiments, glass spheres of different diameters (200 nm, 2.0 µm, 6.2 µm) were also successfully trapped in the plasma.  Furthermore, strongly absorbing carbon nanotube particles, with extremely different material properties in terms of light absorbance and particle morphology, were trapped in the plasma under various operating conditions. These experiments demonstrate that micro dust particles of diverse properties (light absorbance, size, and morphology) can be optically trapped in RF plasmas. 

\begin{figure}[H]
\centering
\includegraphics[width=0.5 \textwidth]{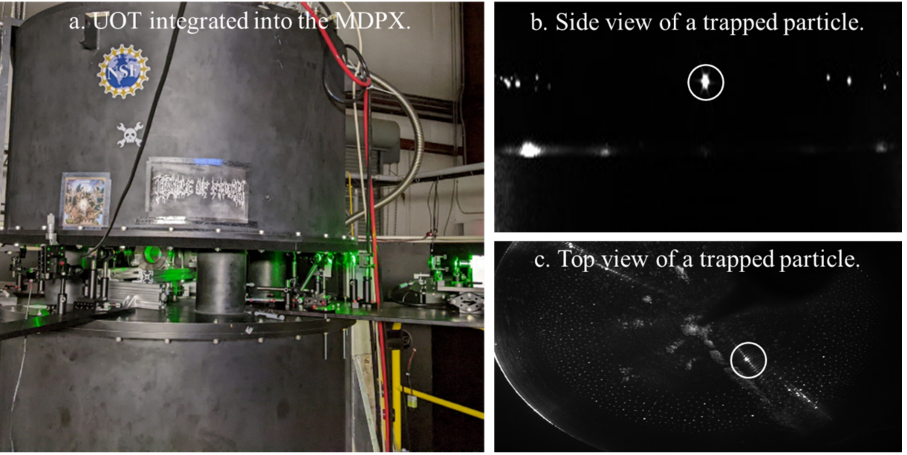} 
\caption{The first optical trapping experiment in the Auburn MDPX facility.  (a) An image of the UOT system that was integrated into the MDPX facility. The upper and lower halves of the cryostat (in black) are visible; (b) A side view of the single particle trapped in the plasma, as marked by the circle; and (c) a top view of the trapped particle in the plasma.}
\label{fig:MDPX}
\end{figure}

There are some nontrivial challenges in integrating an `optical trapping' with a dusty plasma experiment. All optical designs and configurations must be tailored to fit within the predefined space and geometry of the plasma system while ensuring sufficient optical trapping robustness and efficiency. Additionally, we must consider unavoidable mechanical vibrations and potential electromagnetic interference, which affect the optical trapping system. In our first optical trapping experiment in the MDPX system \cite{ref21}, as shown in Fig. \ref{fig:MDPX}(a), single particles were successfully trapped. Specifically, 6.2 µm glass spheres were introduced into the plasma and trapped near the geometric center of the plasma volume above the electrode surface, as shown in Fig. \ref{fig:MDPX}(b, c), in which (b) and (c) are the side and top views respectively. In addition to the optically trapped particle (circled), there are some particles trapped in the sheath, as viewed from the top. During this ten-day experiment, no attempts were made to manipulate trapped particles in the plasma, nor were measurements of plasma parameters carried out using the trapped particle. Future developments are focused on upgrading the optical configuration of the trapping system. Nevertheless, this work demonstrates the feasibility of integrating an optical trapping system into an existing plasma facility and successfully trapping single particles in the plasma.

\begin{figure}[H]
\centering
\includegraphics[width=0.5 \textwidth]{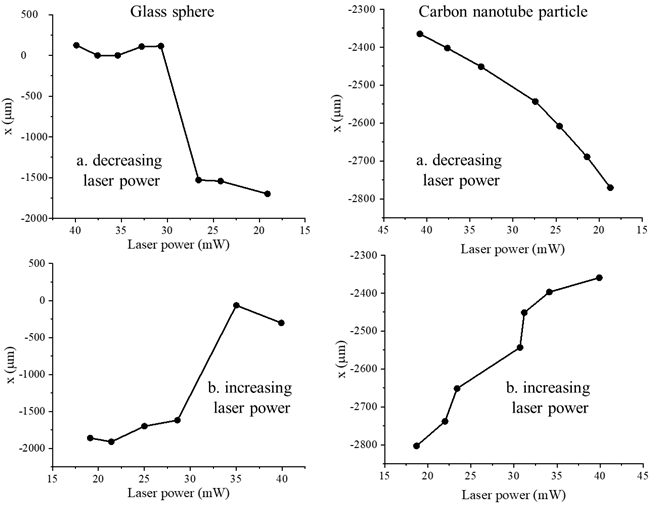} 
\caption{Transport of a single trapped particle in the plasma from one location to another. Transporting a glass sphere (left-a,b) and a carbon nanotube particle (right-a,b) back and forth in the plasma along the x-axis that is perpendicular to the gravitational field.}
\label{fig:transport}
\end{figure}

Smoothly transporting a particle in plasma is crucial for using the single particle as a probe in plasma diagnostics. In the current studies of dusty plasmas using particles levitated in the plasma, the particles cannot be levitated beyond the plasma sheath. However, we use optical trapping to demonstrate that a single particle can be trapped in any location and smoothly transported from one location to another in the whole plasma volume. Figure \ref{fig:transport} shows a single glass sphere being trapped and transported along the x-axis, originating from the center of the trap and perpendicular to the gravitational field. This particle transport was achieved by adjusting the trapping laser power to change the trapping forces. In Fig. \ref{fig:transport} (left-a), as the laser power was decreased incrementally, the trapped glass sphere moved step by step from right to left along the optical axis of the trapping beams (also along the x-axis); similarly, increasing the laser power caused the same trapped particle to move stepwise from left to right, as shown in \ref{fig:transport}(left-b). The right side of the figure shows the results of transporting a nonspherical, carbon nanotube particle back and forth along the x-axis in the plasma.  

Electric field $E$ and particle charge $Q_d$ are two crucial parameters in dusty plasma, which influence key plasma properties such as electron, ion, and neutral atom or molecule densities, electron temperature, electric potential, and sheath structure. In many cases, $E$ at a location where a charged particle resides can be determined given a known $Q$, and vice versa. When a charged particle is optically trapped within a dusty plasma, its subsequent motion is primarily governed by the electrostatic force $F_e$, gravitational force $F_g$, ion drag force $F_{id}$, and neutral drag force $F_{nd}$ when the optical trapping forces are turned off. Over the small spatial region explored by the particle, its charge was assumed to remain approximately constant. Using a charge value obtained from an identical particle under the same RF plasma conditions, we reconstructed the electric field along the particle's trajectory~\cite{refx62}. Because the sheath electric field is strongly axial, the axial electrostatic force nearly balances gravity. As a result, the particle exhibits minimal axial displacement, and the transverse direction dominates its motion, as shown in Fig.~6(a). After release, the particle travels with nearly constant axial velocity, enabling determination of the axial electric field from the balance of gravitational, electrostatic, and ion drag forces. Turning off the optical trapping forces is assumed not to perturb the particle charge. Although the particle undergoes millimeter-scale transverse motion, previous studies indicate that dust charge depends mainly on axial position and is only weakly sensitive to transverse variations~\cite{refx63}. In our measurements, axial displacement is limited to only a few micrometers, making charge variation negligible. The resulting axial electric field is on the order of $10^{3}\,\mathrm{V/m}$, consistent with typical sheath values. Figure~6(a) shows the particle trajectory in the $xz$ plane, while Fig.~6(b) presents the corresponding transverse electric field $E_x$. Two repeated measurements with similar initial conditions $(x,z)$ demonstrate good reproducibility. The extracted electric-field distributions achieve spatial resolution on the order of tens of micrometers and precision better than $1\,\mathrm{V/m}$. The particle's significantly larger displacement along the $x$-axis than along the $z$-axis confirms that $E_x$ is relatively small and nearly uniform, as illustrated in Fig.~6(b). In contrast, $E_z$ dominates the force balance, resulting in the observed minimal axial motion.

\begin{figure}[H]
\centering
\includegraphics[width=0.5 \textwidth]{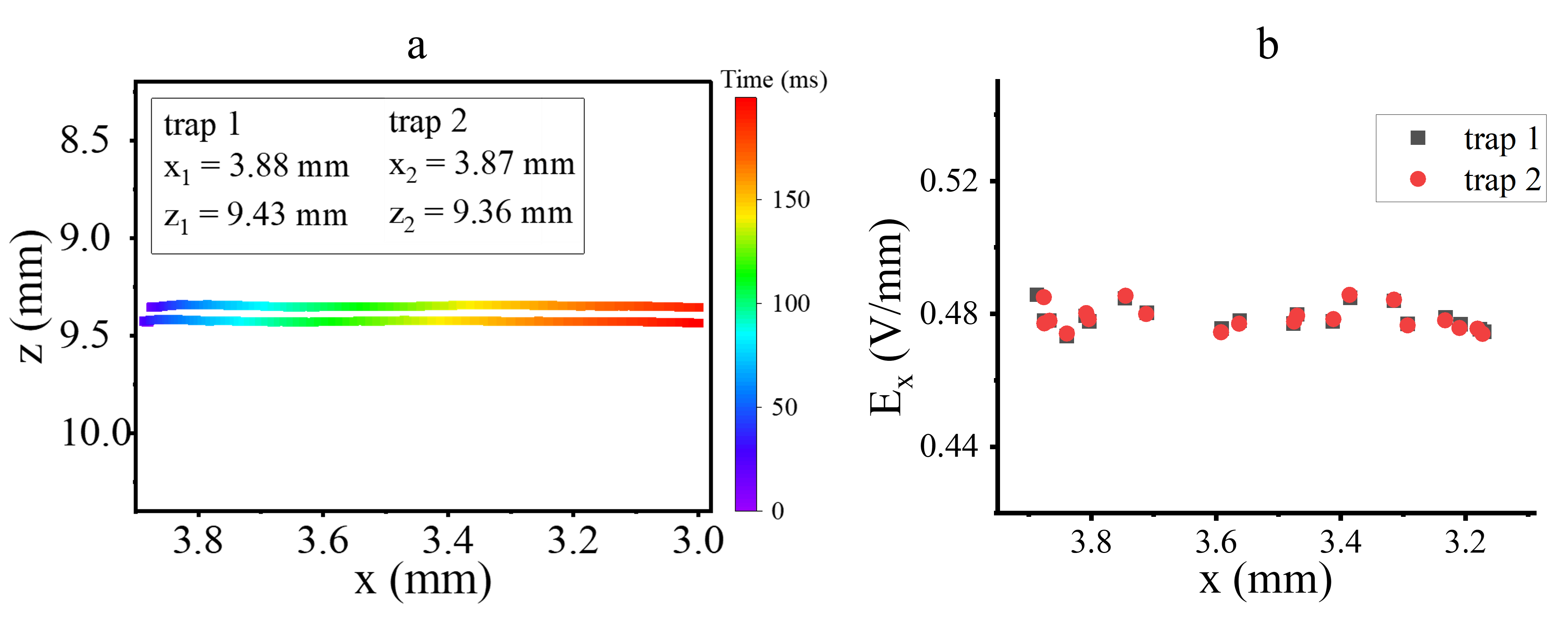} 
%\caption{High precision mapping electric field \( E \) around a localized position using a trapped particle, with a space resolution of micrometers. (a) Recorded motion trajectories of the single trapped particle when the trapping force was off, (b) measured \( E_z \) distribution along the z-axis. The inset in (b) shows that \( E_x \) is nearly the same in the range of 5000 µm.  The color and symbol denote repeated measurements.}
\caption{High precision mapping of the electric field E around a localized position using a trapped particle, with a space resolution of micrometers. (a) Recorded motion trajectories of the single trapped particle when the trapping force was off, with color indicating time progression, (b) measured Ex distribution along the x-axis. The color and symbol denote repeated measurements.}
\label{fig:E field}
\end{figure}

In summary, optical trapping and manipulation of single dust particles in plasma represent an emerging research frontier, with only a few experiments reported so far. The recent experiments demonstrated the capability to trap single particles in several types of low-temperature plasmas operating at either atmospheric or reduced pressure. Current measurements of plasma parameters using a single trapped particle involve using a known parameter to determine another, such as using a known \( Q \) to measure \( E \), and vice versa. In future studies, absolute measurements of plasma parameters, e.g., both \( E \) and \( Q \), are expected, and the study of dusty plasma dynamics using a single trapped particle as a probe is also highly conceivable. Unlike conventional plasma probes such as Langmuir probes, a single-particle probe is nonintrusive and offers high spatial resolution. By overcoming gravitational constraints, a single particle can be trapped anywhere within the entire plasma volume, extending beyond the sheath region where conventional studies using levitated particles as a probe have been limited.

\subsection{Structure dynamics}
Dust can be building blocks of `large' structures ($>$ 100 \textmu m) consisting of hundreds or more dust grains suspended simultaneously in a plasma. In the laboratory setting, one of the most well-known structures is dust crystals, or plasma crystal, or Coulomb crystals (also colloidal crystals, Coulomb solid and other names) discovered experimentally in 1994 by multiple groups~\cite{ChuI:1994, HaTa:1994, TMDG:1994}. The existence of such structures was predicted earlier theoretically~\cite{Ikez:1986}. Two-dimensional (2D) or planar dust Coulomb crystals with particle sizes ranging from 7 to 10 $\mu$m can take on hexagonal and other packing configurations~\cite{ChuI:1994, TMDG:1994}. Three-dimensional (3D) or multilayer Coulomb crystals with face-centered cubic (fcc), body-centered tetragonal
(bct), bcp structures can be formed by smaller fine particles, less than about 2 $\mu$m in diameter~\cite{HaTa:1994}.  Phase transition from Coulomb crystal to liquid has also been observed to depend on particle size and other experimental conditions, such as an external force~\cite{HaTa:1994, Thomas1996,  Lin1996}. In a binary dusty plasmas consist of two different types
of dust particles, a phenomenon called phase separation exist~\cite{SWIR:2009, JiDu:2022}. When external forces, such as gas flows, laser pressure, and ion flow exist, dusty plasma vortex-like coherent motion and other dynamic structures have been reported~\cite{ChBe:2016, STM:2023}. Additional examples of dynamic Coulomb crystals are given below in Sec.~\ref{sec:dataset}.

\begin{figure}[!htb]
    \centering
   \includegraphics[width=0.85\columnwidth]{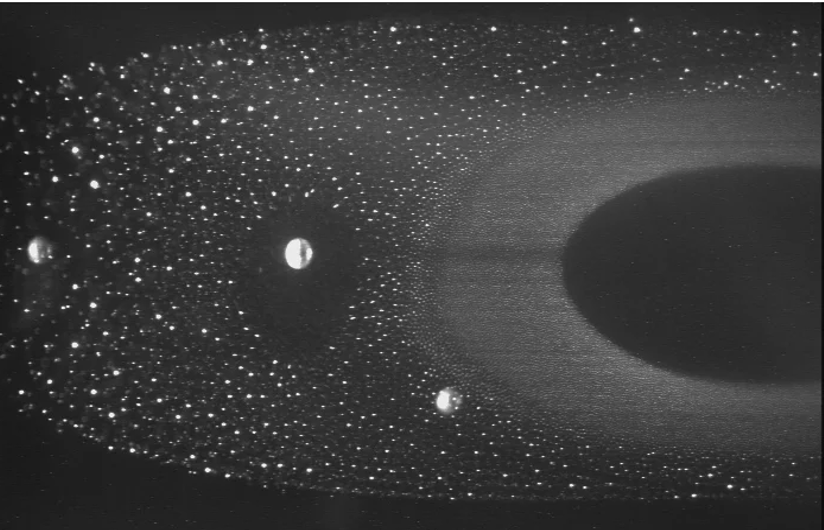}
    \caption{A picture taken during the last PK-3 Plus experiment on the International Space Station. Some ``large" particles of about 1 mm in diameter were induced into the experiment chamber by accident. Interesting interactions between the large particles with the smaller particles were observed. Image credit: The Max Planck Institute for Extraterrestrial Physics (MPE). Additional information may be found in~\cite{Schwabe_2017}.}
    \label{fig:PK3Plus}
\end{figure}

Coulomb crystals and complex plasmas have since been studied in microgravity environment~\cite{Thomas2008, TTKMK:2023}, such as the International Space Station (ISS), which eliminates the influence of gravity and allows long-duration observation of 3D dust cloud formation, crystallization, phase transitions, and other dynamics of Coulomb crystals. The spontaneous self-organization of dust particles into plasma crystals and liquid-like structures and have allowed observation of dust charging, levitation, wave propagation, statistical mechanics~\cite{ivlev2015statistical}. There are qualitative differences in plasma crystals and their dynamics in experimental series, such as PK-3 Plus, Plasma Kristall-4 (PK-4)~\cite{Pustylnik2016}.  One unique feature is the dust or microparticle free void in the center of the Coulomb crystal, as shown in Fig.~\ref{fig:PK3Plus}. The size of the void grows with the discharge power~\cite{Fortov:2005}. Other features include phase separation of complex plasma clouds formed by microparticles of different sizes, crystalline structures along
the central axis, and torus-shaped vortices in different areas away from the central axis.  These microgravity experiments have also helped validate theoretical models for grain charging, dust grain dynamics, and inter-particle forces.
The charged dust particles interactions via long-range Coulomb forces exhibit collective behavior akin to soft matter~\cite{chaudhuri2011complex}.

Dusty plasmas also exhibit a rich spectrum of wave phenomena and nonlinear dynamic behaviors. %Depending on the degree of dust-plasma coupling and charge density, 
Dusty plasmas can support (i) high-frequency plasma and electromagnetic wave modes largely unaffected by dust, (ii) modified wave modes influenced by the presence of charged dust grains, such as ion-acoustic waves~\cite{Fortov:2005}, and (iii) entirely new modes unique to dusty plasma environments, 
such as dust acoustic waves (DAWs)~\cite{RAO1990543}, which can be directly visualized using high-resolution imaging. DAWs and alike make dusty plasmas a unique platform for studying wave dynamics at the kinetic (individual particle) level. Additional examples of unique dusty plasma waves include transverse shear waves, and dust-lattice waves~\cite{ShEl:2009}.

Beyond linear wave phenomena with relatively small deviations from equilibrium, dusty plasmas also exhibit a variety of nonlinear and nonequilibrium phenomena. In strongly coupled regimes, such as two-dimensional plasma crystals, dust grains can undergo collective dynamical transitions between ordered and disordered states. These transitions are often mediated by mode coupling, where energy transfer between longitudinal and transverse acoustic or phonon modes can trigger instabilities~\cite{couedel2010direct, liu2010mode, zhdanov2009mode}. Such mode-coupling instabilities are known to be precursors to melting or reconfiguration of the crystal lattice. Recent studies have uncovered  emergent states in dusty plasmas, pointing to complex phase behavior even under weak external forcing~\cite{kryuchkov2020strange, gogia2017emergent}.
Furthermore, shocks and solitons have been observed in both laboratory and astrophysical dusty plasmas, when compressibility of dusty plasmas as a fluid need to be considered. 

We shall only briefly mention the rapidly growing datasets concerning the larger-scale structures of dusty plasmas found in nature, which are becoming increasingly accessible due to rapid growth in space exploration and astronomical observation. As human activity in space continues to expand, through multi-national space missions, commercial satellite deployments, in-situ observation and sampling of complex dusty plasma environments have become more feasible. Notable examples include noctilucent clouds in Earth’s mesosphere, which contain charged ice particles that exhibit plasma-like behavior; cometary tails, where dust and ionized gas interact with the solar wind and solar radiation; and lunar dust and asymmetric dust cloud around the Moon~\cite{HSKS:2015}, which becomes electrostatically charged due to solar wind and UV radiation exposure. Other natural dusty plasma environments of interest include near-Earth meteoroid streams and planetary ring systems such as Saturn’s rings, where dust-plasma interactions contribute to the dynamics and stability of these structures~\cite{Goer:1989, wahlund2009detection}.% Kempf et al., 2005).

Beyond the solar system, observations of dusty plasma in protostellar disks and  interstellar clouds are improving steadily, thanks to successive generations of powerful space telescopes. The Hubble Space Telescope, the Spitzer Space Telescope, the Chandra X-ray Observatory, the XMM-Newton space telescope and others provided insights into dusty regions around stars and supernova remnants. More recently, the James Webb Space Telescope has significantly advanced the resolution and sensitivity of infrared observations, allowing for detailed studies of dust composition, distribution, and interactions in distant stellar nurseries and galaxies~\cite{GBOW:2024}. These large-scale datasets are now critical in enhancing our understanding of dust dynamics in astrophysical plasmas, complementary to laboratory dusty plasma research.

In terms of data processing and understanding, correlating the wide range of observed multi-scale (both spatial and temporal) structures with their underlying physics mechanisms and elementary processes presents a significant and ongoing challenge in dusty plasma as a growingly interdisciplinary research field. %These structures are governed by complex interactions across multiple spatial and temporal scales. 
One of the primary difficulties lies in bridging the gap between microscopic processes, such as dust charging, dust motion, dust-dust interaction, dust-plasma interaction, and macroscopic phenomena, such as collective structural formations and evolution. Developing models that can accurately capture these processes requires a multi-scale, multi-physics approach. However, such models tend to be computationally intensive, often demanding substantial resources to resolve the different scales simultaneously. Moreover, our understanding of the system is frequently limited by incomplete or imprecise knowledge of initial conditions at the microscopic level, which further complicates efforts to build predictive or interpretative models.

In recent years, the application of machine learning (ML) and artificial intelligence (AI) to the analysis of laboratory dusty plasma structures has emerged as a promising approach to address some of these challenges. These data-driven techniques offer new avenues for inferring latent physical parameters and identifying patterns that may not be immediately evident through conventional analysis. For example, dusty plasma crystals can often be described using a reduced set of dimensionless parameters, such as the Coulomb coupling constant, which characterizes the degree of particle interaction. Many of these parameters are not directly measurable through experimental means, making them well-suited for inference through ML algorithms trained on simulation or experimental data. As the field evolves, integrating ML with traditional physics-based models may offer powerful tools for uncovering the complex relationships between micro-physics and macro-structures in dusty plasmas.

%
%
%
%
%\subsubsection{Other structures}

%
%
%
\section{Dusty plasma models \label{sec:model}}

As an interdisciplinary field of plasma physics and materials science~\cite{Shuk:2001,Piel:2017},  the state-of-the-art understanding about dusty plasmas, whether laboratory, industrial or larger natural systems, such as Saturn's rings or interstellar clouds, is now driven by computation~\cite{279017, ScGr:2013, MVC:2022}. Analytical or semi-analytical models~\cite{Ikez:1986,RAO1990543, Ivlev2011}, which have contributed significantly to the physics understanding and still should be encouraged when feasible, are usually insufficient to capture the large degrees of freedom in problems encountered in dusty plasmas.

Computational plasma physics is also sometimes called `numerical or computational experiments'~\cite{Taji:2004}. Computational dusty plasma physics consists of plasma simulations, dust simulations, and dust-plasma interaction simulations. As computing power, memory and algorithms continue to improve, simulations of a full plasma experiment as large as the ITER scale with high fidelities, which usually combines multiple codes to resolve and connect phenomena across scales, is now a reality. WDMApp~\cite{WDMApp}, for example, couples two advanced and highly scalable gyrokinetic codes, XGC and GENE (or GEM), where the former is a particle-in-cell code optimized for the treating the edge plasma while GENE (GEM) is a continuum code optimized for the core. In another example, Fusion Energy Reactor Models Integrator (FERMI) couples plasma (MHD) simulations with structural mechanics (first wall, blanket designs), neutronics, and fluid flows~\cite{BSSB:2023}.  In practice, many `phenomenal approaches' or reduced order models are adopted, which require isolation of important physics, simplified numerical algorithms, aiming for numerical results that can be directly compared with measurements or observations. Neural networks offer a completely new way to develop pragmatic reduced order models.%Examples are...

\subsection{Plasma models \label{mod:plas}}

\begin{figure}[!htb]
    \centering
   \includegraphics[width=0.95\columnwidth]{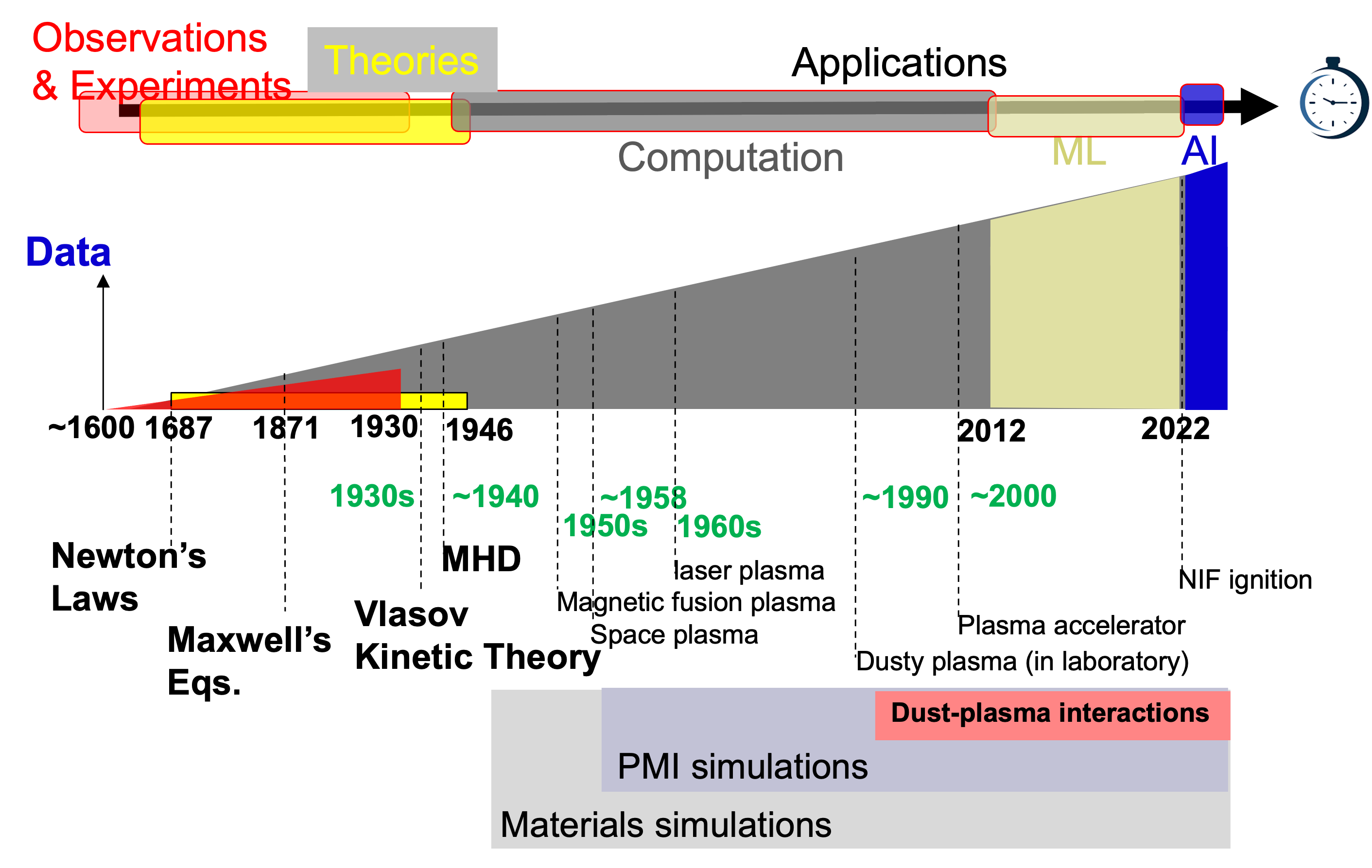}
    \caption{Dusty-plasma modeling in the context of plasma physics theory and simulations, and materials simulations, and the latest advances in data-driven science.}
    \label{fig:THOv}
\end{figure}

As a branch of physics, the basic multi-physics framework for plasmas, which includes Newton's mechanics, classical mechanics, Maxwell's theory of electromagnetism, relativity, thermodynamics, statistical physics and kinetic theory,  had been established by the 1950s for most practical purposes, as summarized in Fig.~\ref{fig:THOv}. Quantum mechanics, which has often been neglected in plasma physics, is growingly incorporated into the existing `classical plasma physics' framework to better understand plasmas under extreme conditions or when the microscopic physics details of individual electrons, atoms or molecules are needed, such as in plasma chemistry. Many advances in plasma physics require modeling of very complex systems in laboratory or nature. Examples, as highlighted in Fig.~\ref{fig:THOv}, are magnetic fusion plasmas, space plasmas, laser plasmas, plasma accelerators, and laser fusion. One of the latest accomplishments happened at the National Ignition Facility (NIF), where nuclear fusion ignition was demonstrated in 2022. The `devils' of plasma modeling have been hiding in the `{\it computational details or high fidelities}' to this day. Whole device modeling of fusion plasmas, such as WDMApp, has recently become possible due to exascale computing~\cite{WDMApp}. Advances in machine learning and neural networks in the 2010s, mostly due to advances outside plasma physics, is a growingly adopted in plasma physics to further enhance the traditional high-performance computing. 

\begin{figure}[!htb]
    \centering
   \includegraphics[width=0.95\columnwidth]{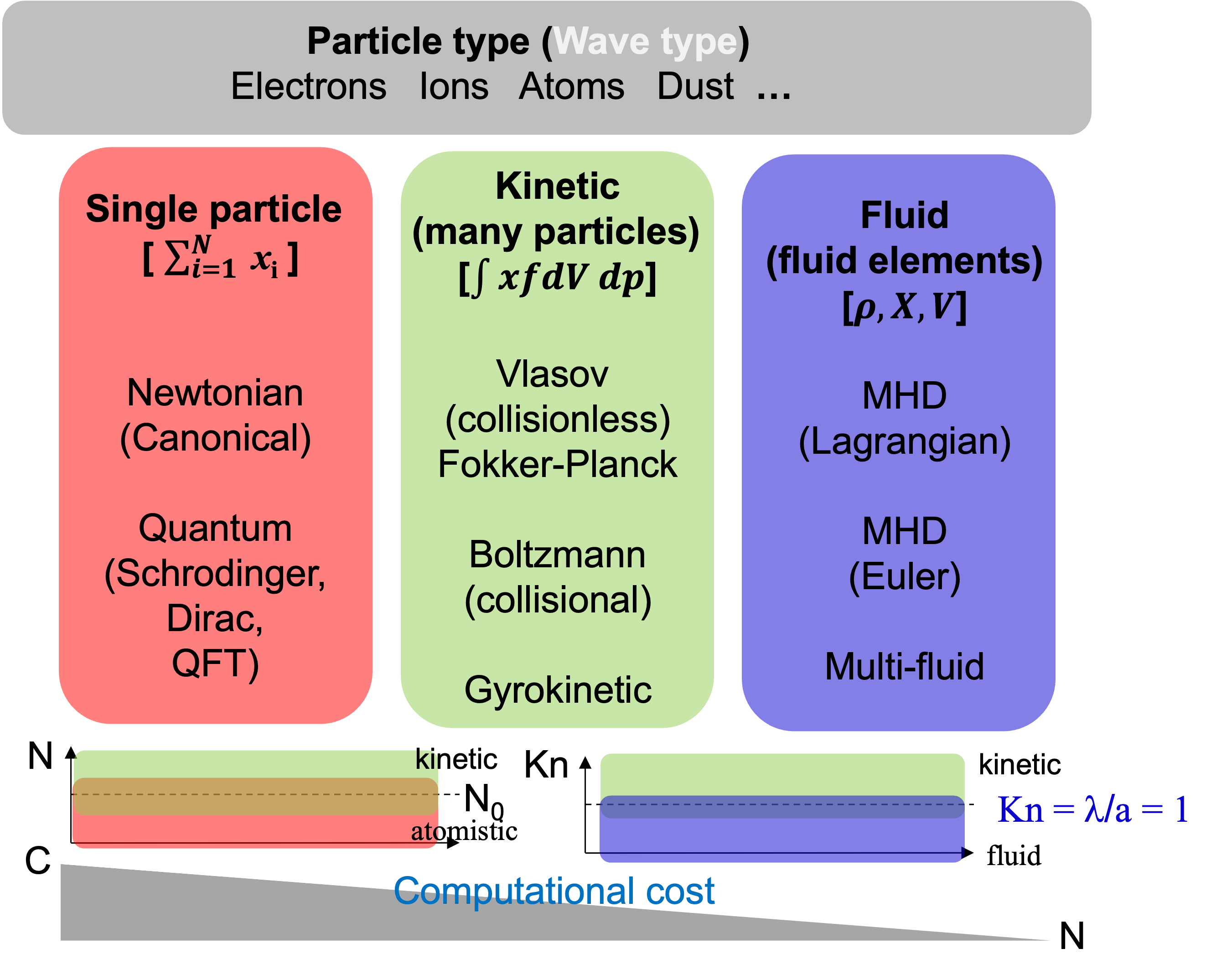}
    \caption{Three main different approaches to plasma and materials modeling depending on the number of atoms (or electrons, ions, dust grains) involved, $N$, and collisional properties characterized by the Knudsen number $Kn$. See text for more details on model selection.}  %d: Eulerian and Lagrangian codes in fluid mechanics.}
    \label{fig:MScale}
\end{figure}

%Dusty plasma modeling includes models for different particle species and spans a wide range of time and length scales. 
In the context of dusty plasmas, one may divide plasma models into reactive and non-reactive plasma models. The main difference between a reactive and a non-reactive plasma lies in the significance of plasma chemistry~\cite{Kush:1988, Holl:2000,Frid:2008}. In a non-reactive dusty plasma, plasma chemistry may be ignored, and dust mass, geometry, and chemical composition stay constant~\cite{PPZS:2021}. We may call such models as `{\it constant dust}' models. Modeling plasma chemistry can capture the material dynamics such as  dust growth and destruction, meanwhile, it requires a large number of chemical species, radicals, chemical reaction cross sections and reaction rates.  %A generic dusty plasma model includes the models for the dust grains, plasmas, and dust-plasma interactions. Besides dust particles, there are electrons, ions, and usually neutral atoms. 
In industrial plasma settings~\cite{Frid:2008}, including high-temperature plasma boundaries, modeling plasma chemistry is almost always necessary. Open source software, such as OpenFOAM~\cite{OFoam:2025}, together with commercial software, such as VSimPlasma~\cite{TechX:2025} and COMSOL~\cite{COMSOL:2025}, is available for the complicated physics and chemistry, and for integrated multi-disciplinary system modeling and optimization~\cite{MaNi:2021}. 
 In many laboratory low-temperature plasmas, non-reactive plasma assumption may be adopted for simplified computation of Newtonian dynamics and structure dynamics, while ignoring materials dynamics and plasma chemistry. For each kind of particle (electron, ions, atoms, dust grains),  atomistic, kinetic, or fluid models may be used, depending on the number of particles ($N$) involved and the collisional properties of the particle with the rest of system, as shown in Fig.~\ref{fig:MScale}. At a more fundamental level, the atomistic or first-principle models simulate individual particle motion and their interactions with others, such as particle-particle interactions, particle - electromagnetic field interactions, and particle-wall interactions. The computational cost for the best atomistic codes scales polynomially with $N$ as $O(N)$, which may not be practical when $N> 10^{12}$ assuming Newtonian dynamics (also other parameters, such as time-step size, duration of the simulation, memory, and communications) on the state-of-the-art supercomputers, such as the Frontier supercomputer in the Oak Ridge National Laboratory~\cite{Fron:2024}. 
 
 Besides computational cost, the selection of atomistic, kinetic or statistical, or fluid or continuum model for a specific type of particle also depends collisional properties of the particle. The collisional properties may be measured by the Knudsen number $Kn$, the ratio of the collisional mean free path $\lambda$ to the characteristic length $a$. $Kn$ for ions interaction with a dust particle with a radius $r_d$ may be defined as $Kn_d = \lambda/r_d$, while $Kn_p = \lambda/L$ for the plasma as whole, with $L$ being the plasma size and $r_d \ll L$ is very common in laboratory and nature. In other words, while ions may be modeled as fluid when the problem is to quantitatively calculate the global plasma properties such as transport properties or velocity field, ion-dust interaction may need kinetic models if $Kn_d > 1$.

Attempts at balancing pragmatic computationally accessible model for a dusty plasma with scientific and logical rigor will necessarily run into the  famous Hilbert's sixth problem~\cite{Slem:2015,Gorb:2018,DeHM:2025}, which aimed at establishing a rigorous mathematical framework for all the laws of physics (which, of course, includes dusty plasma physics and computation), and axiomatizing the laws of physics like what has been done to geometry. From atomistic models to kinetic models, such as the Boltzmann equation, for example, the Bogoliubov–Born–Green–Kirkwood–Yvon (BBGKY) equations hierarchy reduces the comprehensive $N$-particle description in phase space to the single-particle distribution as in the Boltzmann equation~\cite{Harr:1971}. Another reduction is from the kinetic equation to the fluid equations, as in Fig.~\ref{fig:MScale}, through the Chapman-Enskog method for charge-neutral gas. Equivalent approaches to the Chapman-Enskog method, often known as solving the closure problem, remain an active field of research in and outside plasma physics~\cite{KoKa:2025}.

As examples of `atomistic' or `$N$-particle' models, particle-in-cell (PIC) in combination with Monte Carlo collision (MCC), or `PIC-MCC' method, is widely used in simulations of partially ionized gases~\cite{birdsall_particle--cell_1991,Vasu:1995,ABGT:2018,refx63}. The number of particles $N$ that can be simulated continues to increase from 10$^3$ in the 1970s to above 10$^6$~\cite{BiLa:2004,Verb:2005}. In PIC methods, real particles (like electrons and ions) are represented by computational particles, also called macro-particles or super-particles, each representing a large number of real particles (typically 10$^5$–10$^8$)~\cite{Hara_2019}. A weight factor is used to relate the density of computational particles to the physical
density of the electrons or ions~\cite{HoEa:1988}. Axisymmetric cylindrical or 2D PIC simulations have been applied to complex boundary conditions that match those found in an experimental discharge chamber, such as a GEC cell, or to DC discharge tubes, such as the PK-4 experiment onboard the International Space Station \cite{donko_particle_2011,hartmann_ionization_2020}. The MCC technique is based on elementary collisions of electrons, ions and fast neutral atoms with the thermal background gas and is able to describe charged particle and energy transport at the kinetic level for individual particles. The PIC-MCC method provides a fully self-consistent, kinetic level description of the gas discharge along with space and time resolved properties (including electric potential, densities, fluxes, energy distributions, surface processes) across the entire discharge including the walls of the discharge. 
A hybrid-PIC-MCC (h-PIC-MCC) algorithm was described to study the effect of dust charge fluctuation on ion acoustic waves in a dusty plasma~\cite{ChBo:2020}.

Molecular dynamics (MD) are often used to simulate $N$-atom/molecule/ion/dust (microparticle) dynamics. The key idea behind a MD simulation is straightforward by using Newton's laws of motion to trace particle motion~\cite{NeBr:2017,HoDr:2018}. 
While PIC simulations track computational particles and their interactions with fields, MD simulations treat directly real particles and their interactions, including interatomic potentials such as the Lennard-Jones potential, Coulomb forces, or bonded force models. MD simulations do not require a spatial grid for force calculations unless coupled with external fields; forces are computed directly from the particle positions~\cite{FrSm:2002}.
One approach is to adopt existing MD codes for dusty plasmas, {\it e.g.} the microparticles were modeled using the classical MD package
Large-scale Atomic/Molecular Massively Parallel Simulator (LAMMPS)~\cite{ScGr:2013}.  Another approach is to develop customized MD codes for specific dusty plasma problems, {\it e.g}, molecular dynamics simulation of ion flows around microparticles was implemented in a code called molecular asymmetric dynamics (MAD) and its variants~\cite{piel_molecular_2017,piel_molecular_2018,piel_molecular_2018-1}. In another example, Dynamic Response of Ions and Dust (DRIAD) is a MD code for ions and dust~\cite{matthews_dust_2020}, where the dust and ion motion are resolved at different time steps, with $\Delta t =$ 10$^{-4}$ s for dust and 10$^{-9}$ s for ions.

Besides particle-based methods, such as PIC and MC codes, for plasma kinetics~\cite{ABBL:2015}, the other choice is direct computation of distribution functions, which appear naturally in the Boltzmann equation and other kinetic theory~\cite{White_2009}. The hierarchy of collisional frequencies in a dusty plasma (relative to the characteristic time scale of the phenomena of interest, such as dust charging, wave growth, structure formation as described in Sec.~\ref{sec:pheno} above), including electron-electron collisional frequency ($\nu_{ee}$), electron-ion collisional frequency ($\nu_{ei}$), ion-ion collisional frequency ($\nu_{ii}$), electron-neutral collisional frequency ($\nu_{en}$), ion-neutral collisional frequency ($\nu_{in}$), electron-dust collisional frequency ($\nu_{ed}$), ion-dust collisional frequency ($\nu_{id}$), {\it etc.} determines whether a kinetic model is necessary, or when the electron or ion distribution functions are non-Maxwellian and have not reached a global or local thermal equilibrium (LTE)~\cite{Vasu:1995,TsAn:1999}.  We only list a few examples of kinetic models here since PIC-MCC approach appears to be more popular. A kinetic model for both electrons and ions was used to calculate the particle potentials~\cite{FZPS:2007}. The kinetic equation for ions, together with an analytical model for electron distribution, was used in to calculate the electric potential around an absorbing particle~\cite{KhKM:2008}.  

Fluid models of dusty plasmas, which may be regarded as reduced kinetic models, describe the transport of
electrons, ions and possibly other particle species, including dust particles, by
the first few moments of the Boltzmann equation~\cite{HaPi:2005}. Fluid models require the input of transport coefficients and
rate coefficients that depend on the electron, ion energy or phase space distribution function~\cite{PPZS:2021}. Fluid models became more sophisticated by incorporating realistic conditions such as dust charging, ion-neutral collisions, and variable charge distributions. This increased the computational demands significantly, as these models involved solving coupled nonlinear fluid equations to capture the macroscopic behavior of the plasma. The existence of energetic or non-thermal electron (or ion) tails that can excite and ionize gas sometimes motivate `hybrid' simulation approach, when low-energy electrons are modeled with fluid, and energetic electrons modeled with a distribution function or a MC particle model~\cite{DiKB:2009}. Another motivation for hybrid approach is due to the coupling of multi-scale multi-phase physics in dusty plasmas. Addition of magnetic fields to dusty plasma also inject new time and temporal scales into dusty plasmas. 

\subsection{Dust models \label{mod:dust}} 

Models for dust or microparticles in plasmas can broadly be divided into two classes: single-particle (or single-dust) models and collective (or multiple-dust) models. Single-dust models focus on the behavior of an individual dust grain interacting with the surrounding plasma, such as dust charging, dust motion, forces on dust, dust transport, dust growth and destruction. Dust charging has been extensively analyzed using collisionless sheath models, such as  Orbital Motion Limited (OML) theory~ \cite{allen_probe_1992, laframboise_probe_1973}, and collisional sheath models~\cite{WhNM:1985,ShGo:1991,deAn:1992}. Dust motion, influenced by forces such as electric fields, ion drag, gravity, and thermophoresis, is another well studied aspect through both experiments and computation. Dust charging and motion together are necessary to understand dust confinement, levitation, and transport in plasmas. However, dust growth and destruction, which involves accretion or removal of material through plasma deposition or agglomeration, remains relatively under explored, despite its importance in astrophysical plasmas, industrial plasma processing, and high-temperature plasmas such as nuclear fusion. Collective dust models treat a large number of dust grains together, capturing emergent phenomena such as dust acoustic waves, self-organization, crystallization, and collective charging effects. Both classes of models often require coupling of dust physics with different plasma models as discussed in Sec.~\ref{mod:plas}, equations for electromagnetic fields, and may also include interactions with neutral gases, UV and other types of radiation. %Together, these models provide complementary insights into diverse dusty plasma dynamics across scales.

The amount of charge on a grain of dust ($Q_d$) may be estimated by the electric potential of the grain, $Q_d = C_d V_d$, where $V_d$ is the dust potential and $C_d$ the dust capacitance. The simplest model for the dust capacitance is $C_d = 4 \pi \epsilon_0 r_d$ for a spherical conducting grain with a radius $r_d$. $\epsilon_0 $ is the vacuum permittivity. $V_d$ is usually a few times $-k_BT_e/e$ for the electron temperature $T_e$~\cite{deAn:1992}. The negative sign is due to the higher electron mobility than ion mobilities in many plasmas. The dust potential reaches a steady state, $dQ_d/dt = 0$, when there is a balance of different charging currents, $\sum_k J_k =0$.  In laboratory plasmas, one of the simplest cases is when only ion and electron current are important and balance each other~\cite{allen_probe_1992}. In space plasmas, the most important charging mechanisms are plasma (electron and ion) current, photoemission, and secondary electron emission~\cite{Whip:1981}.

\begin{figure}[htb]
    \centering
    \includegraphics[width=0.5\linewidth]{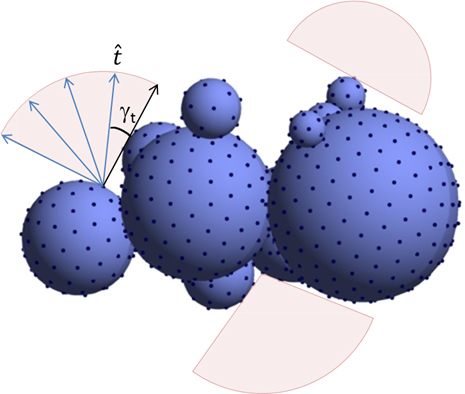}
    \caption{A 2D representation of the open lines of sight to three points on the surface of an aggregate.}
    \label{fig:LOS_illustration}
\end{figure}

In reality, dust charging remains an open problem for practical reasons and there is no `universal formula' that can be used to calculate grain charge accurately for all situations. Besides unique material and structural properties, such as dust electric conductivity (dielectric or conducting), surface conditions such as absorption~\cite{BrFe:2017,MeBG:2024}, material phases (liquid or solid), more complicated dust shapes other than an ideal spherical or cylindrical geometry~\cite{MaSc:2006,IMMS:2008}, plasma parameters and properties, such as electron density, Debye length, ion physics, kinetic effects, dust charging is best dealt with numerical methods. First-principle methods, such as Quantum MD simulations~\cite{DuBr:2018}, involve solution of Schr\"odinger's equation but are not often used in dusty plasmas. The time characteristics of dust charging, such as the relaxation time to reach the steady grain charge
and the charge fluctuations of grains of different sizes~\cite{MVC:2022}, also needs to be analyzed numerically on a case-by-case basis. As an example, Fig.~\ref{fig:LOS_illustration}, % In charge fluctuations.
by employing OML theory with a line-of-sight (LOS) modification, the charging currents to the surface of an aggregate grain can be calculated \cite{matthews_formation_2007, matthews_charging_2007,matthews_charging_2012}. The code is also adapted to model the charging of aggregates or individual dust grains which are in close proximity by removing any charging currents which are blocked by a nearby grain. 

In the OML\textunderscore LOS method, the surface of an aggregate grain is divided into many patches and open lines of sight to points at the center of each patch are determined, allowing a LOS factor to be calculated for each point (see pink shaded areas in Fig.~\ref{fig:LOS_illustration}). Plasma currents to points on the surface of a dust aggregate are assumed to only come from directions with open lines of sight \cite{matthews_charging_2008}.  Plasma current densities, $J_e (\phi)$ and $J_i (\phi)$, are then calculated to each point as a function of the LOS\textunderscore factor and electrostatic potential $\phi$ due to the charge on each patch, $Q_p$. The charge collected on each patch of area $A_p$ during time interval $\Delta t$ is calculated as $Q_p=\Sigma J_{\alpha} A_p \Delta t$, where the sum is over the charging currents $\alpha$. The charge is assumed to stick at the point of contact for a dielectric material, as shown in experiments \cite{goree_particulate_1992}. The charging currents are then updated based on the new grain potential and the process is iterated in time until equilibrium is reached.% \ref{fig:charged_aggregate}. 
The plasma distribution function can be adjusted to reflect the characteristics of laboratory (Maxwellian) or space plasmas (Lorentzian) \cite{matthews_charging_2012}.  In addition, charging currents other than the primary electron and ion currents, such as the photo-ionization current or secondary electron emission, can be added \cite{ma_charging_2013}.

In many plasma environments, the densities of the ions and electrons are small enough, or when the dusty density is sufficiently high, so that stochastic characteristics of the charging process must be taken into account~\cite{MVC:2022}. Dust particles of size 1-10 $\mu$m are strongly influenced by stochastic fluctuations in charge, with the deviations being relatively large compared to the mean charge \cite{shotorban_stochastic_2012, shotorban_nonstationary_2011, matthews_discrete_2018,matthews_cosmic_2013}.  Smaller grains ($a < 1$  $\mu$m) or grains in a tenuous plasma environment are sensitive to single additions of electrons or ions.  A discrete stochastic method (DSM) is used to allow for integer increments of fluctuations of elementary charges collected on the surface of particles. In the DSM, a random time at which a charged particle is added to a particular patch on the aggregate surface is determined as a function of the charging currents to the patches. 

In a laboratory setting, atoms, small molecules and nanoparticles are easily levitated at room temperature, since the kinetic energy corresponds to thermal motion $\displaystyle{ k_BT \sim \frac{1}{2} mv^2 > \max \{V_g, V_1, \cdots\} }$, where $V_g = mgh$ is the gravitational energy, $V_1$ and others may include Van der Waals potential energy, {\it etc.} Van der Waals potential for individual atoms can be approximated by a Lennard-Jones potential of the form, $\displaystyle{ U(r) = 4 \epsilon \left[ (\frac{\sigma}{r})^{12} - (\frac{\sigma}{r})^{6} \right] }$,  typically a few meV. The gravitational potential for a hydrogen is about 103 neV per meter at sea level. For an argon atom, 4 \textmu eV per meter since its mass is about 40 times the hydrogen mass.  Addition of optical, electric, magnetic and other forces leads to many schemes of levitodynamics~\cite{levito:2021}.  While dust levitation in the air is a common experience, dust levitation in plasmas depends on the force balance of different origins, similar to levitation of atoms, molecules and nanoparticles. 

Forces on individual dust particles may be classified using different schemes. At the most fundamental level, the dust motion is determined by gravitational and electromagnetic forces. The fundamental forces of the other two kinds, the strong and weak force, because of their short range ({\it i.e.} sub-atomic), do not come into play in most energy regimes of dusty plasmas ({\it i.e.} non-relativistic for dust motion). Ref.~\cite{Fortov:2005} proposed to classify
the forces into two groups: the first one includes electric charge ($Q_d$) independent forces, such as gravity, neutral drag, and thermophoretic
forces. The second group of forces are $Q_d$ -- dependent or electromagnetic in nature, such as electron drag, ion drag, and
electrostatic forces. We shall mention that triboelectric effects do exist even for charge-neutral fluids such as the air. The calculation of the ion drag force, or the momentum exchange between the dust and ions, can be particularly complicated, and may require sophisticated ion and electron models as outlined in Sec.~\ref{mod:plas}. The collisionality or the Knudsen number determines whether OML-like model can be used. More generally, because of the ion flow and non-Maxwellian distribution of ions, a kinetic theory is necessary for self-consistent calculations of ion drag force~\cite{IZKM:2005}. 

\begin{figure}[!htb]
    \centering
   \includegraphics[width=0.95\columnwidth]{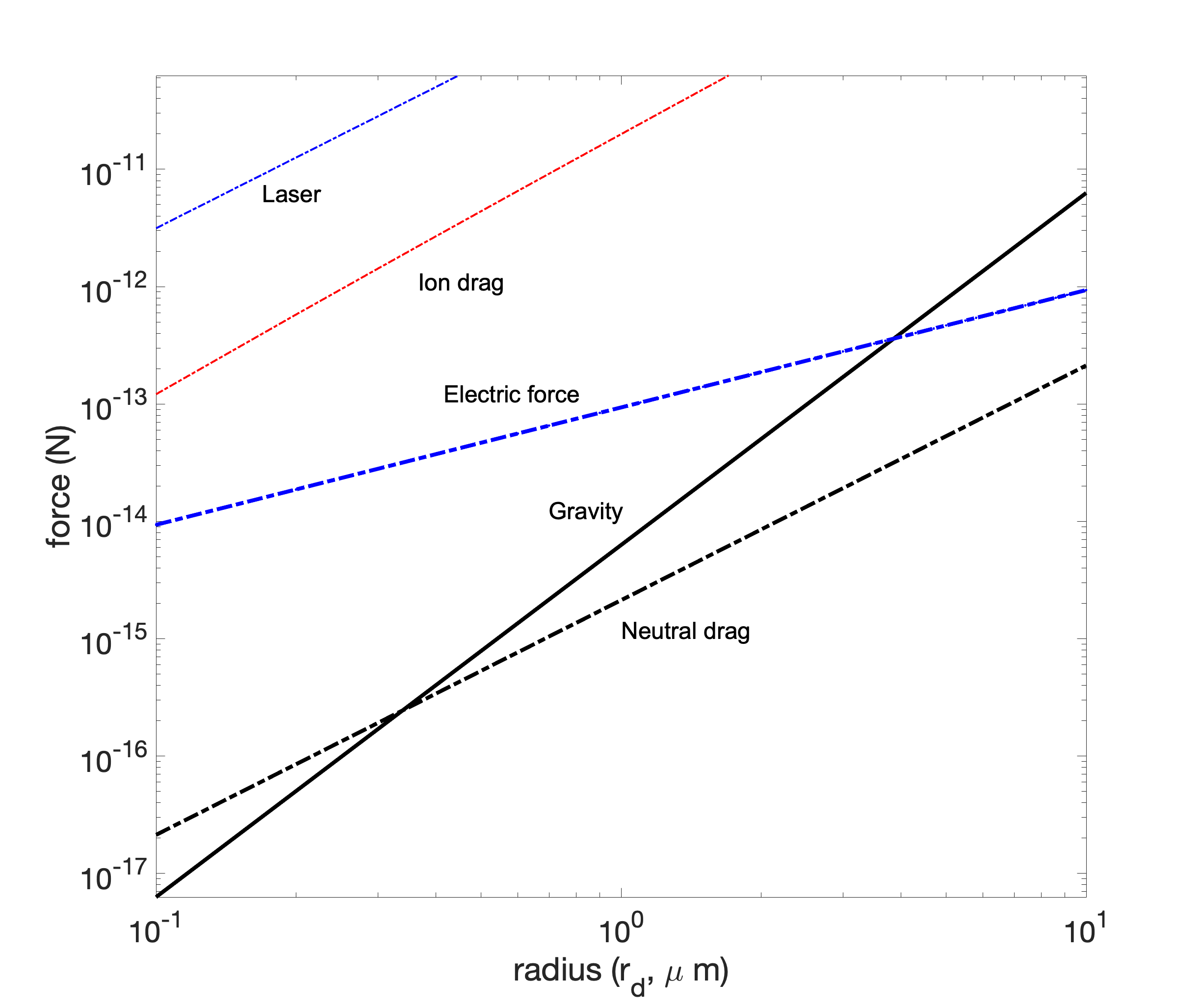}
    \caption{An example of different kinds of forces on a dust grain and their magnitude as a function of dust size. The dust mass density is assumed to be 1.5 g/cm$^3$.}
    \label{fig:Force}
\end{figure}

Another force classification scheme is based on the force scaling with dust dimensions, as summarized in Fig.~\ref{fig:Force}: body-type forces ($\propto r_d^3 \sim V_d$, an example is gravity), surface-type forces ($\propto r_d^2 \sim S_d$, examples include neutral drag force, ion drag force, and laser pressure force) or size-type forces ($\propto r_d \sim l_d$, an example is electric force)~\cite{Fortov:2005,Piel:2017}. New classes of effective forces, with a size-scaling of $r_d^\gamma$ and the exponent $\gamma$ being a non-integer, may also arise. Examples may include dipole-dipole interactions and wakefield effects~\cite{Shuk:2001},
Coulomb forces modified by the surrounding plasma~\cite{melzer2008fundamentals, morfill2009complex}. As a result, the effective forces between particles
can be non-reciprocal and break energy conservation~\cite{melzer2019finite, nikolaev2021nonhomogeneity, kolotinskii2021effect, vaulina2015energy, ivlev2015statistical}.

\begin{figure}
    \centering
    \includegraphics[width=0.95\linewidth]
    {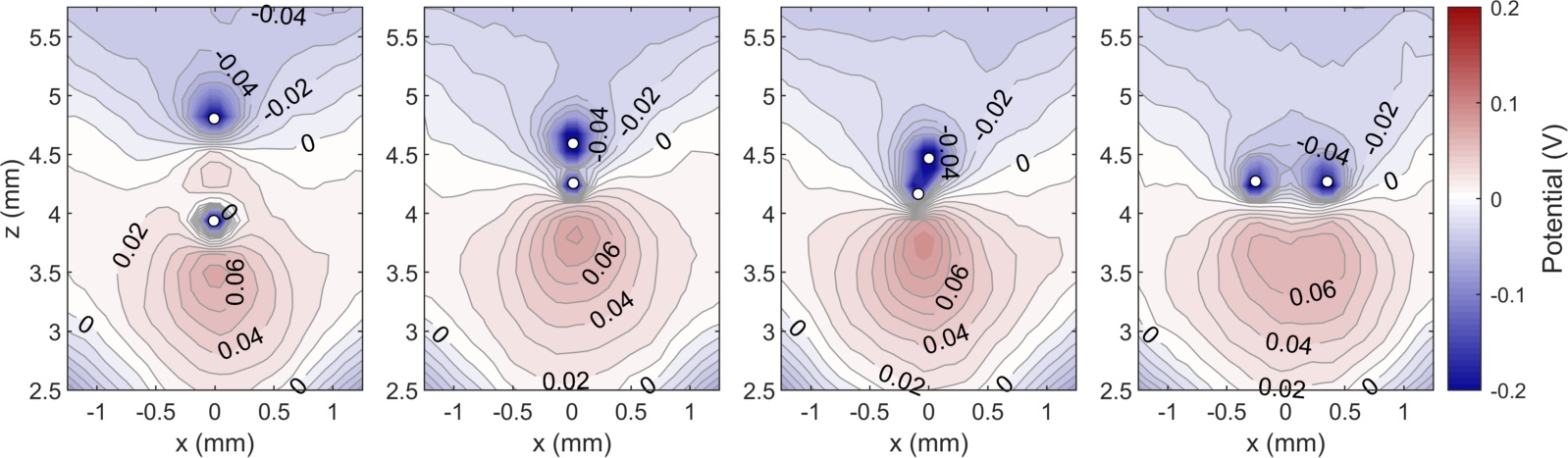}
    \caption{Non-linear combination of ion wake potentials for two particles interacting in a plasma with ion drift speed $v_d=0.4$M. Figure after \cite{matthews_dust_2020}.}
    \label{fig:ion_wakes}
\end{figure}

Different dust motion in a plasma can be coupled directly through the long-range Coulomb force in-between them and indirectly through their interactions with the plasma. When the ratio of the electrostatic energy of the
charged dust grains to the thermal energy, or the coupling parameter $\Gamma$, exceeds $\sim$170, crystallization of dust particles occurs~\cite{Ikez:1986}. Beyond one-component plasmas, $\kappa = a_{WS}/\lambda_D$ ratio may be used~\cite{HaFD:1997}, with $a_{WS}$ being the Wigner-Seitz radius and $\lambda_D$ the Debye length. $\Gamma \sim 170$ corresponds to an infinite $\lambda_D$, or $\kappa$ = 0. Meanwhile, the degree that the plasma charge depletion by many dust particles may be measured by Havnes parameter, $4\pi \epsilon_0 r_dn_dk_BT_e/n_e e^2 \sim n_d Q_d/n_e e $~\cite{HaAM:1990}. Only when Havnes parameter is significantly less than 1, the charge depletion effects of the dust grains may be neglected. Different mechanisms in plasma, such as Debye screening, ion flows, plasma wakes~\cite{couedel2010direct}, exist that can modify and weaken the dust-dust coupling and lead to melting of dust crystals~\cite{Feng2010}, different kinds of dusty plasma waves, and emergent phenomena such as super diffusion and non-Gaussian statistics~\cite{LiGo:2008}. Example of numerical approaches to examine dust-dust coupling include MD simulations~\cite{AKGK:2024}, Brownian dynamics simulations based on the Langevin equations for 20 thousand interacting
Brownian dust particles~\cite{HoPS:2009}, and hybrid particle-PIC-fluid simulations of dust-acoustic waves~\cite{WiMR:1999}. Figure \ref{fig:ion_wakes} illustrates the non-linear interaction of the ion wakes downstream of two particles as they move relative to one another. The charging and charge fluctuations of a dust grain as it moves through the wake of another grain can also be tracked.  

\subsection{Models for dust-plasma interactions \label{mod:int}}

To understand a dusty plasma as a whole, which is necessary to understand a wide array of nonequilibrium dynamical behavior in dusty plasmas, for example, mode-coupling \cite{couedel2010direct,liu2010mode,zhdanov2009mode} in large Coulomb crystals and dynamical transitions between different energy states \cite{kryuchkov2020strange,gogia2017emergent}, requires the integration of  plasma models with dust models as described in Sec.~\ref{mod:plas} and Sec.~\ref{mod:dust} above. Integrated modeling and computation are inherently multiscale, involving physics of electrons, ions, neutral gas atoms or molecules, and dust particles that spans a wide range of length and time scales. Separation of spatial or temporal scales for different particles and scale dynamic range reduction~\cite{TaCh:2005}, if feasible, can significantly simplify modeling complexity without compromising the physical fidelity required for specific applications. For example, to understand the transport of dust in laboratory experiments~\cite{land_experimental_2009} and void formation and void closure  in microgravity,  electrons, ions, and dust may all be modeled as fluids~\cite{land_fluid_2010}. The model solves the plasma equations on sub-RF time-scales and uses these to calculate the charge on the dust fluid in the model, by solving the orbital motion limited electron and ion currents \cite{Douglass_dust_2011}. Forces that act on the dust fluid immersed in the plasma include the electrostatic force, thermophoresis and the ion drag force\cite{land_probing_2010}. For calculation of the latter, the effects of non-linear ion scattering, significant ion drift and ion collisionality are taken into account \cite{land_effect_2006}. By assuming balance with the neutral drag, the dust fluid velocity is obtained and integrated in time (on much larger time-steps). %The model is run until both plasma and dust fluid convergence criteria are met, at which point the final equilibrium solution for the dust and plasma parameters are obtained. 

In systems with flowing ions, the interactions between grains are non-reciprocal due to the ion wake field.  When two dust particles are in close proximity, the resulting wakefield is not simply a linear combination of two separate wakes \cite{miloch_charging_2010}. To complicate matters further, the downstream grain becomes decharged as it enters the wake of the upstream grain, changing the strength of the ion focus.  Thus, wake formation is a dynamic process requiring numerical simulations in order to fully understand the complex behavior. %CASPER researchers have developed a molecular dynamics simulation which self-consistently models the dynamics and charging of dust grains immersed in an ion flow \cite{matthews_dust_2020}. 
The molecular dynamics simulation, DRIAD (Dynamic Response of Ions and Dust), is designed to resolve the motion of both the ions and the dust grains on their individual timescales, while allowing the dust charge to vary in response to the changing ion density in the ion wake field \cite{matthews_dust_2020}. %The DRIAD model employs super-ions with the same charge-to-mass ratio (and therefore dynamical behavior) as a single ion and takes advantage of high-performance GPU processors, greatly improving the overall time efficiency and allowing for larger simulation volumes.
Following the method described by Piel in developing the MAD (Molecular Asymmetric Dynamics) code \cite{piel_molecular_2017}, the forces between pairs of ions and among ions and dust particles are treated in an asymmetric manner. The electric field  $\vec{E} = -\nabla \Phi$ is assumed to arise from the contributions due to the charged dust, ions, and thermal electrons (with temperature $T_e$), which are treated as a fluid governed by the Boltzmann factor  $\displaystyle exp(e \phi /k_BT_e)$, %$\exp⁡(e\phi/k_B T_e)$. 
Ion-ion interactions are treated as Yukawa interactions, 
\begin{equation}
    \Phi_Y(r_i)  = \frac{1}{4\pi \epsilon_0} \sum_{j} \frac{q_j}{r_{ij}} exp(-r_{ij}/\lambda_{De}).
\end{equation}
					
The region near the negative dust grains is depleted of electrons, and thus the force on the ions from the dust is calculated assuming a Coulomb potential,
\begin{equation}
\Phi_c(r_i)  = \frac{1}{4\pi \epsilon_0} \sum_{j} \frac{q_d}{r_{id}}.
\end{equation}

However, there is an asymmetry in the system in that the force acting on the dust particles is calculated from the shielded ion charges (Yukawa potential). This combination allows the nonlinear shielding of the dust grains by the ions to be properly addressed, albeit at the expense of an approximate treatment of the electron shielding. Nevertheless, it provides good agreement with previous results from PIC codes used to model the ion wakefield \cite{hutchinson_ion_2002}, \cite{hutchinson_ion_2005}, \cite{hutchinson_collisionless_2006} [35], [36], [37], and the method has been adapted to model dust charging for irregular structures \cite{asnaz_charging_2018}[38]  and ion wakes in the presence of magnetic fields \cite{piel_molecular_2018}, \cite{piel_molecular_2018-1} [39], [40].
Accordingly, the equation of motion for an ion with charge $q_i$ and mass $m_i$ is 
\begin{equation}
m_i \ddot{x} = -q_i E_i+ F_{in}
\end{equation}
where the electric field $E_i$ consists of contributions from the other ions in the simulation, the charged dust particles, and any electric fields present in the plasma (i.e., the electric field in the sheath of a rf discharge); $F_{in}$ is the ion-neutral collision force. The ion-neutral collisions are implemented using the null collision method \cite{donko_particle_2011}[8], with Ar-Ar$^+$ collision cross sections taken from the Phelps database (LxCat project) \cite{phelps_application_1994} [41].
Dust dynamics are governed by the equation of motion 
\begin{equation}
m_d \ddot{x} = F_{dd}+F_{id}+m_d g+Q_d E-\beta \dot{x} + \zeta r(t).  	
\end{equation}
where $F_{dd}$ is the force between the dust grains, $F_{id}$ is the (Yukawa) force exerted by the ions on a dust grain, $m_d g$ is the gravitational force, $E$ is the confining electric field within the region, $\beta$ is the neutral drag coefficient, and $\zeta r(t)$ represents random kicks from a thermal bath. The force between pairs of dust grains $F_{dd}$ is assumed to be a Coulomb interaction, since screening is mainly provided by the ions.  
Ions will typically reach an equilibrium distribution within one ion plasma period, $\tau_i = 2\pi \sqrt{\epsilon_0 m_i / n_i e^2}$, where $m_i$ is the mass of an ion, $n_i$ is the number density of the ions, and e is the elementary electric charge.  For the range of plasma densities characteristic of the sheath of a rf discharge, the plasma period is typically $\tau_i \approx$ 1 $\mu$s. Similarly, the dust response time is on the order of the dust oscillation period, $\tau_d = 2\pi/\omega_d$, which for typical experimental conditions is on the order of 100 ms.  Fully resolving particle motion in numerical simulations usually requires that the time steps $\Delta t$ used to advance the simulation are on the order of $\Delta t=\tau/100$. As the ions in the simulation are accelerated in the vicinity of a dust grain, the DRIAD code resolves the motion of the streaming collisional ions on even smaller timescales, typically $\Delta t \le \tau_i/1000$.  Very close to the dust grain, the ion time step is modified by factors of two. The simulation advances the ions for $N(\Delta t_i)=100-200$ ion time steps (the exact number used is determined by the plasma conditions and run-time). The ions are then frozen, and the dust is advanced one dust time step, $\Delta t_d = \tau_d /1000 = 10^{-4}$ s, with the appropriate parameters associated with the ions averaged over the elapsed ion time steps.  

While dust-plasma interactions models assume `constant dust' are common for fundamental physics of dusty plasmas, these models only capture electric-charge, momentum, and energy exchanges between the plasma and the dust particles. Such models, including kinetic models, ignore the mass and materials changes between the dust particles and the plasma, and suppress nano-particle generation and plasma chemistry~\cite{Kush:1988, Holl:2000,Frid:2008}. Materials dynamics in reactive dusty plasmas,  using noble gases mixed with silane, acetylene,  methane and other molecules, are important to industrial applications of dusty plasmas. New forces due to asymmetric ablation loss of materials from a dust grain may also arise.
Modeling reactive dusty plasmas is a multiscale, multiphysics frontier central to both natural and engineered systems, from interstellar clouds, planetary rings, solar dust, to fusion devices and plasma-based manufacturing. In addition to charge, momentum and energy exchanges, dust grains in reactive plasmas actively participate in mass exchange processes, including coagulation, chemical reactions, sputtering, mass loss, and re-deposition. These interactions shape the evolution of the plasma and the dust simultaneously, demanding integrated, multiscale multi-physics modeling approaches that couple plasma and dust dynamics with surface chemistry and materials evolution. %While mitigating dust and nanoparticles are important in semiconductor processing applications, nanoparticle generation in reactive plasmas can also be useful to surface coating, thin film deposition and solar energy. 

Since a comprehensive reactive dusty plasma model across all spatial and temporal scales is yet to be realized, most existing models, which may be regarded as reduced models, aim at answering some specific questions driven by applications, measurements, and observations. In semiconductor processing plasmas, a key question is how to mitigate dust formation and contamination. In plasma coating and nanoparticle applications, the key questions, complementary to questions of plasma-induced contamination,  become quantitative understanding of nanoparticle distribution and optimization the rate of nanoparticle production.  In fusion plasmas, the questions are potential core contamination, tritium retention and mitigation, and genesis of dust through the plasma-first-wall interactions. In space and astrophysical plasmas, the questions include how to interpret large structure formation (much larger than individual dust) and evolution  on a  time scale much longer than dust oscillation and motion. 

Despite the diversity of applications, a unifying theme in integrated dusty plasma modeling is the need for cross-scale integration or scale bridging. There are natural separations of spatial and temporal scales. For example, electrons and ions interact on sub-nanosecond and sub-micrometer scales, whereas dust grains, typically ranging from nanometers to microns in size, exhibit much slower dynamics. 
 While such natural separations allow for simplified models, reactive processes such as coagulation, recombination, adsorption, or surface erosion can complicate the picture, particularly in non-equilibrium plasmas. %Increasingly, integration models aim to resolve these coupled interactions across scales, as traditional plasma-only or dust-only models often fail to capture emergent behaviors. 
In particle coagulation modeling~\cite{KoBh:1999}, for example, a more complete approach will require the model-
ing of the particle nucleation, including non-Maxwellian electron distribution functions and particle losses in spatially inhomogeneous plasmas. Validation of integrated models requires good measurement tools, such as novel use of self-excited dust density waves (DDW) as a diagnostic tool to characterize embedded amorphous hydrocarbon nanoparticles of different size and density~\cite{PATG:2022}.
 
%
%=================end Matthews contributions=============================
%
%
%
%

\subsection{Model reduction and acceleration} 

The trend towards integrated multi-scale multi-physics models~\cite{E:2011},  as described in Sec.~\ref{mod:plas}, Sec.~\ref{mod:dust}, and Sec.~\ref{mod:int}, has led to several parallel development, including a.) Adoption of more powerful computational architecture and hardware such as graphics processing unit (GPU) and parallel computing~\cite{HHLZ:2021}. For example, OpenDust is a GPU-accelerated code for force calculation on microparticles in a plasma flow~\cite{KoTi:2023}. A comparative study between the GPU and CPU calculations of dust particle kinetics was conducted
by running each of the GPU and CPU programs for 30,000 timesteps and measuring the time in
milliseconds taken to complete these operations~\cite{AApp:2024}; b.) Reduced order models (ROMs) or Model order reduction (MOR) with higher computational efficiency and cost reduction~\cite{ABGT:2018}; c.) Novel computational algorithms such as Lattice Boltzmann method,  adaptive mesh refinement methods, and stochastic weighted particle control~\cite{CCWX:2025}; d.) Introduction of machine learning models as surrogates of expensive traditional physics or first-principle models; e.) Model acceleration through data compression, such as pruning and quantization in machine learning; and f.) The family of statistical optimization methods, such as entropy maximization, Bayesian optimization methods,  that use statistical inference, information theory, noise reduction, uncertainty quantification (UQ), and active learning to identify and reduce costly or redundant parts of a computational workflow. These methods, often several of them used together~\cite{CWH:2021}, help to meet high computational power requirements in integrated models, allow efficient parallelization in modern multi-core and distributed systems, and efficient handling of large multi-modal datasets.

Data analysis and reduction are also needed to validate physics models, or to learn new physics from the data, or, at the very least, to determine the lowest dimensional latent representation of data and to train NNs to narrowly predict the dynamics of these latent representations~\cite{page2021revealing,colen2021machine,chen2022automated}. Learning new physics from experimental data using machine learning for different physical systems have made significant progress in different sub-fields of physics~\cite{carleo2019machine}, such as colloidal systems~\cite{pineda2023geometric}, and also more complex systems including glassy phase of matter and living organisms in biology~\cite{bruckner2021learning, toner1998flocks, swain2024machine}. People at DeepMind have created an AI called Physics Learning through Auto-encoding and Tracking Objects (PLATO) that is designed to `follow the basic physics laws' in the physical world~\cite{PWBB:2022}. Unfortunately, most of these methods have only been tested on simulated data, where the  ground truth may not be  known, noise is simple, and AI finds the best fit within a limited library of potential functional relationships by balancing the quality of fit against the model complexity \cite{bongard2007automated,champion2019data,brunton2016discovering,daniels2015automated,Lusch2018,bapst2020unveiling,pathak2018model}. Dynamical inference has been rarely used on real, noisy experimental data where the underlying dynamics are not known {\em a priori} \cite{bruckner2021learning,Daniels7226,lai2020vulnerability}.

Here we highlight some progress in ROMs, which are growingly used across the physical sciences for approximating the behavior of complex dynamical systems, from climate models~\cite{FKBW:2015}, systems biology~\cite{Wilk:2018, Yona:2011}, and fluid mechanics~\cite{SRSV:2024} to magnetized plasmas and plasma–material interactions~\cite{RuMM:2023}, with significantly lower computational cost. Mathematical foundations for multi-scale modeling~\cite{E:2011} and ROM are based on the Kolmogorov-Arnold representation theorem~\cite{Kol:1957,Arn:1957}, as a solution to Hilbert’s
13th problem~\cite{Brag:2018}. Kolmogorov-Arnold representation theorem provides a method to approximate multivariable continuous function using a sum of univariable continuous function~\cite{Kol:1957, Arn:1957}.
By capturing the dominant modes or features of a system, ROMs enable faster simulations, real-time control, and efficient parameter sweeps, often while preserving key physical insights that can be directly used to interpret observations. %In both general physics and plasma physics, the demand for ROMs has grown rapidly with the increasing complexity of high-fidelity simulations,.

The toolbox and algorithms of ROMs are growing to enhance the existing MD, PIC~\cite{ReFK:2023}, kinetic and fluid simulations~\cite{RoSB:2022}. Besides dimension reduction through the discovery of scaling laws, proper orthogonal decomposition (POD), dynamic mode decomposition (DMD)~\cite{NTNK:2024}, and Galerkin projection~\cite{ChoC:2019} are used to extract low-dimensional representations from high-dimensional data or equations. When ROMs are used directly to speed up the governing equations, such as the Boltzmann transport equation~\cite{CHOI2021109845}, the ROM is called intrusive~\cite{RoSB:2022}.

Recent efforts have focused on data-driven reduced models, when the governing equation of the evolution may be unknown, particularly those through the use of neural networks~\cite{KBJH:2023}. For instance, the relatively simple artificial neural network (ANN), in today's standard, was trained to reproduce the spectral energy distributions predicted by a hybrid code comprised of the
GALFORM semi-analytical model of galaxy formation, which predicts the full star formation
and galaxy merger histories, and the GRASIL spectro-photometric code, which carries out a
self-consistent calculation of the spectral energy distributions, including absorption and emission of radiation by dust~\cite{ABL:2010}. The ANN had a feed-forward neural network
architecture with 12 input neurons, one hidden layer with 60 neurons, or two hidden layers with 30 neurons each, and an
output layer with one or more neurons. More recently, techniques like autoencoders, neural ODEs~\cite{LJPZ:2021}, and PINNs are being applied to extract low-dimensional dynamics from simulation or experimental data~\cite{PaGR:2024}.

Another key development is the integration of `multi-fidelity modeling', such as fully kinetic simulations~\cite{alves2022data}, where high-fidelity simulations are used to train lower-fidelity surrogates that are faster to evaluate. Such a hybrid approach may be used to discover new mathematical relationships, such as scaling laws, reduced operators, or generate new examples or counterexamples for mathematical hypotheses about a physical system such as a dusty plasma. 

Error estimates, noise mitigation, and UQ or equivalently sensitivity analysis~\cite{SRAC:2008} are essential to any ROM. MultiNest is Bayesian inference tool calculates the evidence, with an associated error estimate, and produces posterior samples from distributions that may contain multiple modes and pronounced (curving) degeneracies in high dimensions~\cite{FHB:2009}.

Applications of ROMs are relatively new to dusty plasmas. Real-time control of dusty plasmas remotely could find near future applications. Meanwhile, there are some known challenges with ROMs, especially in the data-driven approaches. One concern is about generalization outside the training regime, particularly under varying boundary conditions or parameter shifts. Moreover, satisfying known physical constraints, such as conservation of energy or mass, remains a persistent concern in purely data-driven ROMs.  %
\subsection{Machine learning \& AI models \label{sec:MAM}}
%Burton, Nemenman, Wang

Machine learning algorithms, such as neural networks, is growingly used to construct surrogate models in place of traditional models, which are usually based on differential or integral equations. Machine learning, especially deep neural networks address several challenges for traditional physics models: When the physics is not completely clear (i.e. analytically or computationally solvable), the physics is clear but computationally too complex in a multi-scale system, unknown or incomplete initial or boundary conditions, noise corruption of experimental data, or computational errors, correspondingly, there are `surrogate or blackbox' ML models, or `pragmatic machine learning' models at different scales, or machine-learning-augmented physics (computational) models.

Augmenting Physical Models with Deep Networks for Complex Dynamics Forecasting was reported in~\cite{YGD:2020}. APHYNITY framework is a principled approach to augment  {\it incomplete physical dynamics} described by differential equations with deep data-driven models. APHYNITY breaks down the dynamics into two components: a physical component for the dynamics for which some prior knowledge exists,
and a data-driven component accounting for errors of the physical model. Extensions of this work to learn partially known equation of motion using VAE have been reported in~\cite{TK:2021}.  An Symplectic Recurrent Neural Network (SRNN) models the Hamiltonian function of the system by a neural network and furthermore leverages symplectic integration, multiple-step training and initial state optimization to address the challenging numerical issues associated with Hamiltonian systems~\cite{CZA:2019}. 

In dusty plasmas, ML models have shown some promising results. \citet{ding2021machine} were able to learn the nonlinear response function of oscillating dust particles by minimizing a loss function using Bayesian optimization. While NNs were not used in this study, it demonstrated a powerful way to efficiently minimize a physics-constrained loss function, similar to training NNs. \citet{zhe2021machine} used a multi-layer perceptron (MLP) algorithm, the simplest deep neural network, to measure the electron density and temperature from sparse Langmuir probe measurements. In this sense, the model acts as a nonlinear approximator with the plasma pressure and probe voltage as inputs. \citet{yu2022extracting} used simulations of the Brownian motion of one- and two-particle dust systems to infer the environmental and interaction forces acting on particles in experiments. The supervised learning model used hundreds of features extracted from the simulated particle trajectories to predict parameters from data. For two particles, the particle charge and screening length could be measured as a function of plasma parameters. \citet{liang2023determining} used a similar method (training ML models on simulations to predict from experimental data), except the screening length was measured from particle trajectories in dense dust crystals, and a Convolutional Neural Network model (a neural network that respects locality and translational and rotational invariances in its inputs)  was used to extract features and process the data. Finally, \citet{huang2022machine} used simulations of multi-phase dusty plasmas composed of melted and crystalline states to train a CNN model to classify different states observed in experiments. Importantly, the training data only labeled purely crystalline or melted states, and made no judgement for ambiguous states near the phase transition. These examples in dusty plasma science, and many others in AI applied to science more broadly, rely on simulations with known physics to train ML algorithms before interpretation of experimental data.

\begin{figure*}[!htbp]
\centering
\includegraphics[width=.9\textwidth]{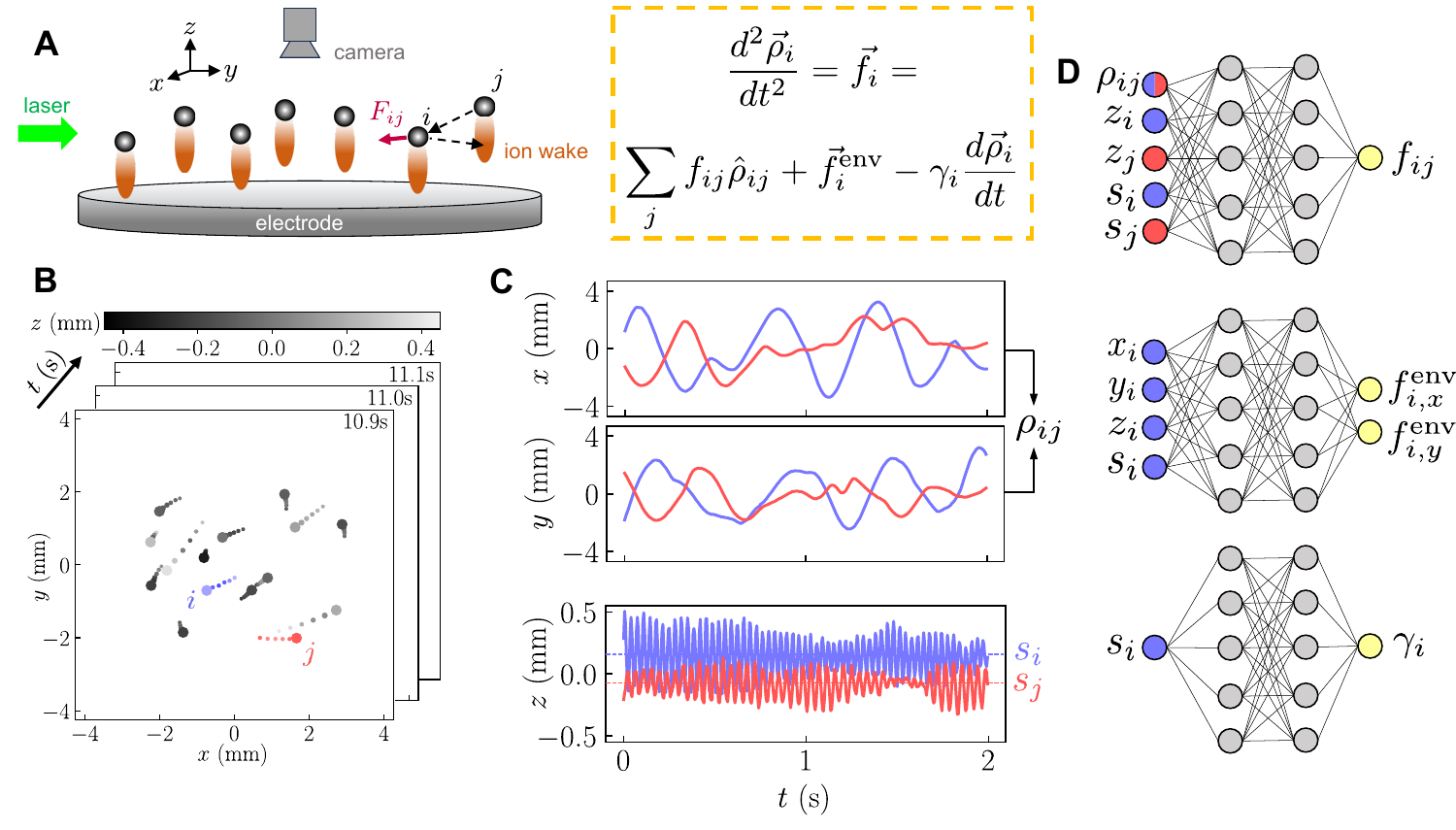}
\caption{Overview of data workflow. ({\bf A}) Particles levitated above the electrode move mostly in the $xy$-plane, with small deviations above and below the plane. The focused ion wake (red) is directly below each particle, and contributes a small attractive part of the total force ($F_{ij}$) on particle $i$. The objective is to infer the horizontal reduced forces on particles using the equation of motion to the right. ({\bf B}) Snapshot of particle positions from a single experiment of 15 particles. The grayscale color indicates the $z$-position, and the tails of each particle represent the previous 5 frames. ({\bf C}) The $x$, $y$, and $z$ position of two particles during two seconds. The particles are marked $i$ (blue) and $j$ (red) in panel (B). {\color{black}The quantity $s_i=\langle z_i\rangle$ is used as a size identifier for each particle.} ({\bf D}) The schematic of the model, which consists of three neural networks trained concurrently (particle interaction $g_\text{int}$, environmental $\vec{g}_\text{env}$, and damping $g_\gamma$). The color of the inputs designates the source (particle $i$ or $j$).}
\label{p1}
\end{figure*}

 Assumptions and simplifications in simulations can obscure new physics in experiments. Put simply, if we do not know about existence of a certain phenomenon, we cannot simulate it. Hence we cannot train a neural network to detect it from data within this class of approaches.  Learning ``new physics'' should be done from experimental data alone. This is more challenging, but possible provided that the AI model structure being inferred is broad enough to allow for new physics, but constrained by physics intuition to allow inference from  data sets of realistic sizes. A recent example of this approach was demonstrated in \citet{yu2023learning}, where three-dimensional particle trajectories were used to train neural networks as universal approximators for the different forces acting on particles. A 3D tomographic imaging method \cite{yu20233d} with a scanning laser sheet at 500 Hz was used to track the particle positions for 1 minute or more (Fig.~\ref{p1}A-C). The NN model accounted for inherent symmetries, non-identical particles, and the need to infer physical forces, instead of focusing on black-box prediction of particle trajectories. The background plasma was considered quasistatic so that interparticle forces only depended on particle positions, and the forces were also pairwise. Overall, the model learned the effective non-reciprocal forces between particles with high accuracy ($R^2>0.99$) for many plasma parameters. 

The model assumed that the horizontal ($xy$ plane) acceleration of each particle was determined by the boxed equation in Fig.~\ref{p1}, where $\vec\rho_i = (x_i, y_i)$, $\vec\rho_{ij} = (x_i - x_j, y_i - y_j) = \rho_{ij}\hat{\rho}_{ij}$, and $f_{ij} = F_{ij} / m_i$, ($F_{ij}$ is the magnitude of the interaction force). The reduced environmental force was $\vec{f}^{\text{env}}_i = \vec{F}^{\text{env}}_i/m_i$, where $\vec{F}^{\text{env}}_i$ is the horizontal environmental force on particle $i$, and the damping coefficient of particle $i$ was $\gamma_i$. In the experiments, the ion wakes beneath each particle resulted in non-reciprocal effective forces ($F_{ij}\neq F_{ji}$, Fig.~\ref{p1}A). Furthermore, the particles were not identical, and required particle-level identifiers. Ideally, this would be the mass of each particle, which is unknown. Yet heavier particles sit lower in the plasma sheath, and thus the mean $z$-position, averaged over an entire time series: $s_i = \left<z_i\right>_t$, was used as a good identifier for each particle. For each of the forces, a separate, fully-connected NN was used to map the input positions and identifiers to their respective forces, and the 3 NNs were trained in parallel (Fig.~\ref{p1}D). To learn each force accurately, the inputs to each term contained minimal overlap (i.e., environmental forces depend on $x_i$ and $y_i$, but interaction forces only depend on the relative distance $\rho_{ij}$). The loss function compared the measured acceleration of each particle to the model acceleration given by the 3 NNs, and was summed over all time points.  

Since the ground truth was not known (there was only experimental data), the model was validated by inferring particle masses in two independent ways, which turned out consistent with each other. In addition to learning the non-reciprocal forces between particles, the model inferred particle-specific properties by comparing the results to known theories. \citet{yu2023learning} found that the particle charge and interaction screening length deviated substantially from expectations, hinting at missing physics in current theories. Although the ML model was tested with simulated data (including non-reciprocal forces), it is the sufficient but not too restrictive physical constraints built into the model and the quality of the fit that make it a reliable tool for investigating dusty plasma, and complex systems more generally. Since the field of AI is developing rapidly, we expect more studies of dusty plasma will utilize ML methods, in both theory and experiment. 

Yet major challenges remain. First, one needs to formalize the notion of ``sufficient but not too restrictive'' physical constraints. Then it would be possible to automate the design and application of NN models to plasma experiments, so that various engineering choices  do not depend on the intuition of a particular researcher. The second challenge is learning field-level descriptions of dusty plasma. Dusty plasmas often have more particles than can be tracked accurately, and particles can be 100 nm or less (too small to individually resolve). For this challenge, one would need to combine methods  that can identify parsimonious continuum equations  \cite{brunton2016discovering,champion2019data,alves2022data,reinbold2021robust,gurevich2019robust} with those that extract low-dimensional predictive embeddings  ({\it aka}, relevant variables) from data \cite{lecun2022path,abdelaleem2023deep}, and then to endow both with the useful physical intuition. A more detailed discussion of specific neural network architectures and their scaling with data size is presented in Sec.~\ref{Sec:dFuion}. 
\section{Different dusty plasma datasets \label{sec:dataset}}

Constructing high-quality datasets is foundational for AI-driven discovery and applications. DustNET aims at systematic dataset development and classification across multiple stages. a.) Existing data curation and standardization. b.) Data annotation and labeling for supervised machine learning, and validation of AI models. c.) Synthetic data and simulation data generation for model interpretation, denoising. d.) Data augmentation and multi-modal fusion for complex model development. Here we give some examples of existing datasets. Different datasets are discussed with an emphasis on laboratory phenomena of materials, dust dynamics, and structures. 

Astrophysical observational datasets may be obtained through a number of open archives and survey databases, which provide large volumes of imaging, spectroscopic, and time-domain data relevant to dusty plasma environments in planetary systems, protoplanetary disks, and the interstellar medium. Examples include the NASA Mikulski Archive for Space Telescopes (MAST), which hosts observations from missions such as the Hubble Space Telescope and the James Webb Space Telescope; the Sloan Digital Sky Survey (SDSS)~\cite{York2000}; the European Space Agency's Gaia mission archive~\cite{Gaia2016}; and infrared survey datasets such as the Wide-field Infrared Survey Explorer (WISE)~\cite{Wright2010}. Additional open data resources are available through the NASA/IPAC Infrared Science Archive (IRSA), the Atacama Large Millimeter/submillimeter Array (ALMA) Science Archive~\cite{ALMA2015}, and the Virtual Observatory framework~\cite{Hanisch2001}, which enables federated access to distributed astronomical datasets. These publicly accessible datasets provide valuable sources of observational data for studying dust formation, transport, and plasma interactions in astrophysical environments.%

\subsection{Magnetized dusty plasmas}
The Magnetized Dusty Plasma eXperiment (MDPX) device~\cite{ref21,Thomas2016}, shown in Fig.~\ref{fig:MDPX}(a), features a rotatable, split-cryostat magnetic-field design with a pair of superconducting coils in each half of the system. The cylindrical bore of the magnet can accommodate plasma chambers up to 0.5 m diameter and 1.57 m long. The “split-bore” configuration and rotational ability adds unique advantages: 1) enabling extensive axial and radial access to the plasma volume, 2) allowing external users to bring their own plasma chambers for use in the MDPX bore, and 3) performing experiments with an arbitrary angle between the magnetic field and gravitational field.

The maximum magnetic field ($B$) was designed to be up to 4 Tesla in MDPX. The device operates regularly with magnetic fields up to 2.5 T. The {\it dust magnetization} condition, defined as the magnetic force ($F_m$) being dominant, $F_m =Q_d v_d B \ge \sum_iF_i$,  is usually not satisfied because of the small charge-to-mass ratio of the charged dust grains. For example, for a spherical dust grain of 1 \textmu m in radius, a total dust charge of $Q_d = 9 \times 10^{-16}$~C (5.6$\times$10$^3$ e),  and dust velocity $v_d$ = 1 m/s, the minimum magnetic field $B \geq m_dg/Q_d v_d = 69$ T. Here we also assume that the Earth's gravity is the other dominant force, and dust mass density $\rho_d$ = 1.5 g/cm$^{-3}$.  For a magnetic field of a few several Tesla, the dynamics of the non-magnetic dust grains can still be significantly impacted by the presence of the magnetic field due to the electron or ion magnetization, magnetic field modification of dust charging, or magnetic dust particles. %Superconducting magnet systems are essential in generating these required large, steady-state magnetic fields. %The MDPX device is located in one of a few laboratories with the necessary technical infrastructure to support these studies. 

\begin{figure}[h!]
    \centering
    \includegraphics[width=0.4\textwidth]{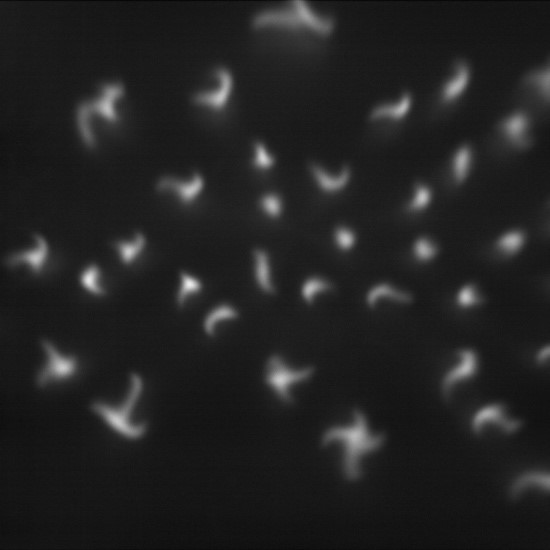}
    \caption{Example of self-organized filamentary structures formed in krypton plasma in the MDPX device. This view is from a camera that is placed on top of the vacuum chamber and looking at visible light emission from the plasma. The view is adjusted so that the filaments appear as the bright shapes embedded in the surrounding plasma. In this view, the magnetic field points downward, into the page. This experiment was performed at a magnetic field (B) of 3.25 T, neutral pressure (P) of 40 mTorr, and RF power of 1 W.}
    \label{fig:filamentary_structures}
\end{figure}

For over a decade, tens of terabytes of experimental datasets,  primarily in the form of images and image sequences (movies), have been produced from MPDX. Self-organized structures of dusty plasmas can be strongly modified by the magnetic fields and plasma boundary conditions. The formation of field-aligned, filamentary structures in the plasma is shown in Fig.~\ref{fig:filamentary_structures}. Experimental studies~\cite{Williams2022} and numerical studies~\cite{Menati2020a,Menati2020b} have shown that the appearance of these structures is correlated with the degree of magnetization of the ions. Turing’s activator-inhibitor model can be employed to describe the formation of filamentary structures in a low-pressure electric discharge exposed to a strong magnetic field~\cite{Menati2023}.

\begin{figure}[h!]
    \centering
    \includegraphics[width=0.4\textwidth]{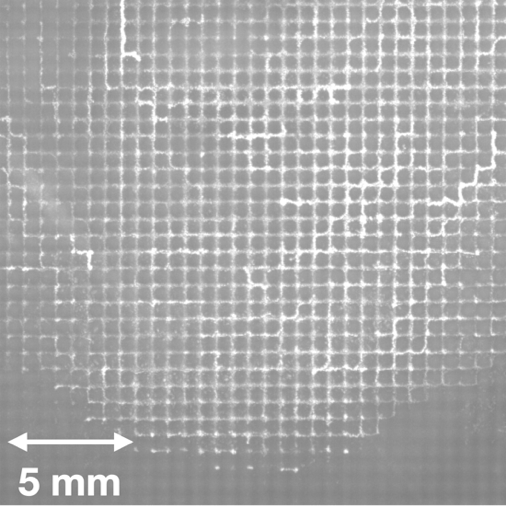}
    \caption{Photograph of imposed ordering in a dusty plasma at high magnetic field.  The image in generated by extracting the maximum pixel intensity from a sequence of 100 images recorded at 12.5 frames/second. This experiment was performed at a magnetic field, B = 2.5 T, neutral pressure, P = 170 mTorr, and RF power, PRF = 2.3 Watt.}
    \label{fig:lattice_structure}
\end{figure}

Another example is the formation of a square-like dusty plasma lattice structure that follows the spatial pattern of a wire mesh that forms the boundary of the plasma. This phenomenon appears in the plasma for conditions of high magnetic field ($B \geq 1\,\mathrm{T}$) and lower neutral pressures~\cite{Thomas2016,Hall2018,Thomas2020}. An example of this is shown in Fig.~\ref{fig:lattice_structure} where 100 images recorded at 12.5 frames/second have been combined to reveal the square-like spatial pattern of the movement of the dust particles.

Additional examples include melting plasma crystals in magnetized plasmas~\cite{Jaiswal2017}, developing pressure~\cite{Kremeyer2020} and coherence imaging spectroscopy~\cite{Kriete2024} diagnostics for the W7-X stellarator fusion device, investigating particle growth in strongly magnetized plasmas~\cite{Couedel2019}, studying laser-plasma interactions~\cite{White2024}, and studying thermal properties of dust clusters~\cite{Kumar2024}.

Due to limitations of typical dusty plasma diagnostics in large magnetic fields, artificial intelligence (AI) and machine-learning (ML) tools may be used to enhance image analysis capabilities in these studies in the near future.
A recent work implemented a CNN to track and categorize the different filament patterns observed in the background plasma~\cite{Avritte2024}.

%\clearpage

\subsection{Dusty plasmas in fusion devices}
%not to include ? ...
Here we emphasize dusty plasmas in magnetic confinement fusion~\cite{Winter_1998,winter2000dust,SHARPE2002153}. Dusty plasmas may also be expected in other high-temperature plasmas, including pulsed power experiments, high-intensity laser facilities and inertial confinement fusion. Due to the ambient high-temperature plasmas, these dusty plasmas involve significant change of dust materials composition and size.

It is safe to assume that there is initially a negligible amount of dust in any magnetic fusion device. The fact that the amount of dust becomes significant only after the presence of high-temperature plasmas implies that dust must be
produced through plasma-wall interactions~\cite{Tanabe.2021}, including neutron-wall interactions and other fusion byproduct-wall interactions~\cite{Wang:2010}. Interests in dust in magnetic fusion devices, most of which involve hard-to-control dust mobilities, come from different considerations, including potential plasma core contamination by high-Z impurities, tritium retention and migration, and first-wall erosion. Meanwhile, controlled dust injection can also be beneficial to fusion and high-temperature plasmas for wall conditioning, tritium breeding, plasma diagnostics~\cite{WanW:2003,TWDW:2006,W:1621292,samarian_2025_dj49n-sdy76}, steady-state plasma operation and control~\cite{Wang_2019}.

Three types of datasets are available: pre-injection dust characterization,  in-situ imaging and other measurements, and post-mortem dust analysis. Pre-injection dust characterization allows quantitative study of dust-plasma ablation physics and dust transport. Post-mortem analysis of dust are necessary to accommodate the versatile dust production processes and richness of dust species in fusion~\cite{winter2000dust,Wint:2004}. Chemical compositions
of dust produced in magnetic fusion devices are determined by the
wall materials exposed to the plasma, particles of the fusion plasma itself,
and particles of fusion byproducts. Coexistence of many dust-production
mechanisms, combined with different kinds of wall materials, can lead to
rather diverse distributions in dust composition, dust size, and dust shape
in magnetic fusion. Magnetic field can also play a role in dust formation
and evolution. Below we give some examples of in-situ imaging of dust ablation and analysis. We recognize that non-fusion laboratory experiments are often useful to supplement the high-temperature dusty plasma experiment data.

\begin{figure}[htb]
\centering
\includegraphics{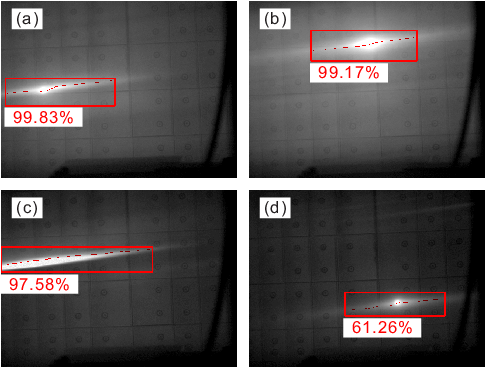}
\caption{\label{fig2} (Color online). Analysis of dust ablation trails using DATA code package from images captured in the EAST discharge $\#70606$. The localized object enclosed by each box is identified and classified as one dust ablation trail, with the evaluated percentage value of $>50\%$. The DATA code package is also able to obtain the skeleton of the identified dust ablation trail, as the thin curve marked there. Reproduced with permission from~\cite{liang2023RSI}. Copyright 2023 AIP Publishing. }
\label{fig:SU2}
\end{figure}

A Dust Ablation Trail Analysis (DATA) code, a deep-learning enhanced package, is being developed to automatically detect and analyze dust ablation trails from experiment images captured in tokamak discharges~\cite{liang2023RSI}. Due to the limited viewing access to tokamaks such as the Experimental Advanced Superconducting Tokamak
 (EAST), it is not practical to obtain a sufficiently large experimental datasets of dust ablation trails with different dust types, orientations, intensity, and sizes. %However, to train the machine learning DATA code package with a sufficient accuracy, a large amount images of dust ablation trails are necessary. 
 Leveraging the similarities between plasma jets and fusion dust ablation trails in the shape and size, laboratory plasma jet images were used to train and validate the DATA code package~\cite{liang2023RSI}. A plasma jet inside a quartz tube was produced from a steady argon gas flow at a rate varying from 3 and $6$  standard liters per minute. Thousands of plasma jet images have been obtained using two cameras by varying relative orientation of the jets to the cameras, relative distances, and the plasma jet size in a controlled setting. %2440 images of plasma jets under various conditions of different relative orientations, locations, and sizes, using two cameras. These obtained thousands of plasma jet images can be used to improve the generalization ability of our DATA code package.

DATA code package consists of three functions for the time being, object localization (or segmentation), image classification, and image characterization. The object segmentation function can automatically identify different objects, including plasma jets from other features in an image. Image classification is able to sort different segmented objects and determines whether they are plasma jets or not. Image characterization portion is used to quantitatively characterize the different plasma jets. Some additional details about DATA are given below or in the Ref.~\cite{liang2023RSI}. 

The object localization function used traditional computer vision technology. In image pre-processing steps, the raw plasma jet images were first converted to grayscale, followed by a Gaussian filter, a adaptive threshold named the Otsu's method~\cite{Otsu1979IEEE}, and an opening operation~\cite{opencv_library}. As a result, the image background was removed from all possible objects, including plasma jets. Then, the object contours were extracted using the {\it findContours} function in the OpenCV~\cite{opencv_library}. %, as marked by the boxes in Figs.~\ref{fig1}(a,b), as a result, we obtain the contour information of each object. 
Next, the object masks were created from the contours, so that each identified object is extracted individually by combining the masks with the corresponding original image. The segmented objects include both plasma jets and other objects, which need classification, as the next step, to separate them. %After these operations, the object localization portion is able to localize and extract all plasma jets in the images, although a large part of these localized objects are not plasma jets. 

A convolutional neural network (CNN) is implemented in DATA to classify the segmented objects and identify plasma jets. %While training the image classification portion, we use the 
Data augmentation techniques, including image rotations with a random angle, horizontal and vertical flips with the $50\%$ probability, were also used. Furthermore, the training data were normalized using Min-Max normalization and Z-score normalization to prevent the CNN from failing to converge and to speed up the convergence of the CNN. After training, the image classification portion identifies plasma jets from these localized objects with the accuracy of $>~99.9\%$.%, as two typical examples shown in Figs.~\ref{fig1}(a, b), clearly demonstrating the excellent performance of our DATA code package. 

Next, the Zhang-Suen thinning algorithm was used in DATA to quantitatively characterize the classified plasma jets~\cite{zhang1984fast}. The principle of the Zhang-Suen thinning algorithm is to reduce the area of the object image without changing its topology, so that the skeleton information of the object is achieved. %As two typical examples shown in Figs.~\ref{fig1}(c, d), the image characterization portion is able to obtain the skeleton information of the identified plasma jet.
The skeleton information can be used for the further investigation of plasma jets, such as the three-dimensional (3D) reconstruction.

The DATA code package has been used to analyze lithium dust ablation trails from the experiment movies captured in EAST~\cite{sun2018inital}. % In the EAST tokamak discharge $\#70606$, lithium dusts are injected into the fusion plasma environment to ablate~\cite{sun2018inital}, so that dust ablation trails are observed, as captured in movies. Using our DATA code package, all dust ablation trails are detected, extracted, and correctly classified. 
Figure~\ref{fig:SU2} shows four identified dust ablation trails with the probability values used to determine whether they are dust ablation tails, or actually plasma jets from our training data. Although the DATA code was trained with the plasma jet images, it is still able to accurately identify dust ablation trails captured in tokamak discharges. The DATA code package is also capable of obtaining the skeleton information of all identified dust ablation trails as the thin curves marked in Fig.~\ref{fig2}, which contain the direction information of these dust ablation trails, and can be used for 3D in-situ plasma diagnostics in fusion and other plasmas~\cite{TWDL:2006,liang2024RSI}.

\subsection{Space microgravity dusty plasmas}
% Knapek, Melzer
%Gissen Niklas Dormagen et al.

Several generations of dusty plasma experiments, PKE-Nefedov~\cite{NMFT:2003}, PK-3 Plus~\cite{Thomas2008}
and PK-4 (Plasmakristallexperiment 4)~\cite{Pustylnik2016} have been carried out in the microgravity environment of the international space station (ISS). 

The interparticle distances typically exceed 100 \textmu m, creating a dilute and transparent system
that allows for effective analysis using cameras~\cite{Stroth.2018}. 
Stereoscopic camera systems for the simultaneous reconstruction of the 3D particle dynamics in a (subvolume
of the) dust cloud have been been employed on parabolic
flights for years, but have not been used on
the ISS, yet. Such a system consists of three~\cite{himpel2011} or four
cameras~\cite{melzer2018} that look into the same region of the particle
system under different viewing angles and record images
at the same instants. The acquired data of all cameras
is then used to reconstruct 3D particle position in the
observed volume~\cite{himpel2011, Wieneke13, Melzer16, Schanz16,WaLW:2016}. A stereoscopic four-camera system is planned for the future microgravity research facility COMPACT~\cite{knapek2022}.

The ISS facilities so far produced a multitude
of data sets on Newtonian particle dynamics and structural formations under controlled conditions: PKE-Nefedov had two monochrome
cameras (overview and high resolution) with recordings
on high-8 tapes. In PK-3 Plus, the video recording was
purely digital, and four cameras were used to record
particles in different spatial resolutions, and the plasma
glow. In the currently operated PK-4 facility, two cameras with larger spatial resolution record particle images,
while a spectrometer and a plasma glow camera yield
information on the plasma. All facilities had or have a
linear stage for the optical diagnostics that allows to scan
through the particle systems to acquire sequential information on the three-dimensional microparticle arrangement. Though the quality of microgravity is excellent
onboard a space station, one disadvantage is that during experiment, the scientists usually see only
low resolution preview images, and the time to download
data from the space station (i.e. hard drive transport to
ground) is long.

Data analysis of microgravity data makes use of the same tools as used for laboratory data: In 2D imaging, particle coordinates are usually found by blob detection algorithms \cite{crocker1996, Feng2007, Ivanov2007, mohr2019}. The coordinates are then used as input for further analysis steps to obtain information on properties of the system, e.g. particle densities by calculating nearest neighbor separations, or structural arrangement by calculating pair- and bond- correlation functions. This gives information on the state of the system (crystalline, liquid, gaseous). Particle dynamics can be studied by tracking particles over consecutive frames and calculate their velocities, yielding velocity distributions and kinetic energies, or information on flows in the system. Waves can be analyzed by means of space-time plots of the recorded intensities in the images \cite{schwabe2007,Menzel10,Tadsen15,melzer2020,knapek2023}, yielding frequencies and wave-numbers that allow to compare wave dispersion relations of e.g. dust-density waves with theoretical models.
For the reconstruction of the 3D particle positions from stereoscopic camera images sophisticated models from fluid physics, such as the multiset triangulation \cite{Zhang98,Wieneke13} or the shake-the-box-algorithm \cite{Schanz16}, have been adapted to dusty plasmas. 

Machine-learning techniques have been used to identify and quantify structures from (2D) video images. Examples include the identification of ``dust strings'' (dust particles that become aligned due to ion flows in the plasma)~\cite{dietz2021phase}. In PK-4 rapid switching of the electric field polarity within the plasma leads to the formation of a string-like dust fluid, a so-called electrorheological plasma~\cite{Ivlev07,dietz2021phase}.  Other example applications of AI and ML include the identification of hexatic phases from the 2D arrangement of particles \cite{Du2024} or the identification of the solid-liquid transition line from supervised and unsupervised learning \cite{Li2024}, see Sec.~\ref{subsec:melting} for further discussions. In addition, for simulated data, it was possible to locally assign 3D crystal structures in a dust system to fcc, hcp, and bcc crystal structures or disordered particles \cite{Dietz17}. This approach requires the input of the 3D particle positions.  

\begin{figure}
  \includegraphics[width=\columnwidth]{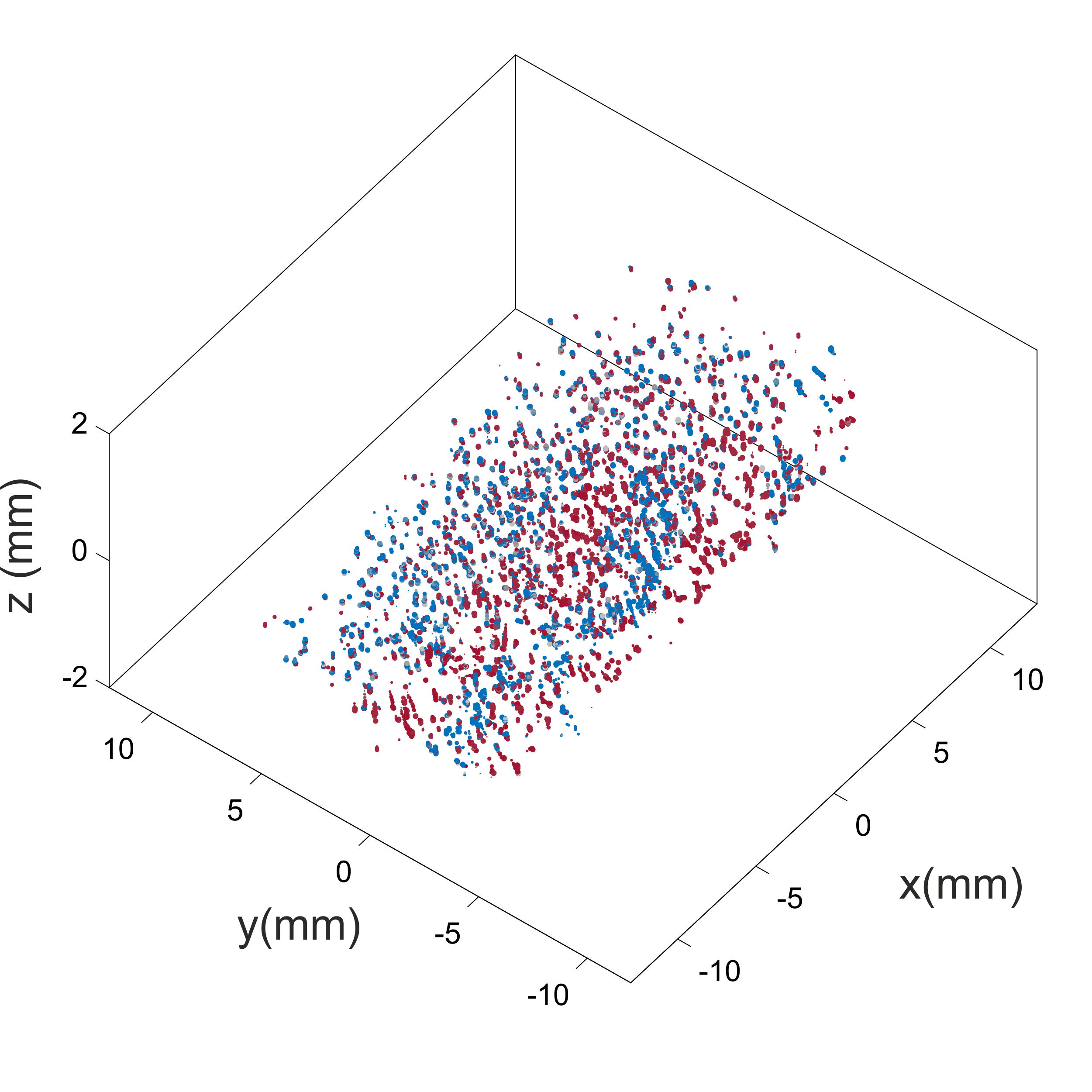}
  \caption{3D particle position reconstructed using a CNN. The particles are colored by their velocity where blue (red) means a velocity in negative (positive) $y$-direction}.\label{fig:cnn}
\end{figure}

Convolutional neural nets (CNNs)  have been applied to retrieve the 3D particle positions from the four stereoscopic cameras used on parabolic flights \cite{Himpel21}. This CNN is trained on artificial camera images calculated from given 3D particle positions. The reconstruction works on a volume grid. The algorithm traces the camera image intensities into the volume field using the camera calibrations and the volume field is fed into the CNN that extracts the 3D positions from this volume-based image intensities. The CNN works similar as a feature extraction network for 2D images, but on a 3D volume grid. The CNN reconstructs the positions in each frame individually. Afterwards a standard tracking algorithm has been applied to track the particles through the image sequence. Figure \ref{fig:cnn} shows the particle positions reconstructed at a specific instant with color-coded ``vertical'' velocity.

Such a machine-learning approach could be employed to reconstruct the particle positions from video images of the stereoscopic setup. Then, only the reconstructed 3D particle positions are transferred from the ISS to Earth instead of full video images. This considerably reduces data traffic, but still allows the experiment controllers on Earth to gain a quick-look impression of the dust configuration. 

\begin{figure}
  \includegraphics[width=\columnwidth]{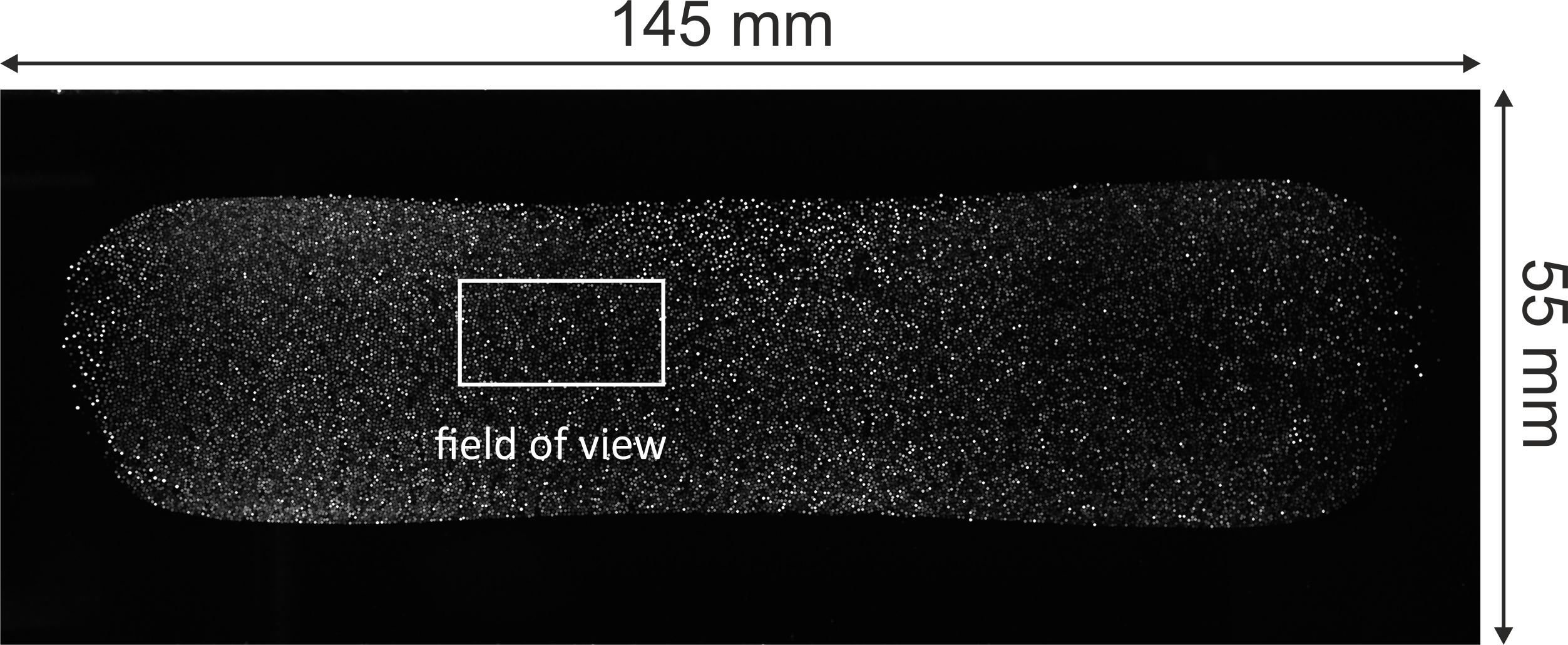}
  \caption{Image of the dust cloud from the overview camera showing that the dust cloud is extended and void-free. The stereoscopic cameras
  image a smaller field of view of the order of  $20\times 10 \times 2$~mm$^3$.}\label{fig:overview}
\end{figure}

ML and AI have been envisioned for experimental data processing and controls aboard the ISS. The Zyflex chamber~\cite{Knapek2021}, being the heart of the COMPACT experiment on the ISS and currently undergoing parabolic flight testing led by the University of Greifswald, is equipped with a center and a ring electrode both on the upper and lower side. The electrodes are driven by a four-channel rf-generator that allows to independently apply rf signals with given amplitude and relative phase. In a pulsed rf-mode, it is possible to confine extended and quite homogeneous dust clouds without central void~\cite{knapek2024} which is a prerequisite for most of the experiments planned with COMPACT, see Fig.~\ref{fig:overview}.

Here, life video images are envisaged to be analyzed by AI techniques with feature detection to judge whether the dust clouds are void-free, whether certain crystal structures or waves or defects are present or whether other desired properties of the dust cloud are realized. Further, AI might then be used to activate particle dispensers to increase and change dust densities and compositions or to set suitable rf conditions. While, an AI control of the experiment on the ISS will certainly not be possible due safety issues such techniques might be implemented on parabolic flights. Nevertheless, for the ISS environment, AI systems might propose rf suitable settings or dispenser activation that can then be set by the experiment controllers.

During the recent parabolic flight campaign in September 2024, a life video feed was analyzed in cooperation with the Mittweida University of Applied Sciences. There, instantaneous dust density maps have been calculated, further developments are planned for future flights.

\subsection{Parabolic flight dusty plasmas} %transient microgravity
% Knapek, Melzer
%Dormagen, Klein

\begin{figure}[!h]
\begin{center}
\includegraphics[angle=90,scale=0.1]{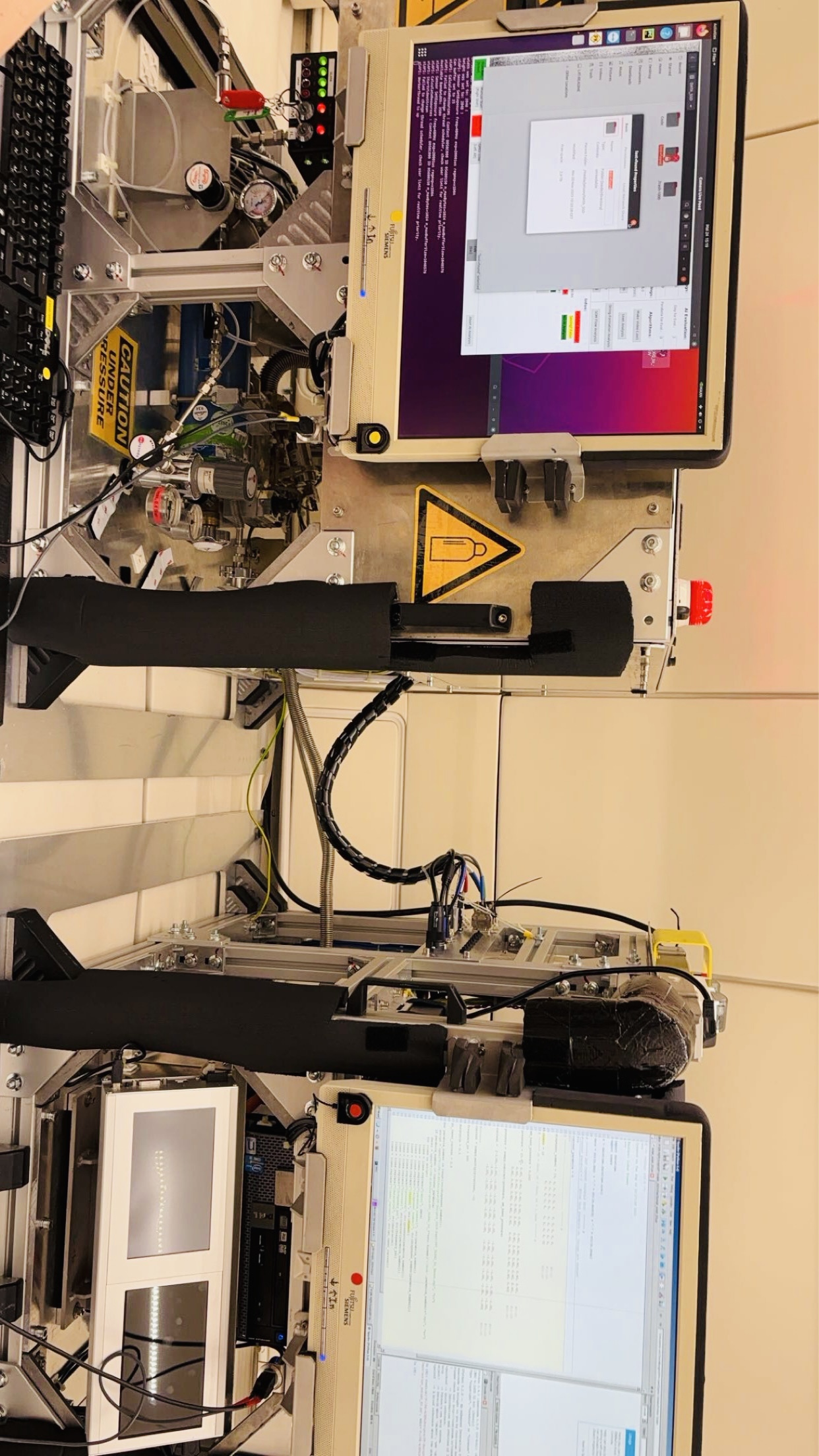}
\caption{The Engineering Model (EM) installed on board the parabolic flight aircraft. The Plasma and Space Physics research group at Justus Liebig University (JLU) Giessen operates both an Engineering Model (EM) of PK-4, which has been adapted for zero-gravity conditions during parabolic flights, and the Science Reference Model (SRM) No. 1, identical to the flight model on the ISS.}
\label{fig:parabolarack}
\end{center}
\end{figure}

While long-duration experiments aboard the ISS remain necessary for observing processes that require extended microgravity conditions, parabolic flights are more accessible to short-duration studies of dusty plasmas in transient microgravity~\cite{thomas2001, thoma2006, Klindworth06,himpel2018,dietz2018,knapek2024}. Transient microgravity can also be achieved on or near the Earth's surface through drop tower and sounding rockets~ \cite{morfill1999}.
Parabolic flights have a reduced gravity on the order of 10$^{-3} \times g$ and significant jitter due to the 
flight maneuver. During parabolic flights, adjustments to the instruments are easier or even in real-time, so that the higher quality data can be obtained. Parabolic flight setups were operated, {\it e.g.},
with plasma chambers of space station facilities (PKE-
Nefedov, PK-3 Plus and PK-4), Fig.~\ref{fig:parabolarack} and in addition with new
designs. The later include new plasma chambers (IMPF-
K2, Zyflex) and new designs for optical diagnostics (improved 2D diagnostics and 3D diagnostics). Due to the
availability of technologically advanced cameras, larger
spatial and temporal resolutions are achieved. This also
increases the size of data sets considerably.

A key distinction between the EM and the flight model, Fig.~\ref{fig:parabolarack},  is the use of a new xiQ camera, which offers higher resolution and increased frame rates. The experimental setup is designed to study complex plasmas in a direct current (DC) discharge within an elongated, U-shaped glass tube filled with either neon or argon. A flow controller regulates the gas flow at one end, while a vacuum pump at the other maintains the desired pressure level \cite{Mitic.2008}. Electrodes at both ends generate a high-voltage electric field, ionizing the gas and creating a low-temperature plasma. The setup is equipped with two particle observation cameras that can move longitudinally and radially within the plasma chamber. The imaging system consists of a CCD camera (Particle Observation 2 or PO2) with a resolution of $1600 \times 1200$ pixels and a CMOS xiQ camera with a higher resolution of $2048 \times 2048$ pixels.
To visualize the microparticles, a \emph{particle observation laser} (PO laser) emitting green light at a wavelength of $\SI{532}{nm}$ was used. The scattered laser light was captured by the cameras, enabling high-contrast imaging of the suspended particles.

\begin{figure}[!thb]
\centering
\includegraphics[width=3.3 in]{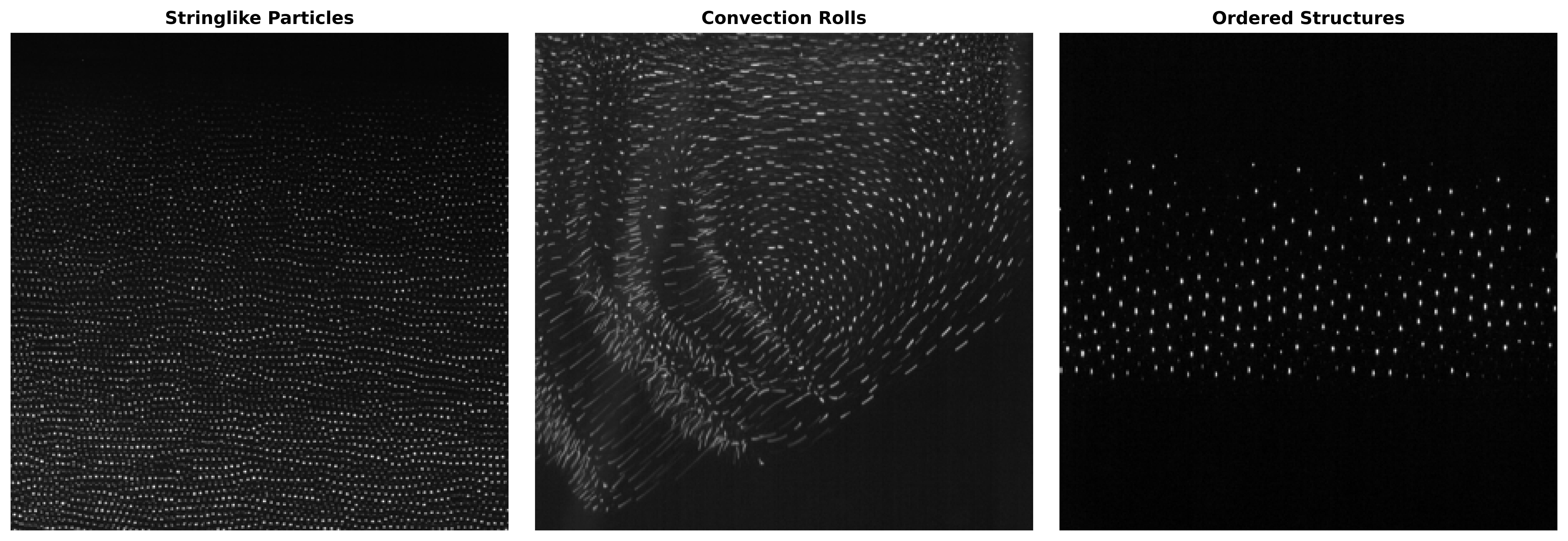}
\caption{Examples of experimental data from parabolic flights (from left to right): string-like structures, convection rolls, and more ordered particle aggregates. Since parabolic flights provide only a limited duration of microgravity, the formation of fully developed crystalline structures is challenging. Crystalline structures typically require longer times to evolve, making them more suitable for investigations aboard the ISS, where microgravity can be maintained over extended periods.}
\label{fig:experimental_data}
\end{figure}

%The trained networks were used to analyze various experimental datasets. 
An example of the parabolic flight experimental data consisting of string formations, convection rolls and ordered structures respectively is shown in Figure~\ref{fig:experimental_data}. These experiments were conducted under different pressures of $\SI{0.4}{mBar}$ and $\SI{1}{mBar}$, a constant plasma current of $\SI{1}{mA}$, polarity switching frequency of $\SI{1000}{Hz}$, and a particle size of $6.8 \mu m$ to analyze the dependence of convective motion on environmental parameters. The collected data consisted of video sequences capturing the rotational and translational movement of the particles within the plasma. These findings contributed to a deeper understanding of how thermal gradients drive convective motion in complex plasmas and how microgravity conditions influence these processes \cite{andy}. Previous parabolic flight campaigns investigated convection rolls induced by thermal gradients in complex plasmas. In these experiments, particles were exposed to a temperature gradient established by the PK-4 thermal manipulator. As the particles moved within the thermal field, convection rolls formed, leading to vortex-like plasma flow patterns.

These experiments were conducted during the 42nd DLR Parabolic Flight Campaign on the PK-4 setup under microgravity conditions. Furthermore, data from previous Parabolic Flight Campaigns in 2022, 2018 and 2016 were used. One of the key investigations focused on ordered crystalline structures within complex plasmas. In these experiments, microparticles were injected into the plasma and stabilized within the camera's field of view (FOV) through polarity switching of the applied electric field. This ensured that the particles remained suspended in the plasma without drifting. A tomographic scan was performed along the y-axis using a camera to capture the spatial distribution of the particle cloud. The collected 2D data was then reconstructed into a 3D representation of the particle positions, which allowed for the identification of crystalline structures within the plasma. These experiments were conducted under various conditions, including different particle sizes of $\SI{3.4}{\mu m}$ and $\SI{6.8}{\mu m}$, pressures ranging from $\SI{0.7}{mBar}$ to $\SI{1}{mBar}$, and polarity switching frequencies of $\SI{500}{Hz}$ an $\SI{1000}{Hz}$.

To study  the formation of particle strings and the electrorheological properties of complex plasmas, particles were injected into the plasma chamber and stabilized within the FOV through polarity switching. Under appropriate conditions, the formations emerge due to the modification of the interparticle potential caused by polarity switching. Different parameters, such as particle sizes of $\SI{3.4}{\mu m}$ and $\SI{6.8}{\mu m}$, pressures ranging from $\SI{0.4}{mBar}$ to $\SI{1}{mBar}$, duty cycle from 0 to 0.5, and polarity switching frequency of $\SI{500}{Hz}$ an $\SI{1000}{Hz}$, were varied to study their effects on string formation.

A crucial part of this experiment involved turning off the plasma at a certain point, causing the particle strings to dissipate. The plasma was then reactivated, allowing researchers to observe the reformation process of the particle strings. These observations provided insight into the dynamic stability and interactions governing the behavior of charged microparticles in a plasma environment.

To study the onset of dust acoustic waves in low-pressure plasma environments, experiments were conducted in which a particle cloud was injected and allowed to move across the observation region at different velocities. By varying the particle velocities up to very high speeds, researchers could analyze the conditions under which dust acoustic waves emerge. These waves are collective oscillations within a dusty plasma that arise due to the interplay between charged particles and plasma ions. The experiment was performed using different particle types and velocities, capturing high-speed video data of fast-moving particles. These observations helped to establish the threshold conditions for dust acoustic wave formation and provided insight into the role of kinetic energy and plasma parameters in wave propagation.

An additional investigation focused on the laser manipulation of microparticles within the PK-4 setup~\cite{PhysRevResearch.2.033404,10.1063/5.0069672}. In this experiment, particles were first injected and stabilized within the FOV using polarity switching. %The plasma was then turned off, and a focused optical manipulation laser was activated. 
A focused optical manipulation laser was activated to produce a shear flow in the system. During the observation of the particle motion, the plasma was turned off. The motion of the particles with and without plasma was compared and a viscosity of the dusty plasma system could be estimated.
By varying the laser power and the particle properties, researchers could examine the forces exerted on individual particles and their response to optical trapping. These experiments provided valuable data on the interaction between light and charged particles in a plasma environment, which has applications in fundamental plasma physics and potential technological applications.

Accurate and efficient detection, tracking, and classification of particles are important to the identification of dynamic structures under different conditions~\cite{Dormagen.2024}. Even slight deviations in positional accuracy can significantly affect the derived physical properties, such as pair correlation functions, distribution functions, diffusion coefficients and structural order parameters, leading to misinterpretations of the underlying physics\cite{Klein.2023}. One example is the investigation of dust-acoustic waves, which are low-frequency waves propagating through the charged microparticles in the plasma \cite{wimmer2024tilted}. The amplitude, phase velocity, and dispersion relation of these waves depend critically on the accurate determination of particle positions. Similarly, the study of electrorheological effects—where external electric fields influence the microstructure and dynamics of particle clouds—requires precise spatial resolution to differentiate between intrinsic and externally induced structure formation \cite{dietz2021phase}. Another example is the study of ion wakes in plasma. In microgravity conditions, the charged dust particles interact with streaming ions, forming asymmetric potential landscapes that influence particle motion. Correctly capturing these particle trajectories is vital for understanding ion focusing effects and the resulting wake structures. If tracking errors occur, the inferred wake patterns may be distorted, leading to incorrect models of plasma-dust interactions.

Traditional methods for particle tracking and structural analysis are computationally demanding and often unsuitable for real-time applications. Machine learning techniques, such as U-Net-based particle segmentation, self-organizing maps for trajectory matching, and PointNet-based classification have emerged as powerful tools for enhancing data analysis in complex plasma physics. ML tools can enhance data processing efficiency, allowing faster and more precise analysis of particle dynamics and crystallization processes. Moreover, real-time data processing through ML has the potential to revolutionize experimental execution, particularly in parabolic flight, space and other resource-limited environments. The growing integration of machine learning into plasma physics research paves the way for future advancements in autonomous experimentation and real-time data-driven insights. In short, data-driven methods enable the following:
\begin{itemize}

\item Higher accuracy in detecting particle positions under noisy conditions.

\item Improved particle tracking performance for analyzing dynamic plasma behavior.

\item Automated classification of structural formations, including string and crystalline structures.

\item Real-time processing capabilities for experiments conducted in microgravity environments, such as those aboard the parabolic flights and ISS.
\end{itemize}

By integrating different data-driven ML techniques, complex plasma research is moving toward a new era of high-precision, automated analysis. The ability to extract reliable and real-time information from experimental data not only enhances fundamental research but also opens new avenues for applications in plasma-based materials processing, space technology, and astrophysical plasma studies.

\subsection{Liquid droplets and aerosols}
The interaction between aerosolized droplets and gas-phase plasma is an important topic in dusty plasma physics, with applications spanning various fields, including biomedical engineering \cite{laroussi2021low}, material deposition \cite{palumbo2020recent}, environmental remediation \cite{yang2021transient}, and food processing \cite{jiang2017cold}. In these applications, aerosols are often intentionally generated by methods such as electrostatic aerosol atomization, ultrasound, and pneumatic techniques, or they appear naturally as targets for treatment. The interactions between aerosols and plasma can serve as sources of reactive oxygen and nitrogen species (RONS) or as carriers of drugs, protecting them from decomposition \cite{sremavcki2021influence}. This feature is particularly beneficial for plasma-liquid systems where the discharge is typically sustained in the gas phase, and the transport of reactive species is limited by the interfacial boundary. Including aerosols in the plasma-liquid system can significantly expand the surface-to-volume ratio, enhancing the transfer of activation energy from the plasma to the liquid, delivering short-lived species, and controlling reactivity in the liquid \cite{stancampiano2019plasma}.

Conveniently, a plasma-liquid system can generate aerosols to various degrees without an external mechanism. When a strong electric field ($>$24.6~kV/cm \cite{taylor1965stability}) is applied to the liquid surface, a Taylor cone forms to balance the electrostatic forces with surface tension, producing minute droplets at its tip where the electric field is strongest \cite{taylor1964disintegration}. This phenomenon has been observed at the liquid cathode of a glow discharge \cite{shirai2020mechanism, bruggeman2008dc, moon2020imaging}, grounded liquid of dielectric barrier discharge \cite{hoft2021role}, and corona discharge \cite{shirai2014atmospheric}. A Taylor cone is an example of an electrohydrodynamic instability, where the ratio of plasma sheath electric stress to surface energy determines the marginal condition for instability \cite{holgate2018electrohydrodynamic}. In such cases, higher sheath density or lower surface tension reduces the voltage threshold for Taylor cone formation, as observed in \cite{shirai2014atmospheric}.

\begin{figure}[htb!]
\centering
\includegraphics[width=.50\textwidth]{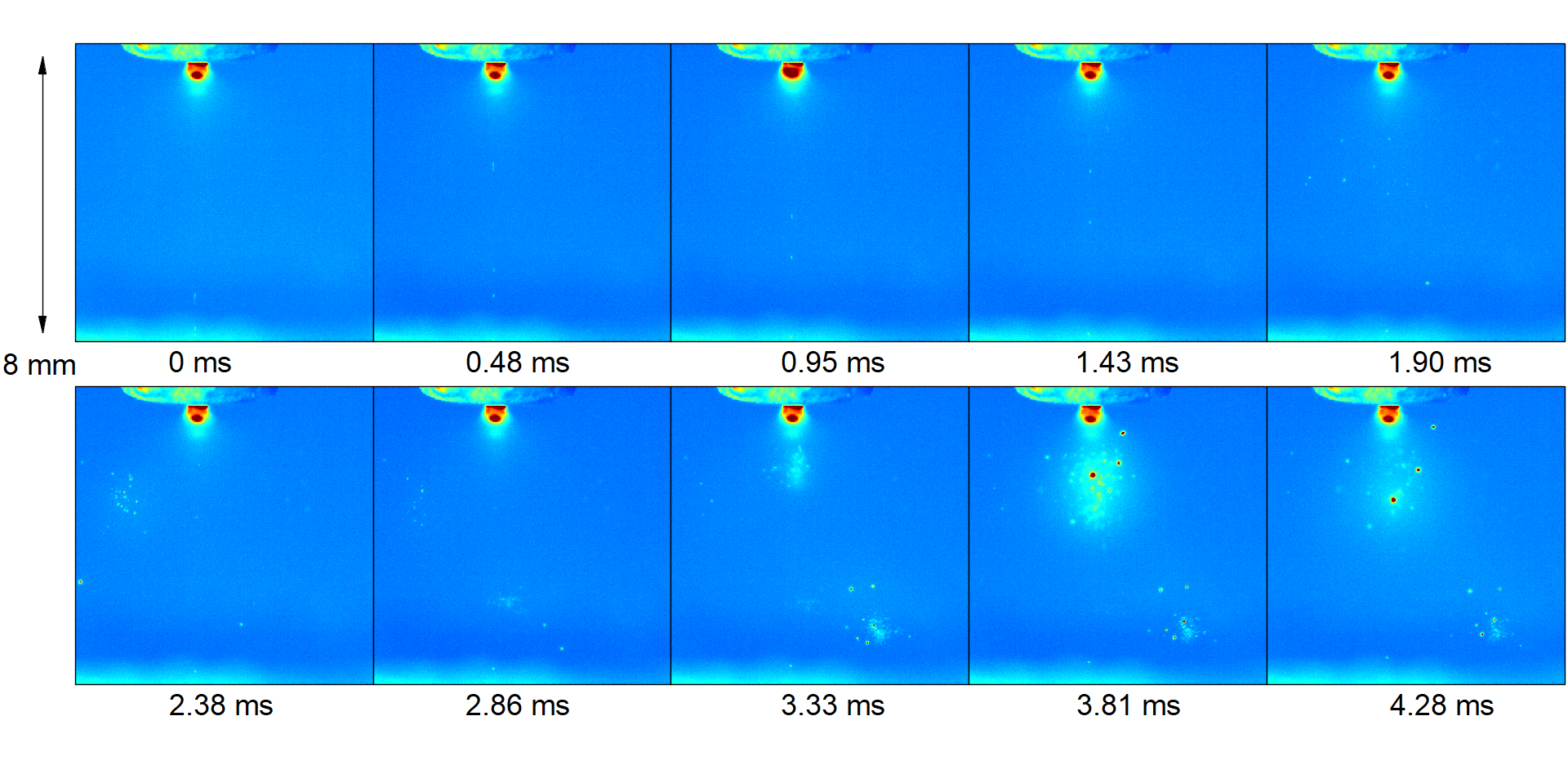} 
\caption{The droplet ejection from a liquid anode of DC glow discharge recorded at 27~kfps. Droplets have an average speed of 3~m/s and a diameter of 85~µm.}\label{droplets_eject}
\end{figure}

Droplets can also be produced readily from bubbles bursting at the liquid surface, a process common in mass transport of sea aerosols into the atmosphere, known as jet droplets \cite{spiel1994number, deike2022mass}. When a gas bubble reaches the liquid surface, its upper surface protrudes from the liquid-air interface, drains, and ruptures due to gravitational forces \cite{spiel1998births}. The shattering of the cavity and capillarity forms a central jet, which then breaks up into soaring droplets through the pinching-off process \cite{macintyre1974chemical}. These droplets have radii ranging from 2 to 500 µm, related to the size of the bubble \cite{kientzler1954photographic}. The speed of the jet scales with the bubble size and can be described by the ratio of gravitational force to surface tension \cite{spiel1994number}. This phenomenon is observed in diverse settings, from effervescent beer to the vast ocean surface, and was even noted in the context of a melted wall of a tokamak device damaging components \cite{shi2011boiling}. In plasma-liquid interactions, jet droplets can be a potential aerosol producer when an intense electric field isn't available, as seen in atmospheric gas-phase glow discharges running on an anodic electrolyte where the anode fall electric field ($\sim$1~kV/cm) is lower than the instability threshold. Evidence of droplet ejection and particle emission after vaporization was reported in \cite{kovach2021particle}.

\begin{figure}[htb!]
\begin{center}
  \includegraphics[width=0.5\textwidth]{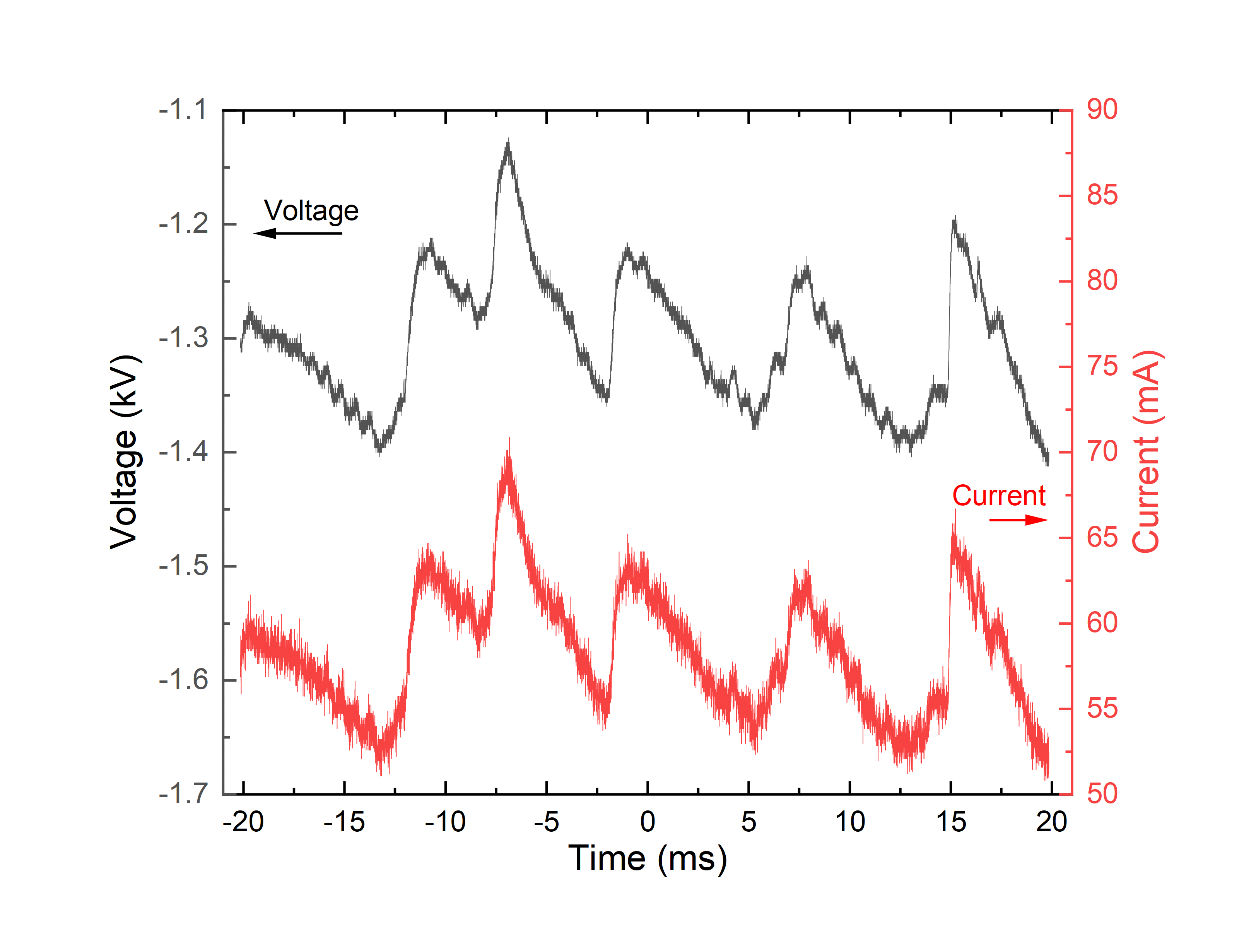} 
  \caption{Waveform of DC discharge interact jet droplets (Liquid anode: NaCl solution with initial conductivity of 14~mS/cm)}\label{droplets_IV_curve}
\end{center}
\end{figure}

\begin{figure}[htb!]
\begin{center}
  \includegraphics[width=0.5\textwidth]{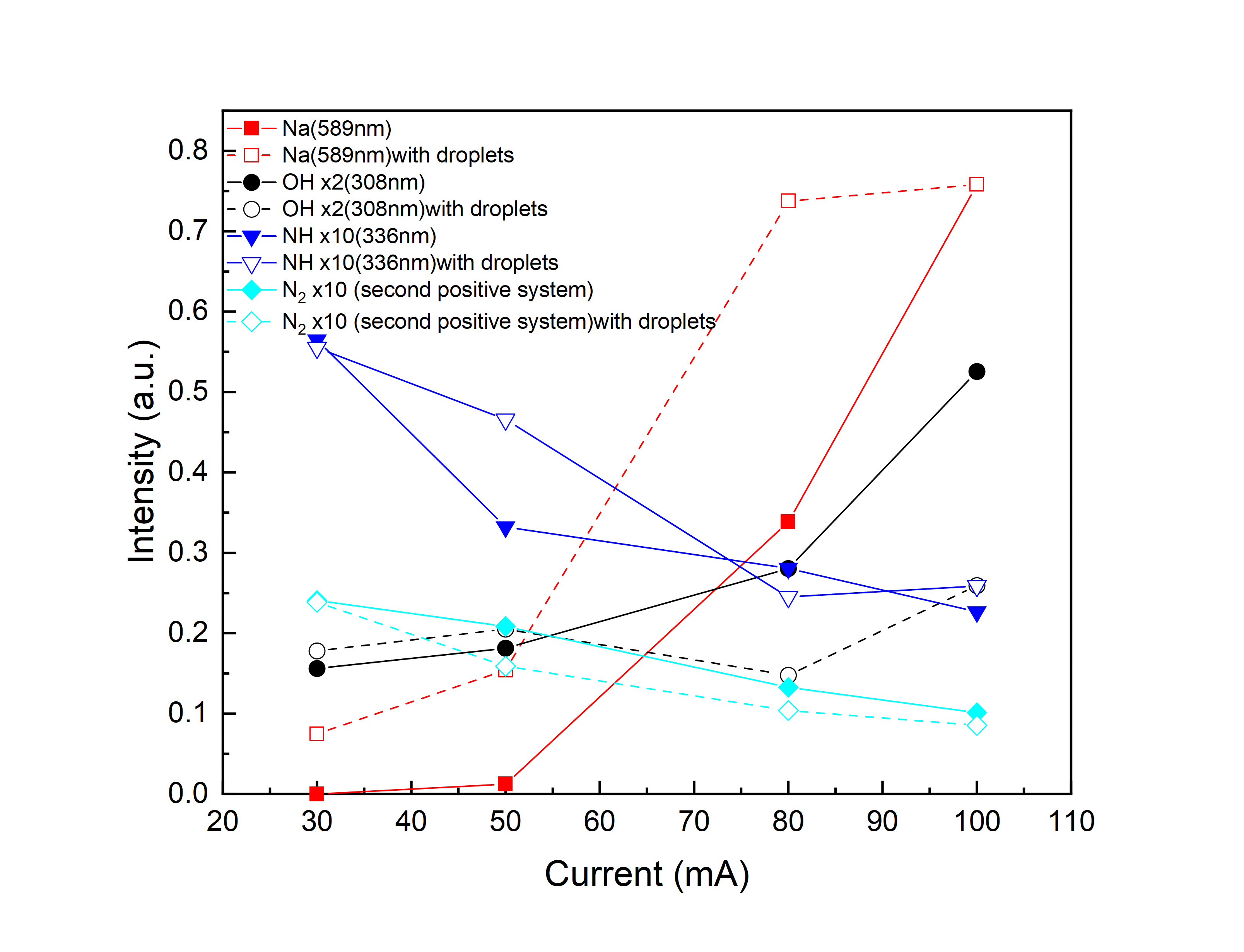} 
  \caption{The impact of jet droplets to plasma species emission (Liquid anode NaCl solution with initial conductivity: 14 mS/cm, some species’ intensities are multiplied to fit the scale).}\label{species_droplets_impact}
\end{center}
\end{figure}

Fig \ref{droplets_eject} shows the sequence of jet droplet ejection events from the liquid anode recorded by a high-speed camera at 27~kfps. The experimental setup is detailed in \cite{yang2022nature}. Compared to Taylor cone droplets from the liquid cathode, the amount of aerosol generated at the liquid anode largely depends on the dissolved gas source, and the droplet average velocity is relatively independent of the local electric field. Statistics show that the initial speed of jet droplets scales with their size, following a semi-empirical equation closely \cite{ghabache2016size}. In an atmospheric glow discharge, exothermic reactions and vibrationally excited molecules induce intensive ohmic heating in the positive column, reaching up to $\sim$2000~K \cite{kovach2019optical} and increasing the liquid phase temperature. The solubility of gas in water declines with temperature, and the released dissolved gas forms bubbles that rupture at the liquid surface. Although droplet ejection is a fluid dynamics phenomenon not unique to this discharge configuration, it can serve as a vehicle to carry solutes and solvents across the interface. Its formation mechanism suggests that it may be feasible to enhance generation by controlling the rate of bubbles formed in the plasma-activated electrolyte. To test this hypothesis, an airstone sparger was installed in the water to generate bubbles and subsequent jet droplets continuously. Fig \ref{droplets_IV_curve} shows the modulation in discharge current and voltage over time as jet droplets vaporize in the plasma. The time intervals among these wave peaks are in milliseconds, with a photodiode showing emission fluctuations synchronized with the IV curve. Each peak implies a sudden rise in plasma conductivity, attributed to increased electron density. Jet droplets also have a significant impact on species emission, as shown in Fig \ref{species_droplets_impact}. Additional droplets enhance OH(A) density at low discharge currents and decrease it at high currents. This reduction is possibly due to a shorter OH(A) lifetime due to collisional quenching by water molecules \cite{nikiforov2011influence}. The emission of the nitrogen second positive system and NH from the reaction between ambient nitrogen and water vapor shows a downward trend due to the reduced air fraction in the plasma. Remarkably, the sodium (589 nm) emission is prominently enhanced by additional droplets from the air stone, especially at low current, where sodium emission is extremely low due to slow mass transport at the plasma-liquid interface.

Overall, jet droplets provide a novel approach to mass transfer from the liquid and control of discharge maintenance by dispersing plasma-activated liquid into the gas phase and chemically and ionically enriching the plasma-liquid interaction. Their interaction with plasma involves complex processes in plasma chemistry, electrohydrodynamics, and thermodynamics, revealing another intricate regime of complex plasma.

%====================================end University of Michigan section=================================
%

\subsection{Phase transition: Melting of two-dimensional dusty plasmas \label{subsec:melting}}
%Cheng-Ran Du section, Mar.27, 2025
Since the discovery of plasma crystals in the laboratory, melting and crystallization in dusty plasmas have always been research topics of great interest~\cite{Thomas1996,Lin1996,Nosenko2009,Hartmann2010}. In contrast to three-dimensional plasma crystals, which have various lattice structures including fcc, bcc, and hcp, two-dimensional plasma crystals only exist as triangular lattice with hexagonal symmetry~\cite{Meyer2017,RubinZuzic2006,Khrapak2011}. Despite of the simpler structure composition, the melting transition in two-dimensional dusty plasmas is more complicated. There exists an intermediate phase between the crystalline and liquid phases, characterized by short-range translational order and quasi-long-range bond-orientational order~\cite{Zu2016,Li2023}. This phase is known as hexatic phase and can be well described by the Kosterlitz-Thouless–Halperin–Nelson–Young (KTHNY) theory based on the unbinding of topological defects~\cite{Strandburg1988}.

On the one hand, melting in the two-dimensional dusty plasmas can be caused by controlling the global parameters such as coupling parameter $\Gamma$ or screening parameter $\kappa$~\cite{Melzer2019,Couedel2022}. These parameters are usually associated with plasma parameters including Debye length, electron density, neutral damping rate. On the other hand, local disturbance such as shocks and shear flows can also induce the melting~\cite{Knapek2007,Feng2010,Jaiswal2019}. In addition,  the lateral wave of a fast-moving particle above or below the plasma crystal lattice leads to heat transport. Kinetic energy is transferred to the lattice particles via collisions with the self propelled extra particle, resulting in the melting of the crystal lattice~\cite{Du2012,Laut2017}.

To investigate the melting process in the experiments, positions of individual particles in the recordings need to be obtained, usually by the particle tracking algorithms~\cite{Feng2007,Ivanov2007}. However, this procedure is rather time-consuming if thousands of particles are involved. Moreover, pixel locking, inhomogeneous laser illumination, and elongated shape of fast-moving particles may significantly reduce the accuracy of the particle tracking, and thus affect the analysis. 

\begin{figure*}[!htb]
	\includegraphics[width=18cm]{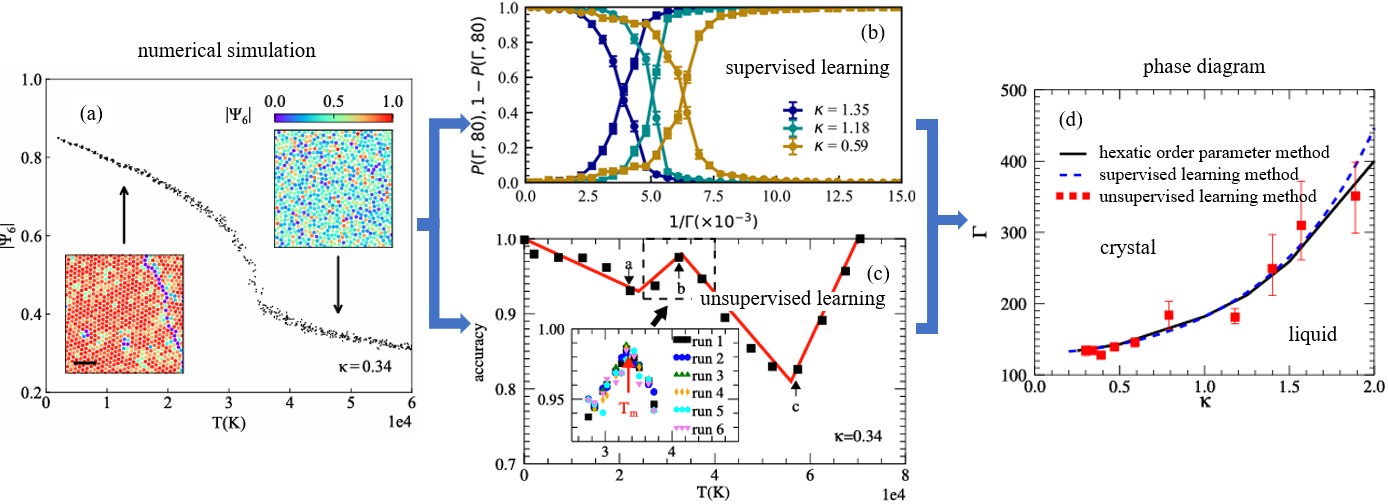}
	\caption{\label{fig_melting} Identification of the melting line with convolutional neural network. (a) Dependence of the averaged hexatic order parameter $|\Psi_6|$ on the kinetic temperature during the melting process. The insets show the spatial distribution of the local hexatic order parameter in crystalline (left) and liquid (right) phase with a scale bar of $2$~mm.  (b) Probability $P(\Gamma,80)$ of dusty plasma being classified as crystalline using supervised learning method as function of the coupling parameter $\Gamma$ for $N=80$. (c) Validation accuracy used unsupervised learning method as a function the assumed kinetic temperature for $\kappa=0.34$. The peak is zoomed in the inset with six independent runs. (d) Melting line in the phase diagram obtained from supervised and unsupervised learning method, compared with that from hexatic order parameter method \cite{Hartmann2005}. Reproduced from \cite{Huang2022,Li2024}. }
\label{fig_melting}
\end{figure*}

Deep learning, particularly convolutional neural network (ConvNet), has been widely applied in the image analysis. As the experimental results in dusty plasma research are often recorded as sequence of images, it is natural to apply ConvNet in the data analysis. Lack of experimental data poses challenges in the application of deep learning. To address this issue, MD ({\it aka}, Langevin dynamics) simulations have been performed. %The equation of motion including the damping from the neutral gas and Brownian motion of microparticles is given by
%\begin{equation}
%  \label{eq:motion}
%  m\ddot{\bm{r}}_{i} +m \nu \dot{\bm{r}}_{i}=-\sum_{j\ne i}\bigtriangledown\phi_{ij} +\bm{L}_{i},         
%\end{equation}
%where $\bm{r}_{i}$ is the position of particle $i$, $m$ is the particle mass, and $\nu$ is the damping rate, which results from the neutral gas. The Brownian motion was included in the equation and the corresponding Langevin force $\bm{L}_{i}$ at a certain kinetic temperature $T$ was defined by $\left \langle  \bm{L}_{i} \right \rangle=0$ and $\left \langle \bm{ L}_{i}(t)\bm{L}_{j}(t+\tau) \right \rangle = 2\nu m k_b T \delta_{i,j}\delta(\tau)\bm{I} $, where $\delta_{ij}$ is Kronecker delta, $\delta(\tau)$ is the delta function, and $\bm{I}$ is the unit matrix. In the simulation, it is commonly assumed that particles interact with each other via Yukawa potential~\cite{Ivlev2011}.
To study the melting process of plasma crystal, a heating process is performed in the simulation, as demonstrated in Fig.~\ref{fig_melting}(a). Initially, the system is equilibrated for a certain amount of time at a low temperature and then gradually heated at a small rate, until the plasma crystal is fully melted. The local structure around particle $i$ could be quantified by the hexatic order parameter $\Psi_{6,i}  = \frac{1}{6}\sum_{k=1}^{6}{e^{j6\theta_{k}}}$, where six nearest neighboring particles were considered and $\theta_{k}$ is the angle between $\bm{r}_{k}-\bm{r}_{i}$ and the $x$ axis. If particle $i$ is located in the center of  a perfect hexagon cell, its $\Psi_6$ is unity, as shown in the left inset in Fig.~\ref{fig_melting}(a).  

\begin{figure}[!htb]
	\includegraphics[width=7cm]{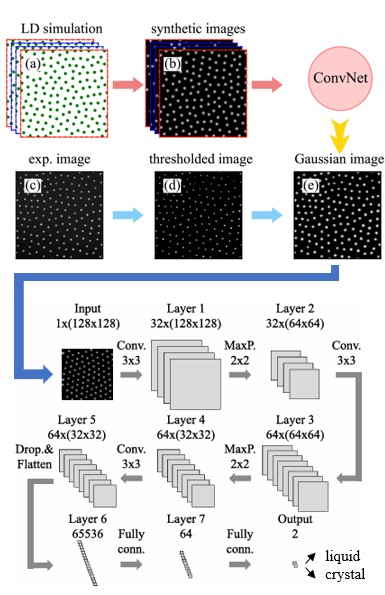}
	\caption{\label{fig_cnn}  Data preparation and structure of a typical convolutional neural network. (a,b) The particle positions are obtained in the numerical simulation and converted to the bitmap images. (c-e) The experiment images are processed with thresholds and Gaussian filter. They are fed to the ConvNet, typically composed of a few convolutional layers, max-pooling layers, dropout layers, fully connected layers, and output layers and then classified into liquid or crystalline phase.  Reproduced from \cite{Huang2022,Li2024}.}
\label{fig_cnn}
\end{figure}

For the purpose of model training and later application in the experiment analysis, sequences of images are prepared based on the simulation results. The particle positions at each time step in the simulation are converted to a gray-scale bitmap image, resembling the image obtained from the video recording in the experiments. Particularly, the intensity distribution of each bright spot in the synthesized image, corresponding to a particle, is assumed to follow the Gaussian distribution. The size is generally much larger than the real size of the particle, due to the diffraction by the particle as well as camera properties such as diffraction by the camera aperture and imperfect lens focusing. However, in the experiments, due to the presence of overexposure and noise in the recording, the brightness distribution of each particle deviates from the Gaussian distribution to some extent. To reduce such discrepancy, a threshold is applied to the images to remove noise, binarize the gray-scale bitmaps, and convolve the resulting binary images with the Gaussian filter. These steps ensure that the algorithm trained with the simulation results can be directly applied to the experiment analysis~\cite{Huang2022,Li2024,Du2024}.

ConvNet is employed to find the melting line, Fig.~\ref{fig_cnn}. ReLU activation function is used for each convolutional layer and the softmax activation function is used for the outpuut layer. Two machine learning strategies can be applied to investigate the melting transition and identify the melting line. In the supervised learning method, the training samples in the form of synthesized images at very high and low coupling are labeled, corresponding to the crystalline and liquid phases. Samples with a thermodynamic state near the critical value are not provided to the training, to avoid ambiguity~\cite{Carrasquilla2017,Li2021}. Once the training is completed, the model is applied to classify the complete data set  covering the full range of the coupling parameter. As the bitmap images are fed to the ConvNet model, it returns a probability $P(\Gamma,N)$ where the images are classified as plasma crystal. Here $N$ is the approximate number of particles included in the bitmap image sample and is selected as $N=80$ for the realization. The crossing of $P(\Gamma,N)$ and $1-P(\Gamma,N)$ provides the coupling strength
of the melting transition $\Gamma_m$.

The trained model can be directly applied to identify the melting transition in the experiments. If an extra particle moves fast above or below the two-dimensional plasma crystal, it attracts or repels the constituent particles of the crystal due to wake effect or Yukawa repulsion, destroying the ordered structure and melting the crystal. Using the ConvNet, the evolution of the localized melting and recrystallization can be monitored. It is also possible to shear-melt the plasma crystal with particle flow driven by optical pressure from manipulation lasers~\cite{VaPV:2021}. Using the same ConvNet, the spatial distribution of the liquid and crystalline structure can be obtained.

\begin{figure}[!htb]
	\includegraphics[width=7cm]{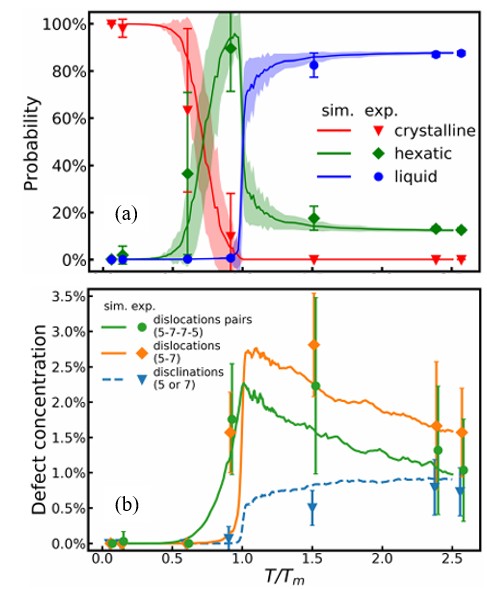}
	\caption{\label{fig_hexatic} Identification of the hexatic phase with convolutional neural network. (a) Probability of dusty plasma being classified as crystalline, hexatic, and liquid and (b) concentrations of dislocation pairs, dislocations, and disclinations as function of reduced kinetic temperature $T/T_m$ in the experiment and simulation. Reproduced from \cite{Du2024}. }
\label{fig_hexatic}
\end{figure}

In the unsupervised learning method, it is not necessary to know the thermodynamic state of any synthesized images. Instead, one can arbitrarily assume a melting temperature and divide the image sequences in a melting process into liquid and crystal accordingly.  If the assumed melting temperature coincides with the true melting temperature, the neural network can successfully recognize the different thermodynamics states of the dusty plasmas in the images based on the structures of the particle positions and thus reach a high validation accuracy~\cite{Nieuwenburg2017}. Otherwise, as ordered and disordered structure are mixed in the training sample, the network is confused and the validation accuracy is low. Two exceptions happen if the melting temperature is assumed to be the highest and lowest temperatures of the training sample such that all the images are virtually regarded in one single thermodynamics state, naturally resulting in a high accuracy in the validation. As result, the overall dependence of the validation accuracy on the assumed melting temperature exhibited a W-like shape and the peak corresponds to the true melting temperature, as shown in Fig.~\ref{fig_melting}(c). In fact, the melting line obtained from the unsupervised learning method agrees with that from the superwised learning, however with relatively bigger error for high $\kappa$, as shown in Fig.~\ref{fig_melting}(d).

In contrast to melting line, identification of hexatic phase is more challenging due to the facts that the temperature range is small and the structure is complicated~\cite{Sheridan2008,Vasilieva2021,Chiu2024}. To tackle this problem, traditional supervised learning is applied. First, training image samples from numerical simulations are labeled as crystalline, hexatic, and liquid phase based on the correlation functions of the positional and bond-orientational order parameters. Second, ConvNet is trained with these labeled sample, but with three-class classification. The classification is determined by the probability. Finally, the trained ConvNet model is applied to the validation dataset as well as the images from the experiments. The results are shown in Fig.~\ref{fig_hexatic}. The hexatic phase is successfully identified. To verify the results, the evolution of topological defects is analyzed.  As result, the temperature dependence of the measured and simulated concentrations of dislocations and disclinations also indicates the existence of the hexatic phase in two-dimensional dusty plasmas, in good agreement with the KTHNY theory. 

The ConvNet is easy to deploy using an open source platform for machine learning and efficient in application. For example, for a training sample containing $2000$ labeled gray-scale bitmap images with a size of $100 \times 100$~pixel$^2$, it takes $5$ minutes to train a model with batches of $100$ images for $20$ epochs on a desktop machine with Intel(R) Core i7-10900 CPU and takes seconds to classify an image from experiments. This is much more efficient than the scheme using traditional particle tracking and order parameter analysis. Such methods can also be extended to classify the liquid and crystalline structure in three-dimensional dusty plasmas in the laboratory and in the payload on board the Space Station~\cite{Thomas2008,Pustylnik2016,Knapek2021}.

%%%%%%%%%=============End Cheng-Ran Du section, Mar. 27, 2025 ============================

%
%
%
%
%
%

\subsection{Dust dynamics in an afterglow plasma}

Chaubey et al.~\cite{ref12} %the same as Chaubey:2021
reported that particles retain a large electric charge in an afterglow plasma, and can have a reversal in the polarity of electric charge from negative to positive. In the late afterglow of that experiment, after electrons and ions had substantially vanished, there was a downward macroscopic electric field in the chamber, and of course a downward gravitation force. The combination of electric and gravitational forces caused the charged particles to fall with an acceleration larger than 1 g from their starting height of 14 mm toward a horizontal lower electrode below. Because the acceleration greatly exceeded 1 g, it was clear that the particles had a large charge with a positive polarity in the afterglow, which was unexpected at the time. By analyzing images to obtain the acceleration of particles, the authors of ~\cite{ref12} were able to use Newton’s Second Law to directly obtain the value of the particle charge, which was about +10$^4$ e. In analyzing their results for plasmas with various voltages on the lower electrode in the afterglow, the authors found that the value of the residual charge increased with the energy of ions flowing due to the macroscopic electric field. This result indicated that those ions remained in sufficient number to charge the particle after electrons had mainly left the region where the particles were located.

In a later paper, Chaubey and Goree~\cite{Chaubey:2022} reported another afterglow dusty plasma experiment, where the positively charged particles were caused to fall more slowly, by applying an upward electric force. This upward electric force was applied using a transistor switch to apply an external dc power supply to the lower electrode, starting 2 ms after the rf power was turned off. Instead of falling to the lower electrode in about 40 ms as in the earlier work~\cite{ref12}, the particles fell more slowly over the course of 190 ms, due to the upward electric force. With this slower downward motion, there was more time for the cloud of particles to expand horizontally, due to Coulomb interparticle repulsion. The circular cloud’s radius expanded from about 11 to 18 mm during that time.

\subsection{Nanometer dust and VSGs}
%Kortshagen papers

Nanometer dust or VSGs can play multiple roles in dusty plasmas~\cite{BKSB:2009,Boufendi_2011}. NPs are usually refer to atomic or molecular clusters less than $\sim$100 nm in size, which are smaller than the wavelengths of visible light, and thus pose significant {\it in-situ} real-time diagnostic challenges through light scattering. Their small size, large surface-to-volume ratio, and versatile structures (spatial distributions of the atoms) are nevertheless key to NP's unique properties, defining their roles in dusty plasmas.  NPs directly participate in materials dynamics, such as seeding micrometer and larger dust growth. NPs changes Newtonian dynamics of dust through modifications of charging, other materials and surface properties of dust.  NPs can also affect the coupling between neighboring dust grains and influence the large structure formation out of a large body of dust particles.  While NPs contaminants must be mitigated in semiconductor processing to make micro-chips, NPs have also found wide and expanding applications in micro- and nano-electronics, catalysis, solar cells, lighting, hydrogen storage, battery, bioimaging and quantum sensing, and medicine~\cite{KSMR:2016, KSLH:2019, C8NR07769J}. 

Correlating the key NP properties, \underline{Co}mposition, \underline{S}ize and \underline{S}tructure (CoSS), with their functions, such as catalytic, quantum mechanical, optical, therapeutical functions, remains a grand challenge in NP applications. Proteins, as naturally occurring NPs, offer illuminating examples about the abundance, complexity, and challenge of the CoSS-function correlations as well as the applications~\cite{pharmaceutics12070604}. The Protein Data Bank (PDB) is a curated open database for the three-dimensional structural data of proteins and nucleic acids~\cite{PDBH:2000}. There are more than 2$\times$10$^5$ listed structures in the PDB as of 2023. Similar databases for NPs may be envisioned~\cite{TrMH:2017}. Besides structures, classification of NPs can be based on the size (1-100 nm), composition, synthesis method, functions~\cite{JoLi:2022}, and applications. There are about $N_{el} \sim$ 10$^2$ kinds of single-element NPs, including the precious metal NPs, such as Au~\cite{SaKh:2021}, Ag, semiconductor nanocrystals of Si and Ge. The type of binary element NPs is $\sim N_{el}^2 = 10^4$, including organic NPs such as C$_x$H$_y$, silicate (Si$_x$O$_y$) particles, wide bandgap semiconductors such as SiC, BN, and titanium dioxide (TiO$_2$)~\cite{RJTM:2024}. Ternary element NPs, estimated to be $\sim N_{el}^3 = 10^6$, may include ternary semiconductors and perovskites.  With the recent introduction of high-entropy NPs~\cite{YDBL:2022}, the types of man-made or artificial NPs in terms of atomic composition are increasing rapidly. %The combination of atomic compositions and structural variation of NPs requires the combination of experiments, computation, and lately information and data science.  

Besides NP classification, there are growing evidences to show that informatics approaches based on machine learning and AI can successfully elucidate and forecast structure–property relationships for nanomaterials, and even guide the experimental discovery of new nanomaterials with desired properties~\cite{TrMH:2017}. 
Dusty plasma community can contribute in several unique ways by introducing data-driven approaches.

One area is to tailor NP  synthesis for targeted applications. In bio-medical applications, for example, a large amount of data regarding the synthesis, physicochemical properties, and bioactivities of nanomaterials now exist~\cite{nano11061599}. A database of inorganic nanoparticles, comprising experimental
datasets from 745 preclinical studies in cancer nanomedicine, were described recently~\cite{MeZC:2024}. Nanoparticle shape
and therapy type were found to be prominent features in determining in vivo efficacy. These datasets can be used to guide NP synthesis, through tuning the atomic composition, size and structures of NPs.  Plasma sources~\cite{MarS_2010,D4NR02478H}, non-thermal plasmas at atmospheric and lower pressures included~\cite{LU20161,WKHU:2019},  complement traditional chemistry (or solution)-based, high-temperature gas-based, such as flame or furnace, biological, and other physical, such as laser ablation, methods in NP synthesis~\cite{KUMARI20231739}. Non-thermal plasmas can produce energetic electrons, negative ions~\cite{BKSB:2009}, atoms, molecules, nano-clusters, and other reactive species to enhance NP production, with materials flexibility (solid, liquid and gas precursors), size control, and industrial scalability. Moreover, plasma processes are environmentally sustainable, with reduced energy consumption, fewer chemical precursors and less waste than traditional wet chemical approaches.
Plasmas synthesis meanwhile may be limited by the kind of gaseous precursors available (some of them are highly toxic), difficult-to-control surface and ambient conditions, sensitive to power or frequency (DC, RF, or microwave) inputs, and limited access to in-situ characterization tools~\cite{Boufendi_2011}. Vacuum and diagnostic equipment for plasma synthesis can also be expensive.%

Another area with a significant growth potential is in using data-driven approaches to optimize or accelerate plasma NP synthesis through integration of NP co-design and modeling with in-situ characterization and synthesis controls~\cite{Etching:24}. The development of high-throughput computational and experimental methods can accelerate the material, size and structure exploration and enable machine-learning tools for performance prediction and guided optimization~\cite{YDBL:2022}. %Plasma-based methods offer unique advantages, including high purity, narrow size distribution, and scalability [2,3]. in advancing our understanding of nanoparticle synthesis in low-pressure plasmas [11]. Their pioneering 
Studies on silicon and germanium nanocrystals have demonstrated precise control over particle size and crystallinity, paving the way for applications in optoelectronics and energy conversion~\cite{kortshagen2009nonthermal,KSMR:2016}. Further advances in in-situ diagnostics and computational modeling would be highly desirable to improved understanding of nanoparticle formation mechanisms and growth kinetics.
Datasets encompassing various plasma parameters, precursor materials, and resulting nanoparticle characteristics would enable the development of predictive models and optimization strategies. %%

\subsection{Solar and lunar dusty plasmas}

Natural dusty plasmas are ubiquitous throughout the solar system and beyond, appearing in environments such as noctilucent clouds in Earth’s upper atmosphere~\cite{LuBB:2021}, the lunar exosphere, Martian dust storms, cometary comae, planetary ring systems, interplanetary dust clouds, protoplanetary disks, and interstellar and intergalactic media. These systems span a wide range of physical conditions, from weakly ionized atmospheric plasmas to strongly coupled dusty environments in astrophysical settings. Understanding dusty plasma processes is therefore essential not only for advancing fundamental plasma physics and astrophysics, but also for enabling space exploration and assessing planetary habitability, including future missions such as NASA’s Artemis program~\cite{NASAArtemis2020}.

Recent advances in ground- and space-based observational capabilities have generated vast and heterogeneous datasets, including high-resolution imagery, spectroscopic measurements, in situ plasma diagnostics, and particle detection data. Ground-based telescopes and observatories enable long-term monitoring of phenomena such as noctilucent clouds, zodiacal light, cometary activity, and protoplanetary disks, particularly through infrared and millimeter observations~\cite{LuBB:2021,Andrews2020}. Space telescopes (e.g., \textit{Hubble}, \textit{JWST}, \textit{Spitzer}, \textit{SOHO}) and heliophysics missions provide complementary high-resolution, multi-wavelength observations of dusty plasma environments. These datasets include high-dynamic-range imaging of planetary ring systems~\cite{PBBB:2005}, spectroscopic measurements of dust composition~\cite{Drai:2003}, polarimetric observations constraining grain size distributions~\cite{KHLG:2004}, and time-resolved imaging of cometary structures~\cite{ABDF:2011}.

The resulting data volumes—often reaching petabyte scale across image archives, spectral cubes, and time-series measurements—necessitate automated methods for detection, segmentation, classification, and feature extraction. Data-driven approaches, particularly machine learning (ML) and artificial intelligence (AI), are increasingly employed alongside traditional physics- and statistics-based techniques to extract physical insight from these complex, high-dimensional observations.

Due to their compactness, ML and AI methods may also allow, in the future, real-time in-situ analysis and reduction of the large amount of raw data available, and autonomous extraction of useful information about the plasma and dust, such as  in-situ characterization of the dust composition, shape and size classification.  Satellites and planetary flybys, including lunar missions, missions to astroids, comets, and planets. provide in situ measurements of Plasma density and temperature, Electric and magnetic fields, Dust particle flux and charge states, Grain velocity distributions
These measurements are often high-dimensional and multi-modal, combining particle counters, Langmuir probes, magnetometers, and imaging systems. Data fusion techniques are essential to interpret such datasets.
See Sec.~(\ref{Edge:AI}) for some possibilities.

\textit{Circumlunar Dust Data.}
In situ measurements from the Lunar Dust Experiment (LDEX) aboard the LADEE mission have fundamentally refined our understanding of the circumlunar dust environment, demonstrating the existence of a persistent, yet highly variable, impact-generated dust exosphere extending from a few kilometers up to $\sim$250~km altitude \cite{HSLD:2014,HSKS:2015}. Lunar dust grains are immersed in a flowing solar-wind gas ambient with a molecular density on the order of 10$^6$ cm$^{-3}$~\cite{Ster:1999}. LDEX directly detected individual grains with radii $\gtrsim 0.3~\mu$m through impact ionization and recorded cumulative signals from smaller particles, establishing typical impact rates of $\sim$1~min$^{-1}$ for charges $q>0.5$~fC and $\sim$0.1~min$^{-1}$ for $q>5$~fC \cite{PZGD:2018}. The inferred number densities at high altitudes are on the order of $(0.4$--$4)\times10^{-9}$~cm$^{-3}$, consistent, within an order of magnitude, with theoretical models of ejecta production driven by micrometeoroid bombardment \cite{PZGD:2018}. Temporal variability correlated with meteor streams---most notably the Geminids---confirms the stochastic impact origin of the observed dust cloud. Critically, both LADEE observations and subsequent theoretical analyses converge on the conclusion that the circumlunar dust population at altitudes of tens to hundreds of kilometers is dominated by hypervelocity impact ejecta, with no observational evidence for a sustained population of electrostatically lofted nanodust at high altitude \cite{PZGD:2018}. These results establish meteoroid gardening and impact vaporization as the principal mechanisms maintaining the lunar dust exosphere, providing a quantitative benchmark for future missions investigating dust transport, surface evolution, and exospheric dynamics.

\begin{figure}[!htb]
\includegraphics[width=3.3in]{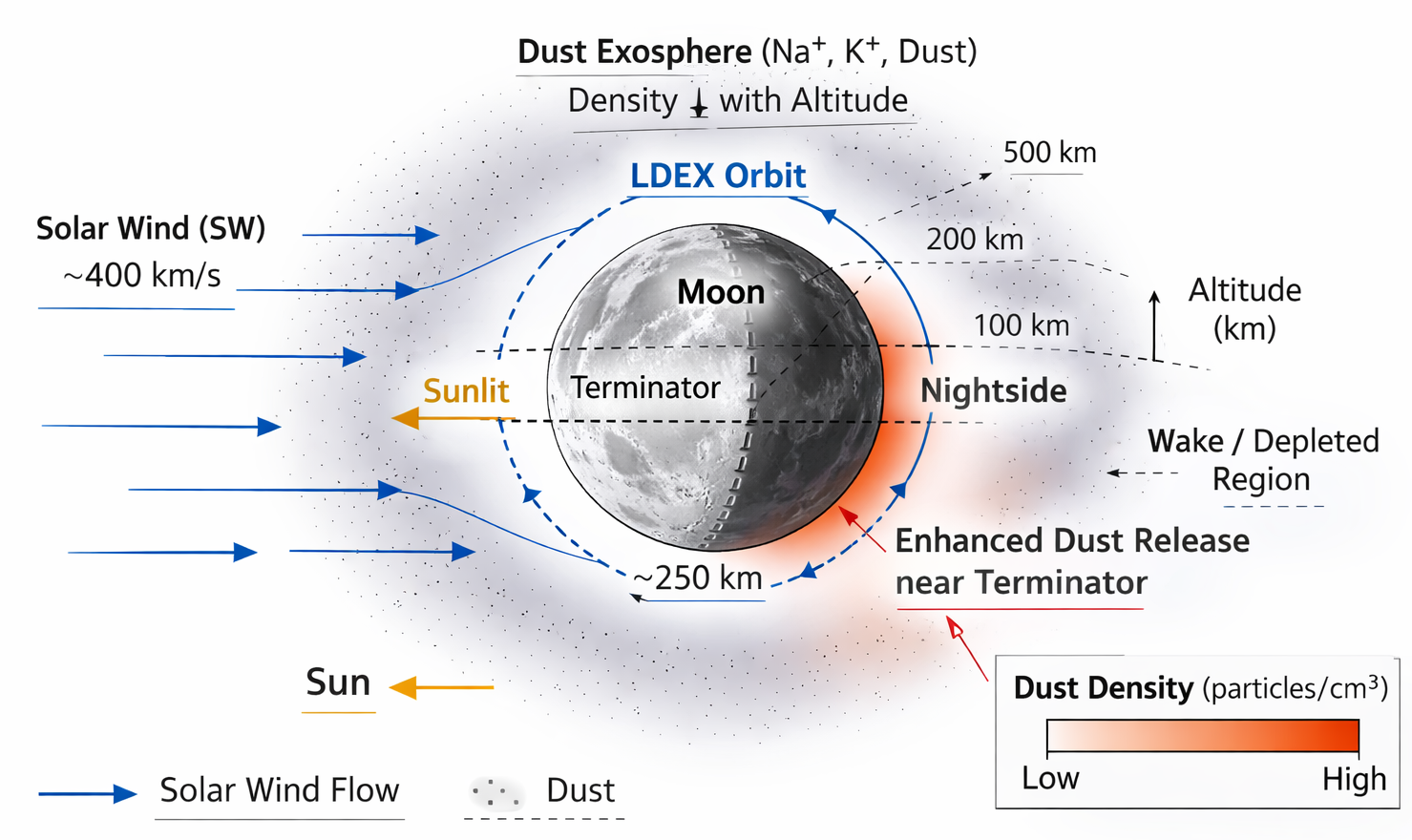}
\caption{A schematic illustration of the circumlunar dusty plasma environment, generated using AI-assisted tools (ChatGPT). Dust grains are produced primarily through micrometeoroid impacts and electrostatic lofting from the lunar surface, forming a tenuous dust exosphere with dust density decreasing with altitude. The Moon is embedded in the solar wind plasma (typical density $\sim 10^6$ cm$^{-3}$ at the Moon), which governs dust charging, transport, and plasma–dust interactions. Enhanced dust activity is expected near the lunar terminator, where strong electric field gradients arise from transitions between illuminated and shadowed regions. Future lunar exploration and sustained human presence may significantly perturb the natural dusty plasma environment through surface operations, resource utilization, and artificial plasma and dust generation.}
\label{fig:lunarSW}
\end{figure}

\textit{Dusty Plasma Data.}
The dusty plasma environment near the lunar surface constitutes a strongly coupled system composed of charged dust grains, photoelectrons emitted from regolith and grain surfaces, solar wind electrons and ions, and, intermittently, magnetotail plasma \cite{PZGD:2018}. On the illuminated dayside, intense photoemission produces a dense near-surface photoelectron sheath, with surface densities $N_0$ reaching up to $\sim10^{5}$~cm$^{-3}$ depending sensitively on the quantum yield and work function of the regolith. Self-consistent kinetic–fluid modeling shows that photoelectrons emitted not only from the surface but also from levitated grains substantially modify the electric field structure and dust charging equilibrium, enabling nanometer- to submicron-scale particles to levitate within centimeter- to meter-scale layers above the surface. Contrary to earlier assumptions, no ``dead zone'' exists near $\sim80^\circ$ latitude where dust lifting would be suppressed; electrostatic levitation is theoretically possible across essentially the entire illuminated surface. At the lunar terminator, the transition from a photoelectron-dominated dayside plasma to a negatively charged nightside environment forms a sheath-like structure with characteristic electric fields on the order of $\sim10^{2}$~V/m, capable of accelerating micron-sized grains to altitudes of tens of centimeters and contributing to the observed horizon glow phenomena. Nevertheless, while electrostatic processes govern near-surface levitation and plasma–dust coupling, meteoroid impacts remain the dominant source of micron-scale particles in the global lunar environment, and LADEE data confirm that high-altitude dust populations are impact-generated rather than electrostatically sustained \cite{PZGD:2018}. Together, these theoretical and observational advances define a coherent picture: electrostatic dusty plasma processes shape the near-surface layer, whereas impact ejecta controls the extended circumlunar dust cloud. An illustration of circumlunar dusty plasma exosphere is generated by ChatGPT, shown in Fig.~\ref{fig:lunarSW}.

\subsection{Synthetic data}
% Dormagen, Klein

Simulations provide a powerful means to generate synthetic datasets for training and validating machine learning models. In addition, generative AI methods are emerging as complementary tools for producing realistic data distributions.

\begin{figure}[!htb]
\includegraphics[width=3.3in]{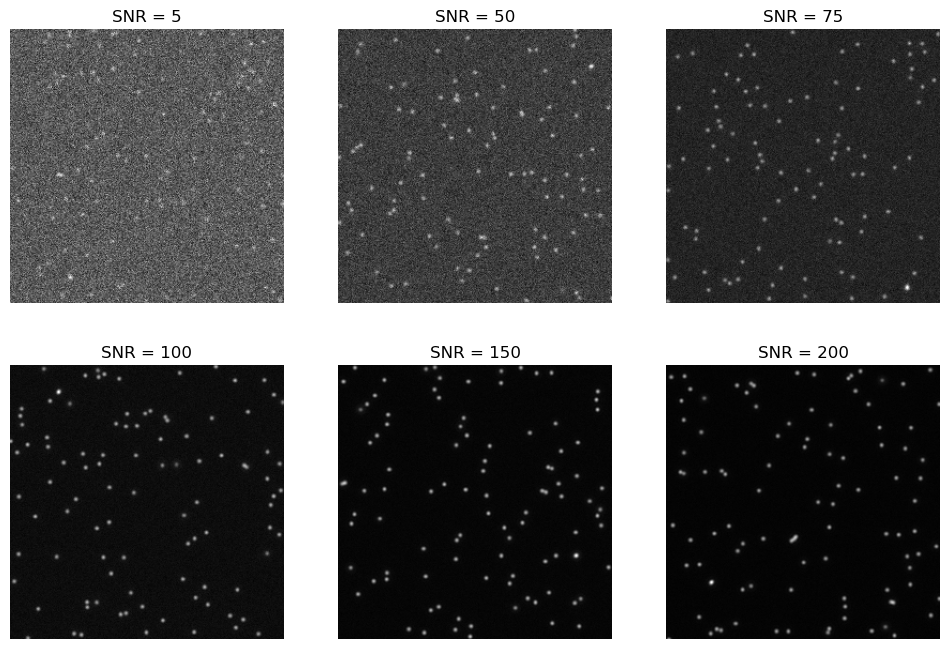}
\caption{Extract of the training dataset: artificial images with different signal-to-noise ratios.}
\label{fig7}
\end{figure}

As an example, synthetic image datasets were constructed using the DeepTrack simulation framework~\cite{midtvedt2021quantitative}, which reproduces particle distributions in complex plasma environments. To improve robustness, images were generated across a range of signal-to-noise ratios (SNR), Fig.~\ref{fig7}, allowing the neural network to learn noise-invariant features. This strategy enhances model generalization and resilience when applied to real experimental data, where noise levels and imaging conditions can vary significantly.

The synthetic datasets consisted of 128×128 pixel images with variations in particle count and noise levels to simulate different experimental conditions. Each image was accompanied by labeled masks marking precise particle positions, serving as ground truth labels for model training. Additionally, lighting conditions were carefully adjusted to resemble the laser illumination used in the PK-4 setup, ensuring seamless transferability to real-world data.

Beyond standard isotropic particle distributions, artificially generated string-like structures were incorporated to train the model to recognize more complex formations. Special labeled masks were created specifically for these structures to improve training effectiveness. To enhance visualization and analysis, the image resolution was increased to 256$\times$256 pixels.

\begin{figure*}[!htb]
\centering
\includegraphics[scale=0.5]{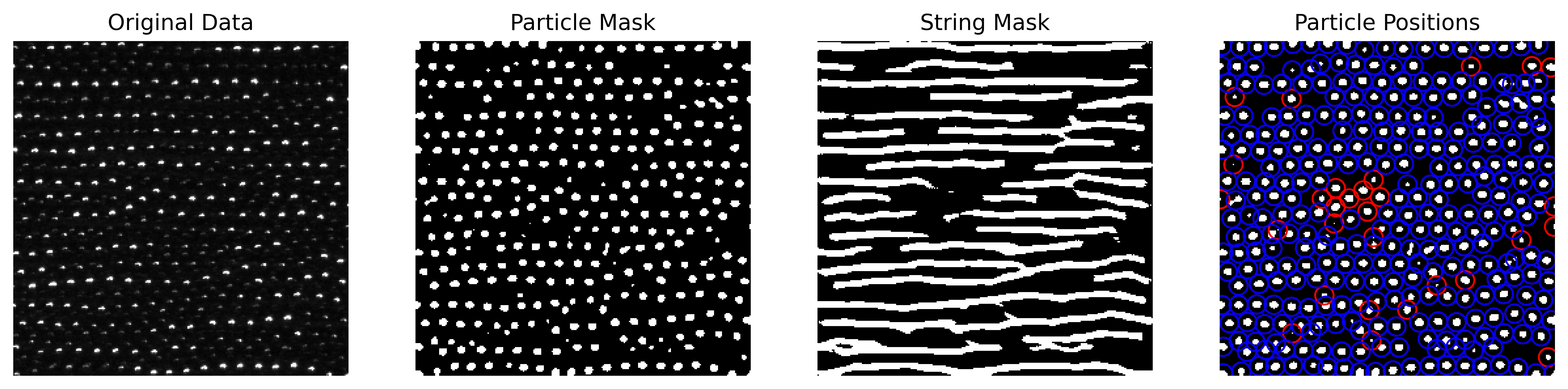}
\caption{Illustration of the original data, the particle mask, the generated string mask, and the analyzed particles. Particles inside strings are marked in blue, while those outside are marked in red.}
\label{fig:strings}
\end{figure*}

\begin{figure}[!htb]
\begin{center}
\includegraphics[width=3.3in]{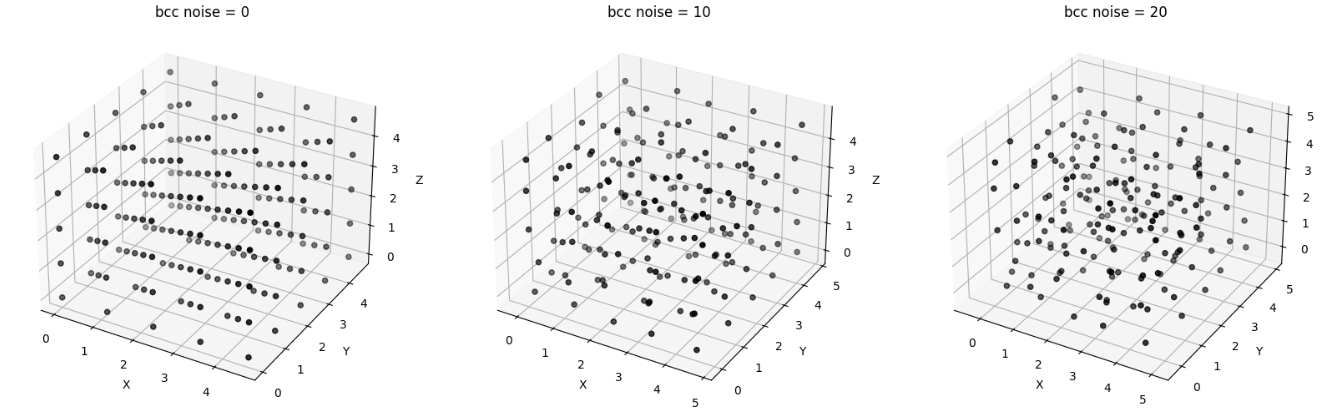}
\caption{Examples of the synthetic data at different noise levels.}
\label{art_data}
\end{center}
\end{figure}

These artificial datasets played a crucial role in training a U-Net model, enabling it to detect particles and structures under varying conditions, including different noise levels and distribution complexities, Fig.~\ref{fig:strings}. By incorporating controlled variations into the synthetic data, the model was trained to generalize across a broad range of scenarios, preparing it for the challenges posed by the noisier and more diverse experimental datasets collected from the PK-4 experiment.
For the study of plasma crystals, three-dimensional data was required, leading to the creation of artificial 3D datasets that accurately represent crystalline structures. These datasets modeled the spatial arrangements of hexagonal close-packed (hcp), body-centered cubic (bcc), and face-centered cubic (fcc) structures \cite{Dormagen.N}. To ensure model robustness across various experimental conditions, noise was artificially introduced at different levels, incorporating drift movements of individual particles to better simulate real experimental conditions.
To introduce noise into the artificial data, Gaussian noise was added to the dataset. This Gaussian-distributed shift mimics the Brownian motion of the particles. %The standard deviation  of the Gaussian noise was gradually increased from  to  in steps of , where  is the average nearest-neighbor distance. 
Consequently, domains were generated for each structure within a specified volume for each noise level, Fig.~\ref{art_data}.

 \subsection{Text data about dusty plasmas}
 \label{sec:text_data_dp}

Text corpora constitute an essential component of any dusty-plasma AI model, noticeably a class of models known as Foundational Models (FoMos), see Sec.~(\ref{sec:FoMo}) for additional details. A FoMo will enable many automation tasks, including but not limited to multi-modal data processing and interpretation, cross-referencing observational results with theoretical models, and automated literature synthesis for multi-scale phenomena. Large language models (LLMs) have emerged as a powerful FoMo framework for extracting structured knowledge, identifying trends, mapping citation networks, and linking observational datasets to governing physical principles. In the context of dusty plasmas, textual data provide not only descriptive and theoretical content but also methodological details about instrumentation, calibration, numerical modeling, and experimental protocols that are difficult to infer from numerical data alone.

The peer-reviewed dusty-plasma literature forms a well-defined domain-specific corpus. A typical scientific article has a digital size of approximately 1--5~MB (including figures, tables, and supplementary material) and it is estimated that the total number of existing dusty-plasma–relevant papers is on the order of $10^4$. The aggregate volume of peer-reviewed text would therefore lie in the range of 10--50~GB. In addition, monographs and edited volumes, estimated to number on the order of $\sim$50 major books in plasma physics, complex plasmas, and related topics, each potentially $\sim$100~MB in digital form, would contribute an additional $\sim$5~GB. While modest compared to general web-scale corpora, this domain-specific literature is dense in physical content and contains high-value structured information (equations, parameter tables, scaling laws, instrument descriptions, and experimental constraints) that is particularly suitable for domain adaptation of scientific LLMs.

Within the broader plasma-physics domain, the literature landscape is substantially larger. Bibliographic databases report approximately 3,369,055 entries associated with plasma-related research. The dusty-plasma subfield itself comprises roughly $\sim$12,000 entries, reflecting its position as a specialized branch of plasma physics. By contrast, literature explicitly labeled as ``data-driven plasma'' accounts for approximately $\sim$1,888 entries, while ``data-driven dusty plasma'' remains extremely limited, with on the order of $\sim$10 entries. This quantitative disparity highlights a major opportunity: the application of data-driven and FoMo approaches to dusty-plasma science is still in its infancy, and systematic text mining and corpus construction can accelerate methodological innovation in this area.

Beyond formal peer-reviewed publications, open-access web resources and scientific repositories provide additional textual and multimodal content. Openly accessible web data (including text and images) are estimated to be on the order of 100--300~terabytes globally, though only a fraction is scientifically relevant. Careful curation is therefore essential to avoid contamination from non-physical or low-quality sources, such as `sci-fi' sources which may violate the fundamental laws of physis. High-value open scientific literature databases include \emph{arXiv} (physics, mathematics, computer science), \emph{PubMed Central (PMC)}, \emph{CORE}, \emph{Semantic Scholar}, the \emph{Directory of Open Access Journals (DOAJ)}, open-access content from \emph{ScienceDirect} and \emph{IEEE Xplore}, preprint servers such as \emph{bioRxiv} and \emph{medRxiv}, and general-purpose repositories such as \emph{Zenodo}. These platforms collectively provide structured metadata, citation graphs, abstracts, and in many cases full-text access suitable for machine ingestion under appropriate licensing constraints.

In addition to corpus acquisition, modern citation cartography and literature analytics tools can significantly enhance corpus construction. Platforms such as ResearchRabbit (\url{https://researchrabbitapp.com/}) enable dynamic mapping of citation networks and identification of thematic clusters, while AI-driven search tools such as SciSpace (\url{https://scispace.com/search?q=dusty+plasma}) facilitate semantic exploration of subtopics (e.g., terminator charging, meteoroid ejecta, photoelectron sheaths, or dust acoustic waves). These tools allow systematic identification of core references, emerging trends, and cross-disciplinary linkages (e.g., between dusty plasmas, nanoparticle physics, and astrophysical dust), thereby improving coverage and balance in training corpora.

For a dusty-plasma FoMo, see Sec.~(\ref{sec:FoMo}), the optimal strategy is not indiscriminate ingestion of all available text, but rather structured curation. The corpus should include (i) foundational theoretical works defining charging models, sheath theory, and dust–plasma coupling; (ii) experimental and in situ mission reports detailing instrumentation and calibration; (iii) numerical and simulation studies describing parameter regimes and scaling laws; and (iv) emerging data-driven or machine-learning studies. When combined with tokenization and domain-adaptive pretraining strategies, this curated textual corpus can provide a scientifically grounded language backbone that complements the image and time-series modalities of a multi-modal dusty-plasma foundation model.

In summary, while the dusty-plasma textual corpus is modest compared to general web-scale datasets, it is sufficiently rich, structured, and well-cited to support domain-specific language modeling. The relatively small size of the explicitly data-driven dusty-plasma literature further underscores the importance of systematic text mining and foundation-model integration as a means to catalyze the next phase of computational and data-centric advances in dusty-plasma research.

\section{DustNET: data and algorithm fusion \label{Sec:dFuion}}

\begin{figure}[!htb]
    \centering
    \includegraphics[width=0.85\columnwidth]{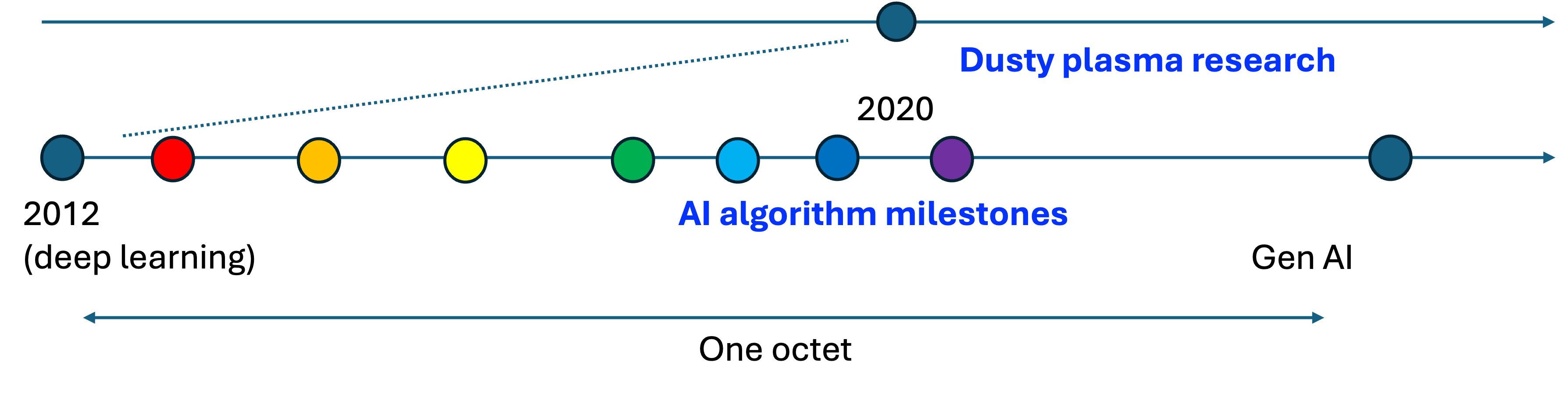}
    \caption{Evolution from `black-box' machine learning models, such as CNNs, to generative AI models, such as LLMs, happened over a period about 10 years. Such advances may be summarized in terms of {\it an octet}. An octet of advances include advances in 1. physical system selection (application targets, such as dusty plasmas), 2. algorithm (neural network architecture), 3. sensing hardware, 4. sensing method, 5. datasets, 6. data pre-processing (noise rejection), 7. modeling and interpretation, and 8. uncertainty quantification (UQ). Dusty plasma can ride on such advances.}
    \label{fig:TimeLine}
\end{figure}

Rapid advances in ML and AI tools allow a growing number of traditionally manual tasks in dusty plasmas to be automated, facilitating the transition from AI-assisted and AI-accelerated research towards AI-driven research. Recent examples of AI-enabled tasks include automatic code generation, automatic experimental data collection, automatic data include natural language text processing, and automatic experimental design. Such transition may allow, eventually, {\it fully automatic} AI discoveries, such as integrating the multi-scale multi-physics of the diverse dusty plasma phenomena using a small set of fundamental laws of physics, or applications such as radioactive dust mitigation in nuclear fusion devices. Here we first highlight the advances in tools related to dust and particle tracking, and introduce the concept of advances in octet, as illustrated in Fig.~\ref{fig:TimeLine}.
Besides dusty plasmas, particle tracking has been widely used in many areas, in time-lapse microscopy~\cite{meijering2012methods}, in imaging processing~\cite{azuri2021role},  in cell biology~\cite{cheng2022review}, and in particle physics~\cite{chenouard2014objective}. %

Early numerical approaches to particle tracking, implemented on CPU-based systems, established many of the fundamental principles still in use today. These algorithms, such as  Kalman Filter, provide a systematic framework for track seeding, building, and fitting. In~\cite{shuang2014troika}, particle tracking is addressed through a combination of signal-to-noise ratio (SNR) enhancement, particle identification, and nearest-neighbor trajectory association. More advanced techniques, such as multiple-hypothesis tracking (MHT)~\cite{jaqaman2008robust}, offer improved accuracy by maintaining competing trajectory hypotheses, although at significantly higher computational cost, motivating the development of more efficient heuristic alternatives.

For three-dimensional (3D) particle tracking velocimetry (PTV), classical approaches such as PTV in three-dimensional flows~\cite{maas1993particle} employ rigorous photogrammetric models to reconstruct particle positions with high precision. Subsequent developments introduced probabilistic matching strategies, such as the algorithm in~\cite{baek1996new}, which estimates particle trajectories between frames based on match likelihood, providing a computationally efficient alternative to deterministic methods. Iterative refinement techniques, including the relaxation method~\cite{ohmi2000particle}, improve particle matching by progressively resolving ambiguities, and have been further extended in~\cite{rubbert2020iterative} to handle dense particle fields in 3D PTV.

More recent advances explore alternative imaging modalities and tracking frameworks. For example, tomographic X-ray PTV~\cite{makiharju2022tomographic} enables volumetric flow measurements in optically opaque systems, while \cite{zhang2013robust} introduces a multi-task tracking (MTT) approach based on particle filtering and sparse representation. In this framework, particles are modeled as linear combinations of dictionary templates, allowing interdependencies between particles to be exploited for improved tracking robustness and efficiency. Together, these developments illustrate the evolution from deterministic, geometry-based methods toward probabilistic and data-driven approaches capable of handling increasingly complex and high-density particle systems.

\begin{figure*}[!htb]
    \centering
   \includegraphics[width=1.85\columnwidth]{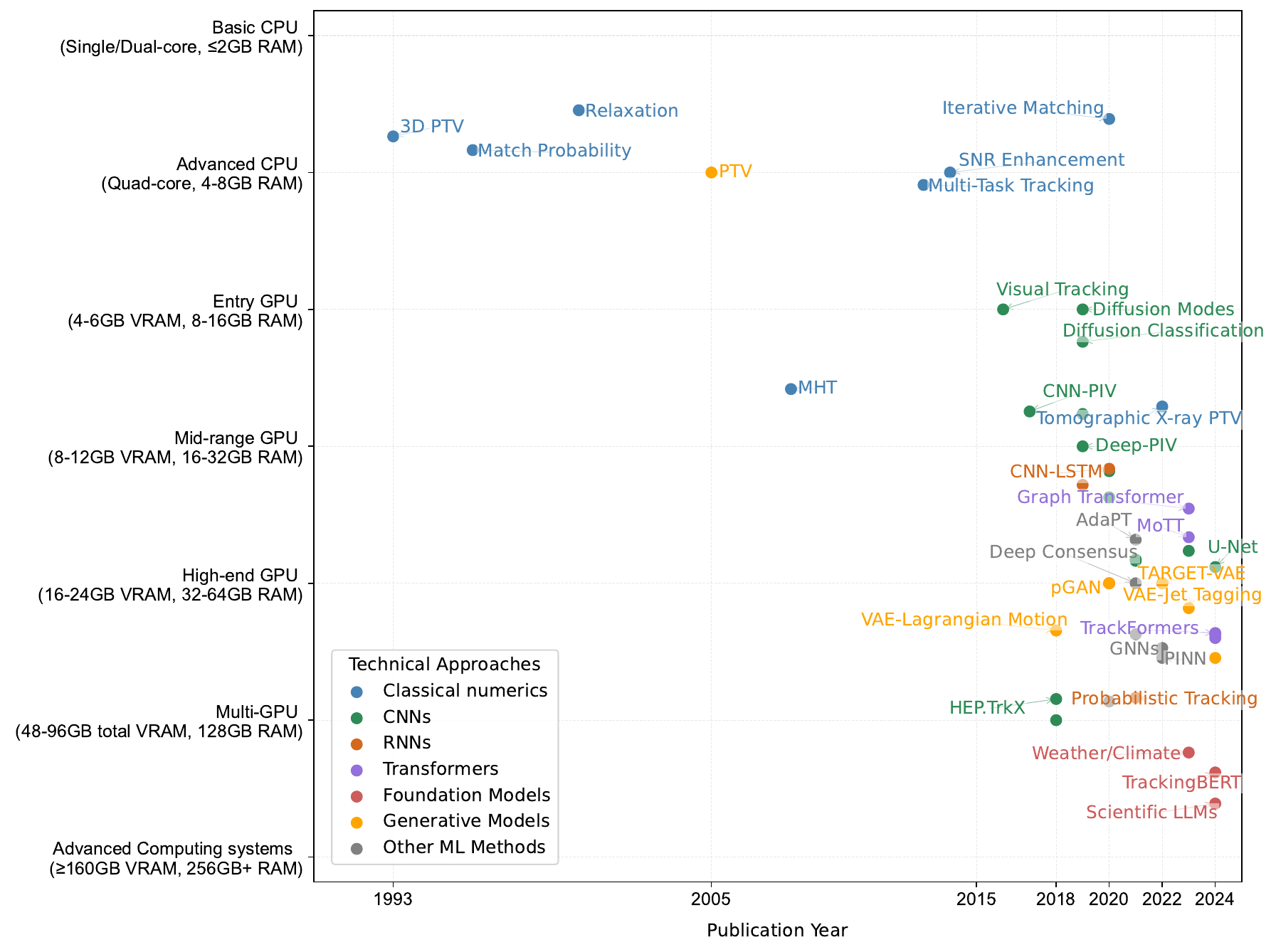}
    \caption{The state-of-the-art CNN models and their current applications to dusty plasma image processing. Figure adopted from~\cite{DLHS:2020}.}
    \label{fig:refs_algorithm}
\end{figure*}

Classical numerical methods are typically lightweight and can be implemented on basic CPU systems (2–4 GB RAM). These include foundational techniques such as 3D particle tracking velocimetry (PTV)~\cite{maas1993particle}, match probability methods~\cite{baek1996new}, and relaxation-based matching algorithms~\cite{ohmi2000particle}. More advanced statistical and optimization-based methods require moderately increased computational resources, often involving enhanced CPUs (4–8 GB RAM) and, in some cases, entry-level GPU acceleration. Examples include multiple-hypothesis tracking (MHT)~\cite{jaqaman2008robust}, variational approaches to PTV~\cite{ruhnau2005variational}, and signal-to-noise enhancement techniques~\cite{shuang2014troika}.

Recent machine learning approaches exhibit substantially higher computational demands, reflecting their increased modeling capacity and flexibility. Entry-level GPU implementations (4–6 GB VRAM) are sufficient for methods such as visual tracking~\cite{zhang2016robust} and diffusion-based trajectory classification~\cite{kowalek2019classification}. Mid-range GPU systems (8–12 GB VRAM) support deep learning–based flow and tracking models, including Deep-PIV~\cite{rabault2017performing} and hybrid CNN–LSTM architectures~\cite{yao2020deep}. At the high end, transformer-based tracking models such as TrackFormers~\cite{caron2024trackformers} and scientific large language models~\cite{zhang2024scientific} may require advanced computing infrastructure with large GPU memory ($\gtrsim$160 GB VRAM) and distributed training capabilities.

In the following subsections, we review machine learning approaches, such as convolutional neural networks (CNNs), recurrent neural networks (RNNs), transformers, and foundation models. 
The computational hardware requirements of these approaches span a wide range and have evolved significantly with increasing model complexity. A summary of these technical approaches and their associated computational requirements is given in Fig.~\ref{fig:refs_algorithm}.

%%%==================

\subsection{Particle tracking with CNN}
Deep learning models are pervasively used in particle tracking related tasks. However, the current practice is to design and train one deep learning model for one task with supervised learning techniques. The trained models work well for tasks they are trained on but show no or little generalization capabilities. Different deep learning models may be unified with a large language model. 

Convolutional Neural Networks (CNNs) \cite{o2015introduction} are effective for particle tracking and image analysis due to their ability to hierarchically extract spatial features through layered convolutions and pooling. 
In particle tracking, CNNs automate detection and extract motion characteristics, proving particularly effective in dense and dynamic environments. For example, \cite{kowalek2019classification} employs CNNs to classify diffusion modes in single-particle trajectories, outperforming traditional methods in discerning complex stochastic behaviors. Similarly, the HEP.TrkX Project \cite{tsaris2018hep} applies CNNs to track particles in dense data environments at the LHC, demonstrating scalability in reconstructing charged particle trajectories when paired with LSTMs.
In fluid dynamics, CNNs have enhanced Particle Image Velocimetry (PIV) by replacing traditional methods, improving velocity field extraction \cite{rabault2017performing}. CNNs are also used for super-resolution mapping of molecular diffusivity in \cite{park2023machine}, where diffusion coefficients are derived from single-molecule images. \cite{cai2019particle} employs CNNs for 2D velocity field extraction in fluid mechanics, demonstrating CNNs’ adaptability to motion estimation tasks. In \cite{granik2019single}, CNNs classify diffusion types in single-particle trajectories, distinguishing patterns like Brownian motion and continuous time random walks, critical for understanding particle dynamics.
High-resolution particle tracking further benefits from CNNs across diverse applications. \cite{newby2018convolutional} demonstrates CNNs’ effectiveness in automating submicron-scale particle tracking in 2D and 3D. In \cite{zhang2016robust}, a two-layer CNN is applied to visual tracking without extensive offline training, underscoring CNNs’ robustness for generalized imaging tasks. In dense environments, U-Net variants \cite{Dormagen.2024} are adapted to track micron-sized particles in plasma clouds, achieving high accuracy. \cite{bozek2021markerless} leverages CNNs for segmenting and tracking honey bees on natural backgrounds, proving their applicability in high-resolution video analysis.
Additional CNN applications include extracting probe orientation from optical data \cite{hu2020single} and refining particle distribution in volumetric PIV through geometry-informed features \cite{gao2021particle}. CNNs are also utilized in \cite{lagemann2019deep}, where architectures like Inception-based, residual, and densely connected networks perform PIV tasks with comparable accuracy to standard tools. Finally, in \cite{puzyrev2020machine}, a CNN-based method using Mask R-CNN for 2D localization and 3D matching in granular gases demonstrates CNNs’ flexibility in high-dimensional particle tracking.

\subsection{GNN model \label{subsec:gnn}}
Graph neural networks are sets of geometric deep-learning algorithms that learn from graphs. A graph represents a set of nodes (pieces of data), and the relationships between them, known as edges. GNNs were introduced in the early 2000s, physicists didn’t start using them to analyze experimental data until 2019. Unlike convolutional neural networks, which work only with data that’s presented in a regular grid, GNNs can analyze datasets that have irregular, 3D geometric shapes.
A novel variant of GNN for particle tracking called Hierarchical Graph Neural Network (HGNN) was described~\cite{LCF:2023}. The Exa.TrkX project has applied geometric
learning concepts such as metric learning and graph neural networks (GNNs) to HEP particle tracking~\cite{JMC:2021}. The training dataset was the TrackML dataset. 

\subsection{RNN model}
Recurrent Neural Networks (RNNs) \cite{zaremba2014recurrent} are designed for sequential data processing by maintaining a hidden state that captures information from prior inputs, making them suitable for tasks requiring temporal context. However, standard RNNs struggle with long-term dependencies due to the vanishing gradient problem.
Long Short-Term Memory networks (LSTMs) extend RNNs by incorporating a memory cell and three gates: forget, input, and output, which regulate information retention, updates, and discarding.

In image data and particle tracking, LSTMs are often paired with Convolutional Neural Networks (CNNs), where CNNs extract spatial features from image frames, and LSTMs learn temporal dynamics to predict particle positions. \cite{yao2020deep} combines CNNs and LSTMs to automate particle tracking without prior assumptions on particle dynamics or extensive parameter tuning. \cite{arts2019particle} segments single-particle trajectories into tracklets with LSTMs, followed by moment scaling spectrum analysis to classify mobility classes. \cite{spilger2020recurrent} uses a deep recurrent network for tracking in fluorescence microscopy, incorporating both past and future trajectory information, while \cite{spilger2021deep} introduces an RNN-based probabilistic tracking method, applying Bayesian filtering to account for aleatoric and epistemic uncertainty.

\subsection{Transformer}
Transformer \cite{wolf2020transformers} is a deep learning architecture that employs an attention mechanism to efficiently model dependencies in sequential data. 
Transformers differ from recurrent models in that they compute relationships in parallel rather than sequentially, allowing for greater computational efficiency and effectiveness in capturing both local and global dependencies. This makes them well-suited for tasks like particle tracking that require understanding intricate patterns over time.

In dusty plasma applications, using neural networks to model physical systems through function fittings (also called fitnet, FFNN, curve fitting network) are found in~\cite{BSRI:2020}.  Lately, such as the transformer architecture (x11), autoencoders
(x12), and adversarial networks (x13) have revolutionized
how one understands the ML process~\cite{LeTu:2024}.  The transformer idea appears to be absolutely
essential for AlphaFold.

A Transformer-based model called Motion Transformer Tracker (MoTT) \cite{zhang2023motion} is introduced for single-particle tracking in fluorescence microscopy, leveraging attention to learn particle behaviors from past and hypothetical future trajectories to estimate matching probabilities. \cite{caron2024trackformers} presents TrackFormers, employing a Transformer encoder-decoder model to assign hits to track candidates at the LHC, focusing on balancing computational efficiency and accuracy by treating hit assignment as a translation task. In \cite{mishra2024comparative}, a proof-of-concept study shows that Transformers significantly accelerate multiple particle tracking in dense environments, outperforming traditional methods like Multiple Hypothesis Tracking (MHT) for random motion estimation.
\cite{qu2022particle} proposes the Particle Transformer (ParT) for jet tagging in particle physics, using attention to model pairwise particle interactions, achieving state-of-the-art results compared to prior methods like ParticleNet. In \cite{wang2024transformer}, a Transformer-based approach for multiple-object tracking (MOT) is presented, incorporating anchor-based queries and template matching to improve detection, data association, and training speed for joint-detection-and-tracking (JDT) tasks. Lastly, \cite{kamiya2023single} introduces a Graph Transformer-based model for single-particle tracking, accounting for relationships among particles to improve trajectory prediction in molecular analysis.
\subsection{Additional Generative Models}

%[Symmetry] ]Physicists have also begun to work with generative  models since 2017, which are even more advanced than CNNs and GNNs—"most notably the transformer, which is at the center of state-of-the-art AI applications in industry and is the backbone of most foundation models.” The difference between foundation models and general deep learning neural networks is scale: Foundation models are trained on datasets more massive than any available in the past, and they are also often trained without any pre-existing knowledge of what is in the data.

\begin{figure}[!htb]
    \centering
   \includegraphics[width=0.85\columnwidth]{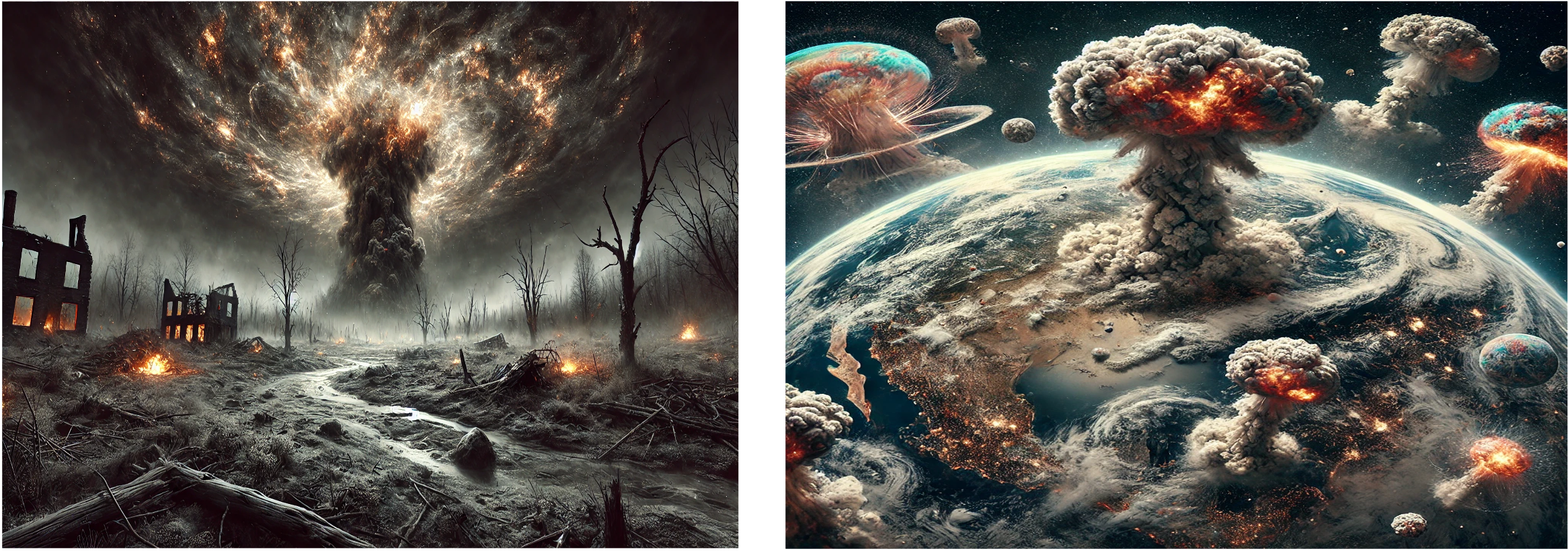}
    \caption{ Visualizations of a post-apocalyptic landscape under a nuclear winter with dusty plasmas swirling in the sky. The ground is barren and covered by dust fallouts from above. The scene, generated by DALL$\cdot$E 3,  is eerie and haunting, capturing the desolate aftermath of a global dusty plasma.}
    \label{fig:apocalyptic}
\end{figure}

Besides transformer, deep generative models, such as Variational Autoencoders (VAEs), Generative Adversarial Networks (GANs), and diffusion models, have shown significant potential in generating realistic data samples. Generative models are useful when it is difficult to do experiments.  Mushroom clouds associated with thermonuclear explosions are probably the largest man-made dusty plasmas to date.  A few examples of computer-generated visualizations of a post-apocalyptic landscape in a nuclear winter with dusty plasma swirling in the sky is shown in Fig.~\ref{fig:apocalyptic}.

\textbf{Diffusion models} \cite{cao2024survey} have been effectively utilized in various domains, including image generation, video generation, time series generation, and graph generation. They operate by progressively adding noise to data through a forward process until the data becomes almost pure noise, followed by learning a reverse denoising process to reconstruct the original data. During training, the model learns to predict the added noise at each step, allowing step-by-step denoising that ultimately yields realistic data. The reverse process involves iteratively applying parameterized transformations $p_{\theta}(x_{t-1}|x_t)$, starting from random noise to generate samples that match the target distribution.
In \cite{li2024generative}, a diffusion model was introduced to generate single tracer trajectories in three-dimensional turbulence, successfully passing most statistical benchmarks across different time scales.

\textbf{GANs} \cite{goodfellow2020generative} have gained significant attention for their impressive performance in image generation, requiring mid-range to high-end GPU systems (16-24GB VRAM).
A GAN consists of two networks: a Generator ($G$) and a Discriminator ($D$). For image data, $G$ maps random noise $z$ to realistic images $G(z)$, while $D$ distinguishes between real images $x$ and generated ones. The training process involves optimizing the following loss function:

\begin{widetext}
\begin{equation} \min_G \max_D V(D, G) = \mathbb{E}{x \sim p{\text{data}}(x)} \left[ \log D(x) \right] + \mathbb{E}_{z \sim p_z(z)} \left[ \log \left( 1 - D(G(z)) \right) \right], \end{equation}
\end{widetext}

where $D$ is trained to maximize its classification accuracy, while $G$ aims to minimize $D$'s ability to differentiate real from generated images.
In \cite{martinez2020particle}, the pGAN model is introduced—a particle-based generative model for simulating pileup events at the LHC. This work explores how a GAN can generate particle four-momenta from proton collisions, effectively abstracting from the irregularities of typical detector geometries.

\textbf{VAEs} are generative models that encode data into a latent space and then decode it to reconstruct the original data. For image datasets, the encoder maps images $x$ to a latent distribution $\mathcal{N}(\mu, \sigma^2)$, from which a latent vector $z$ is sampled. The decoder then reconstructs the image from $z$. The training objective is to minimize reconstruction error while ensuring that the latent space follows a standard Gaussian distribution, achieved through a combination of reconstruction loss and Kullback-Leibler (KL) divergence.
In \cite{liu2018variational}, a VAE is applied to simulate particle motion in a chaotic recirculating flame, successfully generating 3D trajectories that align with experimental data. The work in \cite{nasiri2022unsupervised} introduces TARGET-VAE, an unsupervised model that disentangles latent variables into components such as content, rotation, and translation, providing effective representations across various imaging domains. Similarly, \cite{cheng2023variational} applies VAEs to anomalous jet tagging at the Large Hadron Collider, where the VAE encodes low-level jet constituents, enabling anomaly detection through latent space analysis. \cite{ruhnau2005variational} presents a novel variational approach for particle tracking velocimetry by combining discrete particle matching with continuous regularization terms, enabling reliable velocity evaluations in two- and three-dimensional particle tracking velocimetry (PTV) images.

\subsection{Other ML methods and regularization}

Besides neural network architectures and dataset design, the formulation of loss functions plays a central role in defining machine learning (ML) methods. Incorporating physics, geometry, statistics, and scaling constraints into loss functions has led to a rich class of approaches that extend beyond purely data-driven learning. In reinforcement learning, the loss function is replaced by a reward formulation, which similarly guides model optimization and enables learning from sparse, delayed, or partially observed signals~\cite{SuBa:2018,Szep:2010}. 

Novel loss-function formulations are especially important in settings where labeled data are limited or unavailable. For example, unsupervised and self-supervised learning methods based on photometric consistency, bidirectional flow constraints, and spatial smoothness have been successfully applied to particle image velocimetry (PIV) without requiring ground-truth labels~\cite{zhang2020unsupervised}. As discussed in Sec.~\ref{sec:MAM}, embedding physical knowledge directly into the learning objective further improves model fidelity and interpretability. Physics-informed approaches, for instance, reconstruct dense velocity and pressure fields from sparse PIV/PTV measurements by incorporating the Navier--Stokes equations into the loss function~\cite{wang2022dense}. Such constraints restrict solutions to physically admissible regimes, thereby enhancing generalization, robustness, and consistency with governing laws of physics.

The diversity of loss function designs reflects the broader ML landscape, where physics-based constraints, geometric structure, and statistical regularization can be integrated in a complementary manner. These developments motivate a shift toward hybrid learning paradigms that combine data-driven flexibility with physically grounded priors. In this context, geometric and topological learning provide particularly powerful extensions for processing structured dusty plasma data. For example, AdaPT~\cite{dingel2021adapt} provides an application for tracking spherical magnetic particles using intensity-based localization and hybrid linking strategies. Deep Consensus Networks~\cite{wollmann2021deep} employ feature pyramid architectures and Bayesian regularization for robust object detection in microscopy images. Graph-based approaches have been successfully applied in high-energy physics, where GNNs are used to reconstruct particle trajectories from detector hits, naturally handling complex measurement ambiguities~\cite{duarte2022graph}. Autonomous experimentation frameworks in electron and scanning probe microscopy~\cite{kalinin2021automated} demonstrate how machine learning can guide adaptive data acquisition and analysis. 

Geometric and topological learning offer a complementary framework for advancing algorithms within DustNET~\cite{BBCV:2021,BRVS:2025}, especially for particle tracking and 2D-to-3D inversion problems. Dusty plasma data are intrinsically structured: particle positions form evolving point clouds, trajectories define graphs in space--time, and collective dynamics often exhibit manifold-like organization. Conventional deep learning models, typically designed for Euclidean grids, do not fully exploit these underlying geometric and topological structures. In contrast, geometric deep learning introduces inductive biases based on symmetry, invariance, locality, and topology~\cite{GCZB:2025}, enabling more efficient and physically meaningful representations of dust grains and complex plasma systems. 

An integrated framework combining physics-informed loss functions, geometric learning, and data-driven modeling thus provides a powerful approach for analyzing dusty plasma systems. Such methods can improve tracking accuracy, enhance reconstruction fidelity, and enable cross-regime generalization, ultimately bridging experimental observations with interpretable and predictive models.

In particle tracking applications, dust grains can be naturally represented as nodes in dynamic graphs or point clouds, with edges encoding spatial proximity, interaction strength, or temporal correspondence. Graph neural networks (GNNs), Sec.~\ref{subsec:gnn}, and equivariant architectures~\cite{Kond:2025} leverage permutation invariance and geometric symmetries to improve robustness, particularly in dense, noisy, or partially observed environments. Topological learning methods further extend this framework by incorporating higher-order relationships, such as clusters, filaments, and crystalline ordering, through tools including simplicial complexes and persistent homology~\cite{MBGY:2014}. These approaches enable the extraction of global structural features that are not readily captured by purely local or convolutional methods.

Geometric and topological learning is also central to 2D-to-3D reconstruction, where the goal is to reconstruct three-dimensional particle trajectories from projections or time-sequence 1D or 2D dusty plasma data. By embedding geometric constraints, such as epipolar geometry~\cite{Zhan:1996,Zhang98,WaLW:2016}, manifold smoothness, and physical conservation laws, into neural architectures, models can learn mappings from observation space to latent 3D configurations. Neural operator frameworks and geometric deep learning methods further generalize this approach by learning continuous mappings between function spaces, enabling reconstruction of volumetric dynamics from sparse or incomplete measurements.

Within the DustNET framework, these techniques support unified representations across modalities and scales. By combining geometric and topological priors with data-driven learning, DustNET can enhance tracking accuracy, enable robust 3D reconstruction, and reveal topological signatures of collective behavior, thereby bridging experimental observations with physically interpretable models.

%In parallel, a growing body of machine learning work has focused on particle tracking and image-based datasets across related domains. 

%%%%%============= 
%
%
%
%
%
%
%
%
%
%
%

\subsection{AI foundation models for dusty plasmas}
\label{sec:FoMo}

The dusty-plasma community will likely need several types of foundation models (FoMos), tailored to different scientific objectives. FoMos became popular around 2022, with the introduction of ChatGPT and other Large Language Models (LLMs). LLMs are AI engines for text processing, including text generation (1D sequential data). Other FoMos like DALL$\cdot$E generate images (2D data). 
Transformer, which is at the center of state-of-the-art AI applications in industry, is the backbone of most FoMos. The difference between FoMos and general deep learning neural networks is scale: FoMos, with billions or more even trillions of tunable parameters~\cite{HPB:2024,dHTG:2025}, are trained on datasets more massive than any available in the past, and they are also often trained without any pre-existing knowledge of what is in the data. The size of a one-trillion parameter model is at least 14 terabytes, which is more than 200 times the memory of one MI250X GPU (64 gigabytes) in supercomputers like Oak Ridge National Laboratory's Frontier.

Here we discuss two classes of FoMos. One class comprises LLM FoMos for processing of scientific literature (dusty plasma domain specific), generating new text, extending the discussion in Sec.~(\ref{sec:text_data_dp}). These models support automated literature mining, symbol and equation extraction, cross-paper synthesis, scientific reasoning, and examination of scientific knowledge evolution and lineages through citation-tracking and statistical analysis. A second class of FoMos, for automatic multi-modal data processing and interpretation, are inherently more complex. Such models must integrate not only textual inputs but also scientific data such as time series, 2D images, sequences of images or movies, 3D and even higher-dimensional (tensor) datasets. Designing these multi-modal FoMos requires architectures capable of learning across heterogeneous data types while preserving physics consistency and capturing complementary information across different modalities.

Table~\ref{tab:fomo_models_quant} summarizes representative candidate models that can be directly used or adapted, {\it e.g.}, via retrieval-augmented generation (RAG), to construct a dusty-plasma scientific text FoMo. The comparison emphasizes criteria most relevant for scientific FoMo deployment, including reasoning capability, context length, openness of weights, customization flexibility, and suitability for domain-adaptive pretraining.

\begin{table*}[!hbt]
\caption{Quantitative comparison of candidate LLMs for a dusty-plasma text FoMo (indicative values).}
\label{tab:fomo_models_quant}
\centering
\small
\begin{tabular}{l c c}
\hline
\textbf{Model} & \textbf{Architecture} & \textbf{Parameters} \\
\hline
ChatGPT 5.2 (GPT-5 class) & Transformer (decoder) & ? \\
Claude 4.5 & Transformer (decoder) & ? \\
DeepSeek (frontier) & Transformer (dense / MoE) & $\sim10^{10}$--$10^{11}$+ \\
LLaMA-3 (8B / 70B) & Transformer (decoder) & 8B / 70B \\
Mixtral (8×7B MoE) & Transformer (Sparse MoE) & 56B total ($\sim$12B active/token) \\
\hline
\end{tabular}

\vspace{2mm}
\begin{minipage}{\columnwidth}
\footnotesize
\textbf{Hardware notes:} Frontier closed-weight models (GPT-5 class, Claude, DeepSeek frontier variants) require distributed GPU clusters (A100/H100 class) and support context windows exceeding 100k tokens. 
LLaMA-3 8B fits on a single 24--48GB GPU; 70B requires 2--4×80GB GPUs. 
Mixtral MoE activates only a subset of parameters per token, improving efficiency (48--96GB GPU RAM typical for inference and text generation).
\end{minipage}

\end{table*}

Alongside frontier class models such as GPT-5.x and Claude-4.x systems, we include the DeepSeek open model family, which has demonstrated strong performance in mathematical reasoning and code-intensive benchmarks, as well as open-weight architectures such as LLaMA-3 and Mixtral (a Mixture-of-Experts model by Mistral AI)~\cite{JSRM:2024}. The listed metrics represent indicative ranges based on publicly available information and typical deployment configurations; exact specifications vary by version and access tier. All models considered are transformer-based large language models, reflecting the dominance of transformer architectures in scientific language modeling. Open-weight entries correspond to commonly available public checkpoints.

The candidate systems are primarily decoder-only autoregressive transformers. Contemporary scientific LLMs do not rely on variational autoencoders as primary generative backbones, though hybrid integrations may incorporate retrieval modules, tool-use frameworks, or symbolic reasoning engines. Mixture-of-experts (MoEs) variants {\it e.g.}, Mixtral and some DeepSeek configurations, activate only a subset of parameters per token, improving computational efficiency relative to dense models.

Frontier closed-weight systems (GPT-5–class, Claude, DeepSeek frontier variants) are estimated to operate in the $10^{11}$–$10^{14}$ parameter regime. In contrast, open-weight models such as LLaMA-3 (8B or 70B) and Mixtral (with $\sim$12B active parameters per token) permit full weight-level adaptation and domain-specific retraining. For dusty-plasma applications, models in the 1B–10B range are a good starting point, particularly when paired with RAG for citation grounding and factual control.

Computing costs scale approximately linearly with parameter count: 7–8B models run on a single 24–48GB GPU, whereas 70B-class dense models typically require 2–6 $\times$ 80GB GPUs. Frontier systems are hosted on distributed high-memory clusters. Domain-adaptive pretraining for 7–70B open models may require multi-node GPU resources when token budgets exceed $10^6$~\cite{HBMB:2022}.

All listed models support autoregressive token generation. Open-weight systems commonly provide 8k–32k context windows, while frontier systems often exceed 100k tokens, facilitating full-paper ingestion and multi-document synthesis.

A key design consideration is the relationship between model scale and domain token volume. Empirical scaling results suggest that effective adaptation requires token counts comparable to, or several times larger than, parameter counts. For 1–10B dense models, $\sim10^{8}$ curated tokens are typically sufficient to reshape domain priors. For 30–70B models, $\sim10^{8}$–$10^{9}$ tokens are advantageous, particularly for improving equation-level reasoning and cross-paper synthesis. MoE architectures reduce token demands relative to equally large dense models due to smaller active parameter counts.

Frontier closed-weight systems generally rely on fine-tuning or RAG rather than full pretraining; in these cases, $\sim10^{7}$–$10^{8}$ high-quality domain tokens may suffice when combined with retrieval pipelines. Importantly, effective token volume depends more on curation and regime balance than on raw corpus size. For dusty-plasma literature, on the order of tens of gigabytes, careful redundancy control and physics-informed data-sets enable meaningful specialization even at mid-scale model sizes.

In summary, 1--10B parameter models can be trained from $\sim10^{8}$ curated tokens through RAG and adaptation of existing open LLMs, a significant reduction in token counts from trillions, as required for FoMo development from scratch. 30–70B models from $\sim10^{8}$–$10^{9}$ tokens; MoE architecture through `dense-equivalent' budget model, with active parameters comparable to 13B dense systems; and frontier closed-weight systems from $\sim10^{7}$–$10^{8}$ tokens when coupled with RAG. A hybrid strategy, combining frontier reasoning capacity with reproducible open-weight domain adaptation, may provide a balanced and scientifically robust pathway for developing a dusty-plasma text FoMo.

%==============The next FoMo============================

A second class of dusty-plasma FoMo is designed to support multi-modal data acquisition, processing, and interpretation, particularly for image and time-series measurements. Such a FoMo may be conceptualized as a scientific representation learner rather than a text-centric assistant~\cite{DHWG:2024}. We tentatively refer to this system as the \emph{Dusty-plasma Unified Sensing, Testing and Multi-modal Analysis Platform} (DUST-MAP). Its primary objective is to ingest heterogeneous observational and simulation data, including optical 2D dust images and movies, 1D and 2D plasma density and temperature measurements, electric and magnetic field time series, active probing diagnostics, and numerical simulations, and to learn latent representations that generalize across physical regimes, instruments, and observational conditions.

DUST-MAP can leverage advances in foundation models across language, time-sequence data, vision, and cross-modal learning, such as BERT and masked-language models in NLP~\cite{DCLT:2019}, Vision Transformers (ViT)~\cite{DBKW:2021}, and contrastive multi-modal frameworks such as CLIP~\cite{RKHR:2021}. Domain-adapted systems such as TrackingBERT~\cite{HMCL:2024} and SciBERT, which was trained on 1.14 million full-text scientific papers comprising 3.1 billion tokens, demonstrate the effectiveness of large-scale pretraining followed by scientific specialization. Generalist transformer architectures such as Flamingo~\cite{ADLM:2022} and Gato~\cite{RZPC:2022} further illustrate how heterogeneous modalities can be integrated within unified learning systems. Open ecosystems such as Hugging Face, hosting more than 735,000 models and 168,000 datasets (as of June 2024), provide infrastructure for adapting these advances to domain-specific FoMos. A summary is given in Table.~\ref{tab:table3c}.

\begin{table*}[!t]
\caption{Examples of frontier AI foundation models for multimodal and scientific data processing. Parameter sizes and training details are approximate and may vary by version.}
\label{tab:table3c}
\centering
\scriptsize
\setlength{\tabcolsep}{2.5pt}
\renewcommand{\arraystretch}{0.9}
\begin{ruledtabular}
\begin{tabular}{lccccc}
\textrm{Name} &
\textrm{Developer} &
\textrm{Params} &
\textrm{Dataset} &
\textrm{Hardware} &
\textrm{Modalities} \\
 & & \textrm{[B]} & & & \\
\colrule

AlphaFold & DeepMind & $\sim$100--200 & PDB, UniProt & TPU/GPU clus. & protein struct. \\

ANI-1~\cite{SIR:2017a,SIR:2017b} & Roitberg et al. & $\sim$0.1 & DFT (GDB-17) & GPU & mol. potentials \\

BERT & Google & 0.11--0.34 & BooksCorpus, Wiki & TPU & text \\

BioBERT~\cite{LYK:2020} & DMIS Lab & $\sim$0.34 & PubMed, PMC & GPU & biomed. text \\

BLOOM & BigScience & 176 & ROOTS (multi.) & 384$\times$A100 & text \\

ChemBERTa~\cite{CGR:2020} & IBM & $\sim$0.1 & MoleculeNet, PubChem & GPU & chem. text \\

Claude & Anthropic & $>$100 (est.) & web + curated & GPU clus. & multimodal \\

CLIP & OpenAI & $\sim$0.4 & img.-text (400M) & GPU & image + text \\

DALL$\cdot$E & OpenAI & $\sim$12 & img.-text pairs & GPU & image gen. \\

DBRX & Databricks & 132 (MoE) & curated web/code & GPU clus. & text \\

DeepSeek-V3~\cite{DeepSeekV3:2024} & DeepSeek AI & 671 (37 act.) & 14.8T tok., multi.+code & H800/A100 clus. & text + code \\

Dolly & Databricks & 12 & instr.-tuning data & GPU & text \\

Fuyu-8B & Adept & 8 & multimodal web & GPU & image + text \\

Gemini & Google & $>$100 (est.) & multimodal web & TPU/GPU & multimodal \\

GPT-4~\cite{gpt4:2024} & OpenAI & $>$1000 (est.) & web + code + RLHF & GPU clus. & multimodal \\

Granite & IBM & 3--20 & enterprise data & GPU & text \\

Jurassic-2 & AI21 Labs & 7--178 & web corpora & GPU & text \\

LLaMA~\cite{TLIM:2023} & Meta & 7--65 & web corpus & GPU & text \\

Luminous & Aleph Alpha & 13--70 & multi. corpora & GPU & text \\

Mistral & Mistral AI & 7 & web + code & GPU & text \\

Mixtral & Mistral AI & 46 (MoE) & web + code & GPU clus. & text \\

MPT-7B & MosaicML & 7 & curated web & GPU & text \\

Palmyra & Writer & $\sim$5--20 & business text & GPU & text \\

Phi-2 & Microsoft & 2.7 & synthetic + curated & GPU & reasoning \\

RedPajama & Together & 7--65 & RedPajama & GPU & text \\

SciBERT~\cite{BLC:2019} & AI2 & $\sim$0.11 & Semantic Scholar & GPU & sci. text \\

SciFive~\cite{PAT:2021} & Google/NIH & $\sim$0.77 & PubMed & TPU/GPU & biomed. text \\

Sora & OpenAI & undisclosed & video datasets & GPU clus. & video \\

Stable Diffusion~\cite{StabilityAI:2022} & Stability AI & $\sim$0.9 & LAION-5B & GPU & image \\

StarCoder & HuggingFace & 15.5 & code datasets & GPU & code \\

T5 & Google & 0.22--11 & C4 & TPU & text \\

Titan & Amazon & undisclosed & enterprise/web & GPU & multimodal \\

\end{tabular}
\end{ruledtabular}
\end{table*}

What distinguishes DUST-MAP, however, is its incorporation of dusty-plasma physics into the representation-learning process. Rather than learning purely statistical correlations across modalities, DUST-MAP is designed to embed physically meaningful constraints, such as force balance, charging dynamics, sheath structure, dust–plasma coupling strengths, and multi-scale transport processes, directly into its training and latent space. Analogous to BERT’s masked prediction objective, DUST-MAP would employ masked reconstruction, cross-modal contrastive learning, and self-supervised forecasting, but augmented with physics-informed regularization or inductive biases that enforce consistency with governing equations and conservation laws. By capturing complementary information across modalities while respecting plasma-physical invariants, DUST-MAP aims to learn representations that generalize not merely across datasets, but across regimes and physical scales, thereby enabling scientifically robust interpretation rather than purely data-driven pattern recognition.

Next, we discuss the experimental generation, a fundamental form of physics-imbedded method, of 1D time-series tokens for training DUST-MAP. Experimental tokens may be complemented or modulated by simulations. Extrapolations to 2D image tokens and other higher dimensional tokens are similar. %We may express token requirements in terms of data acquisition rates and tokenization parameters. 

The number of raw samples from a single channel, such as a Langmuir probe, for a duration $T$ is
\begin{equation}
N_{\mathrm{samples}} = \frac{T}{\Delta t}.
\end{equation}
Here $\Delta t$ is the sampling cadence (time interval for each sampling point). The raw data will be segmented for a fixed window width ($W$) and stride ($S$) between consecutive windows. The number of windows ($N_{\mathrm{ws}}$) is
\begin{equation}
N_{\mathrm{ws}} = \left\lfloor \frac{N_{\mathrm{samples}} - W}{S} \right\rfloor + 1.
\end{equation}
If each window is converted into $P$ patch tokens (for example, by splitting a waveform window into $P$ equal-length segments or by waveform decomposition), the total number of time-series tokens per channel is approximately
\begin{equation}
N_{\mathrm{ts}} = P \, N_{\mathrm{ws}}.
\end{equation}
For $N_{\mathrm{ch}}$ parallel channels yields
\begin{equation}
N_{\mathrm{1D}} = N_{\mathrm{ch}} \, P \, N_{\mathrm{ws}}.
\end{equation}

As an illustrative example, we use a Langmuir probe for token generation in a dusty plasma,  $\Delta t = 1$~\textmu s for an experimental duration $T=10$~s, with a window width $W = 10^3$ samples for 1 ms and stride $S = 10$ for 10~\textmu s. This produces about $ 10^{6}$ windows per experiment. If each window is tokenized into $P=8$ patches across $N_{\mathrm{ch}}=4$, the resulting token count is on the order of $N_{\mathrm{1D}} \sim 3 \times 10^{7}$ tokens per 10-s experiment. To generate 10$^9$ tokens, it will take about 400 s or 40 experiments. In practice, effective token rate may be less, say, a reduction to 10\%, then the experimental time may need to increase by a factor of 10 accordingly.

For 2D image modalities, let an image of spatial resolution $H \times W$ pixels be divided into windowed patches of size $w_H \times w_W$.  For a stride of $S^{(2)}=S_H \otimes S_W$ in the two directions, and $P=1$ for each window, the number of image tokens per frame is
\begin{equation}
N_{\mathrm{ts}}^{(2)} = \left( \frac{H - w_H}{S_H} + 1 \right) \times \left( \frac{W - w_W}{S_W} + 1 \right).
\end{equation}
For example, a $1024 \times 1024$ calibrated dust-scattering image tokenized into $16 \times 16$ windows yields $4096$ tokens per frame for $S^{(2)}=4 \otimes 4$. If $N_{\mathrm{img}} = 10^4$ images are acquired per each 10-s experiment at a movie rate of 10$^3$ per second per camera,
\begin{equation}
N_{\mathrm{2D}} = 4096 \times N_{\mathrm{img}} \approx 4 \times10^7.
\end{equation}
The redundancy of images at a high movie rate of 10$^3$/s (in the current dusty plasma context. Higher framerate of 10$^6$ Hz is commercially available at a higher camera hardware cost) may be expected for low-spatial resolution images, which may reduce the effective token rate compared with 1D time-series waveform token generation. Meanwhile, imaged and movies carries complementary spatial information to probe data.

Metadata and contextual parameters contribute comparatively few tokens per experiment, in particular, for fixed experimental settings. Therefore the following approximation may be justified
\begin{equation}
N_{\mathrm{meta}} \ll   N_{\mathrm{1D}},  N_{\mathrm{2D}}.
\end{equation}

Combining different modalities, meta-data and other inputs, the total number of tokens generated per experiment is
\begin{equation}
N_{\mathrm{exp}} = N_{\mathrm{1D}} + N_{\mathrm{2D}} + N_{\mathrm{meta}} \approx N_{\mathrm{1D}} + N_{\mathrm{2D}}.
\end{equation}

These estimates show that 10$^2$ to 10$^4$  experiments can supply token volumes ($\sim$ 1-10 B) comparable to those used in training domain-specific scientific FoMos using the current measurement tools.  Careful tokenization and measurement methods will nevertheless be needed to reduce data and token redundancy from repetitive measurements. Multi-modal dusty-plasma FoMos, such as DUST-MAP, thus become feasible through experiments and measurement data alone, which naturally provide information-dense, physics-imbedded, and scalable token streams.

\section{Applications of DustNET \label{sec:Apps}}
%AI foundation model may be used to predict plasma behavior, diagnose plasma states, or optimize experimental parameters.

Having outlined the conceptual framework, datasets, and methodological foundations of DustNET, we now turn to its potential scientific and technological applications. The DustNET ecosystem, combining curated multi-modal datasets, physics-informed machine learning, and emerging AI foundation models, provides a unifying platform for addressing long-standing challenges in dusty plasma research. In particular, DustNET enables systematic exploration across a wide range of problems, from the origin and fundamental physical properties of dust, to its interactions with plasma environments, and the emergence of collective structures across multiple spatial and temporal scales. 

In this section, we organize these applications into several interconnected themes. We first examine open problems related to the origin, physical nature, and dynamics of dust, followed by dust--plasma interactions and environmental coupling. We then discuss collective dynamics, self-organization, and multiscale structure formation. Building on these foundations, we consider broader scientific and technological applications, including implications for space, astrophysical, and laboratory systems. Finally, we highlight cross-cutting opportunities enabled by data-driven discovery, the role of edge AI agents in real-time diagnostics and control, and the reciprocal impact of dusty plasma physics on the development of next-generation AI systems. Together, these directions illustrate how DustNET can serve as both a scientific tool and a conceptual bridge between plasma physics and data-centric methodologies.

\subsection{Open problems in dusty plasmas}

The open problems in dusty plasmas can be organized into four classes: (i) the origin, fundamental properties and dynamics of dust particles, (ii) the interaction of dust with plasma environments, (iii) collective dynamics and emergent structures in dusty plasmas, and (iv) scientific and technological applications. Together, these themes define a research frontier spanning laboratory plasma physics, astrophysics, planetary science, and emerging data-driven methodologies.

\subsubsection{Origin, Physical Nature, and Dynamics of Dust}
The first class of open problems concerns the origin, formation mechanisms, intrinsic properties of dust particles and their temporal evolution. Dust may form through bottom-up processes, such as nucleation and condensation in cooling plasmas, or through top-down processes, including fragmentation, sputtering, or collisional disintegration of larger bodies. Understanding the relative importance of these mechanisms in different environments—laboratory plasmas, planetary atmospheres, cometary comae, protoplanetary disks, and the interstellar medium—remains an active area of research.

Equally important are questions regarding dust composition, morphology, and internal structure, which influence charging, optical properties, and plasma coupling. Determining the age and evolutionary history of dust grains, including processes such as aggregation, surface chemistry, irradiation, and erosion, presents further challenges. Accurate dust dating and compositional diagnostics are essential for reconstructing the history of astrophysical systems and planetary environments. 

Here are a few problems in this class:

$\dagger$~Birth of dust in the early universe.
When did the universe first become dusty? What astrophysical environments—supernovae, Population III stars, or early galaxies—produced the first dust grains?

$\dagger$~Bottom-up formation of dust through nucleation and growth.
Under what plasma conditions do nanometer-scale clusters nucleate and grow into stable grains in laboratory and astrophysical plasmas?

$\dagger$~Top-down production of dust via fragmentation and erosion.
What roles do collisions, sputtering, radiation damage, and mechanical breakup play in generating dust populations from larger bodies?

$\dagger$~Dust composition, morphology, and internal structure.
How do mineralogy, porosity, and surface chemistry affect charging, optical properties, and plasma coupling?

$\dagger$~Dust age and evolutionary history.
Can reliable “dust chronometers” be established to determine the age, exposure history, or origin of grains in astrophysical and planetary environments, on the large scale, and in parallel, on the laboratory time scale for dust growth?

\subsubsection{Dust–Plasma Interactions and Environmental Coupling} 
The second class of problems focuses on how dust interacts with surrounding plasma environments. Dust particles modify the plasma through charge exchange, electron and ion absorption, secondary emission, and collective electric field effects, while the plasma in turn governs dust charging, transport, and stability. These interactions depend sensitively on plasma parameters such as density, temperature, flow velocity, magnetic fields, and radiation.

In natural environments, dusty plasmas occur in planetary rings, comet tails, interstellar clouds, and the solar system's debris population. In technological environments, they appear in semiconductor manufacturing plasmas, plasma processing reactors, and fusion devices where dust can affect confinement, wall erosion, and safety. Understanding how dust interacts with plasma turbulence, electromagnetic fields, and radiation—particularly under non-equilibrium conditions—remains an important challenge. 

Here are a few problems in this class:

$\dagger$~Charging mechanisms in vastly different plasma environments.
How do photoemission, secondary electron emission, ion collection, and radiation fields determine dust charging under non-equilibrium conditions?

$\dagger$~Dust transport in plasma flows and electromagnetic fields.
What governs dust motion in magnetized plasmas, shocks, sheaths, and turbulent plasma environments?

$\dagger$~Dust interaction with radiation and energetic particles.
How do ultraviolet radiation, cosmic rays, and high-energy ions modify dust charging, erosion, and chemistry?

$\dagger$~Dust–plasma coupling in natural environments.
How does dust affect plasma behavior in planetary rings, cometary atmospheres, protoplanetary disks, and interstellar clouds?

$\dagger$~Dust in technological plasmas.
How do dust particles influence plasma processing, semiconductor fabrication, and plasma–material interactions in fusion devices?

\subsubsection{Collective Dynamics, Self-Organization, and Multiscale Structure} 
A third class of open problems involves the collective dynamics and emergent structures that arise in dusty plasmas. Because dust particles can carry large electric charges and interact strongly through long-range Coulomb forces, dusty plasmas often exhibit complex behaviors such as plasma crystals, phase transitions, waves, turbulence, and self-organized patterns. These phenomena provide a unique platform for studying strongly coupled systems, nonequilibrium statistical mechanics, and phase transitions at experimentally accessible scales.

Key unresolved issues include the role of dust in plasma instabilities, the formation and stability of ordered structures, the coupling between microphysical processes (charging, collisions) and macroscopic structures (dust clouds, rings, or voids)~\cite{Goer:1989}., and the transfer of energy and momentum across multiple scales. In astrophysical contexts, these processes may also influence planet formation, disk evolution, and dust transport in cosmic environments. On an even larger (or small?) scale, are there any observable effects on dusty plasmas by dark matter and dark energy?

Here are a few problems in the class:

$\dagger$~Formation of plasma crystals and ordered structures.
What determines the formation, stability, and phase transitions of strongly coupled dust systems?

$\dagger$~Dust waves, instabilities, and turbulence.
How do dust acoustic waves, streaming instabilities, shock waves, and turbulence arise and propagate in dusty plasmas?

$\dagger$~Multiscale coupling in dusty plasma systems.
How do microscopic processes (charging, collisions) connect to macroscopic structures such as dust clouds, rings, or voids?

$\dagger$~Dust aggregation and planetesimal formation.
How do electrostatic forces, turbulence, and collisions influence the early stages of planet formation in protoplanetary disks?

$\dagger$~Self-organization and pattern formation.
What universal principles or scaling laws, if exist, govern pattern formation and nonlinear dynamics in multi-scale dusty plasmas?

\subsubsection{Applications and Broader Scientific Implications} 
The fourth class of problems concerns the applications and broader implications of dusty plasma research. Dusty plasmas provide powerful experimental analogs for fundamental physics problems, including strongly coupled many-body systems and nonequilibrium phase transitions. In astrophysics, they play a central role in star and planet formation, interstellar chemistry, and the dynamics of cosmic dust populations.

At the same time, dusty plasmas pose practical challenges and opportunities in applied science and engineering. These include dust contamination in semiconductor processing, dust generation and mitigation in fusion reactors, and dust dynamics in space environments relevant to satellite operations, planetary exploration, and potential lunar or Martian settlements. Understanding dust behavior may also contribute to managing space debris or designing technologies for climate change, extraterrestrial resource utilization and exploration.

A few problems in the class are:

$\dagger$~Dust mitigation in fusion reactors.
How can radioactive or metallic dust be detected, controlled, or removed in next-generation fusion devices?

$\dagger$~Dust hazards and opportunities in planetary exploration.
What roles do charged dust particles play in lunar and Martian environments relevant to human exploration and colonization?

$\dagger$~Space debris and dusty plasma interactions.
How do charged dust particles interact with spacecraft surfaces and debris in near-Earth space?

$\dagger$~Dusty plasmas as analog systems for fundamental physics.
Can dusty plasmas serve as experimental platforms for studying strongly coupled matter, phase transitions, and non-equilibrium statistical mechanics?

$\dagger$~Astrophysical implications of dust–plasma interactions.
How do dusty plasmas influence star formation, cosmic chemistry, and the evolution of galaxies? 

$\dagger$~Are there observable effects on dusty plasmas from quantum vacuum fluctuations, dark matter, or dark energy?

\subsubsection{A Cross-Cutting Challenge and Opportunity: Data-Driven Discovery}
\begin{figure}[!htb]
    \centering
    \includegraphics[width=0.85\columnwidth]{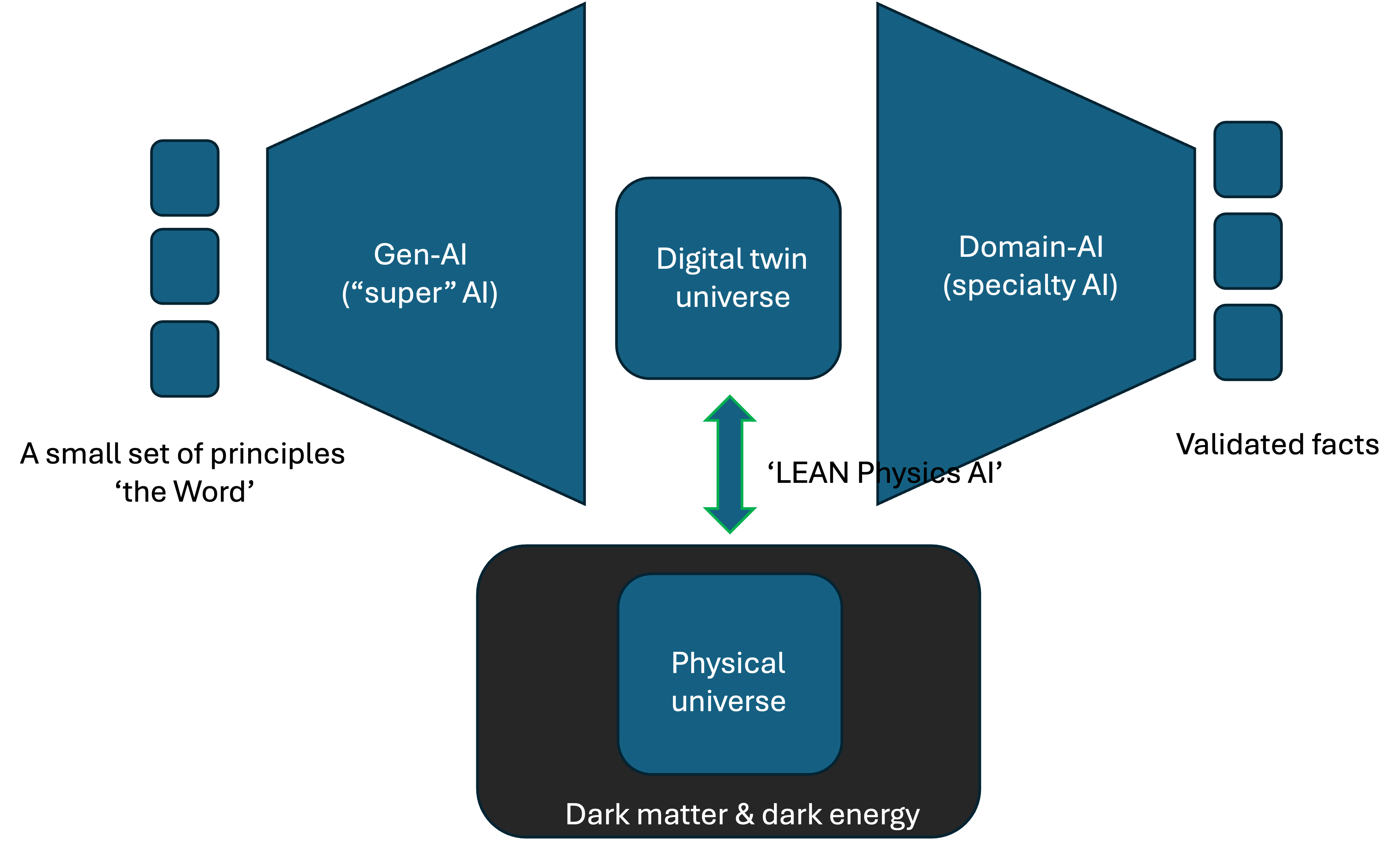}
    \caption{Integrated AI-enhanced framework for dusty plasma research .}
    \label{fig:vision}
\end{figure}

Across all of these areas lies a broader, overarching challenge and opportunity: can the rapidly developing data-driven methods, including machine learning and artificial intelligence, help accelerate and unify our understanding of dusty plasma phenomena across scales? Given sufficiently rich observational and experimental datasets, AI-based approaches may assist in identifying hidden correlations, extracting governing patterns, and linking laboratory measurements with natural observations. A central question is whether such models can ultimately produce physically interpretable insights that remain consistent with independently verifiable measurements and established physical laws. Addressing this challenge will likely require the integration of physics-informed modeling, multimodal data analysis, and large-scale scientific datasets within emerging foundation-model frameworks, as illustrated in Fig.~(\ref{fig:vision}). In the technology and applications frontier, one interesting question is whether AI can help to design `smart dust' that are superior in property and performance to `natural dust' without intelligence.

\subsection{Edge AI Agents for Real-Time Applications}
\label{Edge:AI}

Real-time or \emph{edge} applications of artificial intelligence are becoming increasingly important in many areas. In this context, ``edge AI'' refers to the deployment of machine learning algorithms directly on experimental platforms or embedded hardware, enabling rapid analysis and decision-making close to the data source. Such capabilities are particularly valuable for diagnosing nanometer-scale dust particles, monitoring in-situ charging dynamics, and detecting transient phenomena in laboratory or space-based dusty plasmas. Real-time diagnostics capable of measuring the temporal evolution of dust particle density and size at nanometer scales~\cite{BBB:2023} remain a key experimental challenge. Continued advances in edge sensing, high-speed imaging, and embedded AI inference are therefore needed to characterize the coupling between dust particle properties and plasma behavior with sufficient temporal resolution.

The growing complexity of AI tools also motivates the development of \emph{AI agents} for automated dusty-plasma experimentation and analysis~\cite{DHWG:2024}. An AI agent differs fundamentally from a multi-modal foundation model (FoMo) such as DUST-MAP, which primarily functions as a representation learner. DUST-MAP encodes heterogeneous inputs—including text, images, time-series measurements, and numerical simulations—into structured latent spaces that capture statistical regularities and physically meaningful patterns across plasma regimes. Its principal role is inference and interpretation. In contrast, an AI agent acts as an autonomous decision-making system that interacts with its environment, selects actions, invokes computational tools, and adapts its strategy based on feedback. In a dusty-plasma experiment, for example, a FoMo might identify sheath transitions, dust waves, or anomalous particle trajectories from multi-modal measurements, whereas an AI agent could use those inferences to dynamically adjust diagnostic settings, schedule instrument modes, trigger higher-cadence data acquisition, or launch targeted simulation runs. Conceptually, the FoMo provides perception and structured understanding of the system, while the AI agent provides planning, control, and goal-directed behavior. This distinction is analogous to that between a learned world model and an autonomous operator that acts upon the information extracted from that model.

An important challenge in developing edge AI agents is ensuring that machine learning systems remain consistent with established physical principles. Incorporating scientific knowledge—such as conservation laws, plasma transport models, and statistical constraints—into machine learning workflows has become an increasingly active area of research. A comprehensive survey on integrating physics-based modeling with machine learning was presented in~\cite{WJX:2020}. Several approaches have been proposed to achieve this integration. Physics-guided neural networks (PGNNs)~\cite{DKWR:2017}, for example, combine outputs from physics-based simulations with observational data in hybrid neural-network architectures, improving predictive performance while preserving interpretability. Physics-informed neural networks (PINNs)~\cite{RaPK:2019} extend this idea further by embedding governing equations directly into the training objective through physics-based loss functions. In this framework, neural networks are trained not only to minimize prediction errors but also to satisfy known physical constraints, thereby producing models that remain consistent with established plasma theory even when labeled data are limited.

For dusty-plasma systems, physics-informed AI agents may provide a pathway toward adaptive, closed-loop experimentation. Such agents could continuously assimilate experimental measurements, update predictive models of plasma–dust interactions, and propose new experimental configurations or control strategies in real time. When combined with edge computing architectures and multimodal foundation models such as DUST-MAP, this approach may enable a new generation of intelligent plasma diagnostics capable of autonomously exploring complex parameter spaces and uncovering previously inaccessible regimes of dusty-plasma physics.

\subsection{Dusty plasma physics for AI \label{sec:p4AI}}

The state-of-the-art AI models, such as ChatGPT~5.x, have been trained primarily on large corpora of internet-scale text and images. While such datasets have enabled remarkable advances in language and reasoning tasks, the next stage of AI development is expected to rely increasingly on data grounded in the physical world. Dusty plasmas represent a particularly rich and underexplored domain in this context. The concept of a \emph{world model}, advocated by Yann LeCun and others~\cite{HaSc:2018}, refers to AI systems capable of learning predictive representations of real physical environments. Because dusty plasmas exhibit observable particle dynamics, strong coupling, and multi-scale interactions that can be directly measured in both laboratory and natural settings, they offer an ideal testbed for contributing to such world models.

\begin{figure}
    \centering
    \includegraphics[width=0.95\linewidth]
    {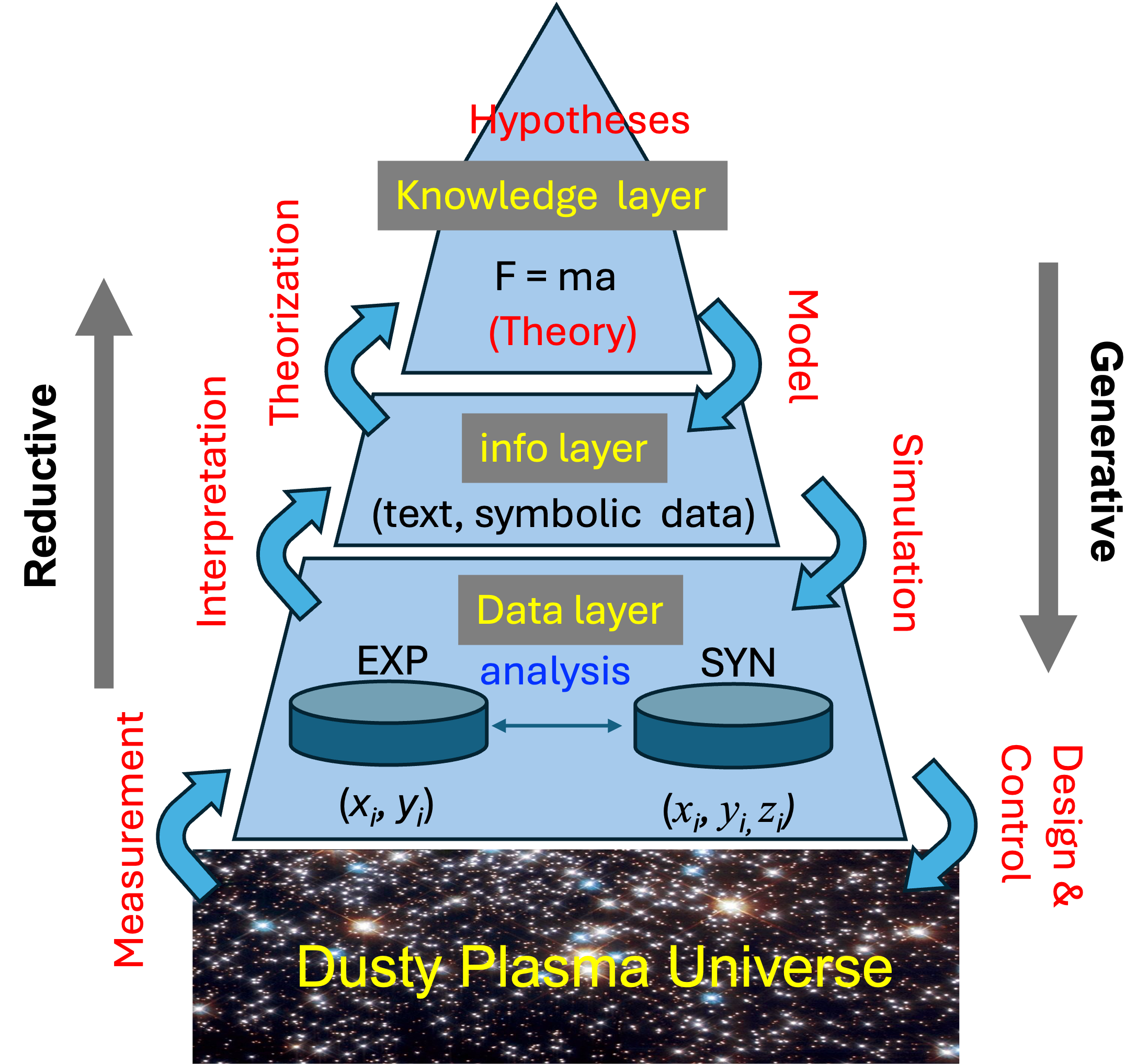}
    \caption{A summary of different machine learning tasks for physics, including dusty plasmas and applications.}
    \label{fig:MLST}
\end{figure}

The DustNET framework can contribute to these developments by providing structured datasets and experimental workflows that capture the physics of dusty plasmas across multiple modalities. The foundation of a dusty-plasma world model is a comprehensive suite of measurements, ranging from high-speed optical imaging of dust grains to plasma diagnostics and numerical simulations, potentially integrated through edge AI and AI-agent–driven experimentation. Figure~\ref{fig:MLST} summarizes several typical workflows in modern data-driven physics. At the most basic level, machine learning is already widely used to process large experimental datasets efficiently through tasks such as denoising, deblurring, compression, classification, and pattern recognition. Additional examples of such applications are discussed in Sec.~\ref{sec:dataset} and Sec.~\ref{Sec:dFuion}.

Dusty plasmas offer a unique advantage for data-driven modeling: individual dust grains are often directly observable through high-speed optical imaging. As a result, two central problems in dusty-plasma image analysis are the determination of particle positions in each frame and the automatic tracking of particles across successive frames in a time-resolved image sequence~\cite{WXKW:2020}. Because imaging cameras typically record two-dimensional projections of inherently three-dimensional particle motion, similar challenges arise in fluid mechanics diagnostics such as particle tracking velocimetry (PTV), particle image velocimetry (PIV), and laser speckle velocimetry (LSV). These related fields have therefore developed a wide range of algorithms for particle-based data analysis that can be adapted to dusty plasma studies. Recent work shows that deep neural networks can significantly improve inference accuracy and robustness in such tasks, although these gains may come at the cost of increased computational or hardware requirements~\cite{DLHS:2020}.

Beyond data analysis, AI is increasingly used to enhance researcher productivity in computational workflows. Tools such as GitHub Copilot and large language models including ChatGPT and Claude can assist in writing and debugging code, generating numerical analysis routines, and producing documentation in programming and markup languages such as Python, \LaTeX{}, or Lean. Such AI-assisted coding environments can accelerate the development of simulation frameworks, data-analysis pipelines, and visualization tools, thereby improving human efficiency and enabling researchers to focus more directly on scientific interpretation and discovery.

In the long term, integrating dusty-plasma experiments, simulations, and AI models may help construct physics-grounded world models capable of predicting plasma–dust dynamics across scales. Such models could enable automated experimental design, real-time adaptive diagnostics, and improved understanding of strongly coupled plasma systems in both laboratory and astrophysical environments.

\section{Summary \label{sec:Summ}}

Dusty plasmas pervade the universe, from laboratory experiments and industrial systems to planetary environments and astrophysical media, yet their behavior remains difficult to capture within a unified predictive framework. As observational and experimental capabilities expand, dusty plasma research is entering a data-rich era in which traditional modeling approaches alone are no longer sufficient. This transition calls for a new paradigm that integrates physical understanding with data-driven discovery, enabling systematic interpretation across scales, environments, and modalities.

DustNET represents a foundational step toward this paradigm. As a community-driven, multi-modal data ecosystem, DustNET aims at standardize, curate, and integrate heterogeneous datasets spanning laboratory measurements, space missions, astrophysical observations, and synthetic data. Beyond serving as a repository, DustNET establishes a common infrastructure for training, benchmarking, and validating machine learning models, thereby enabling reproducibility, cross-domain generalization, and collaborative progress. In this sense, DustNET plays a role analogous to ImageNet for computer vision and deep learning, but with an explicit emphasis on physics-informed learning and scientific interpretability.

DustNET forms the data backbone for a broader AI-enabled framework built around multi-modal foundation models and autonomous agents. Foundation models, such as DUST-MAP, aim to learn unified representations of dusty plasma systems by integrating images, time-series measurements, simulations, and scientific text within a common latent space. These models provide a “world model” of dusty plasmas, capable of capturing multi-scale, multiphysics behavior in a manner that is both data-driven and constrained by physical principles. Complementing this capability, AI agents can operate on top of such models to enable closed-loop experimentation, dynamically adjusting diagnostics, guiding simulations, and optimizing experimental conditions in real time.

From a modeling perspective, we anticipate a convergence of reduced-order physics models, multiphysics simulations, and machine learning surrogates into hybrid frameworks that are both computationally efficient and physically grounded. Advances in geometric learning, physics-informed loss functions, and uncertainty quantification will further enhance model robustness and interpretability. At the experimental level, integration with edge AI systems and advanced diagnostics will enable real-time, three-dimensional, and multi-modal characterization of dusty plasma dynamics, including particle tracking, structural evolution, and emergent phenomena.

Together, these developments form a pathway toward a unified, data-centric, and physics-informed framework for dusty plasma science. By bridging fundamental plasma physics with artificial intelligence, DustNET, DUST-MAP, and AI agents collectively provide a platform for addressing long-standing open problems while enabling new capabilities in space exploration, fusion research, and astrophysical discovery. More broadly, dusty plasmas offer a uniquely rich testbed for next-generation AI systems grounded in physical reality, positioning the field as both a beneficiary of and a contributor to the future of scientific machine learning and AI.

\begin{acknowledgments}

LANL work was performed under the auspices of the U.S. Department of Energy (DOE) by Triad National Security, LLC, operator of the Los Alamos National Laboratory under Contract No. 89233218CNA000001, including DoE Office of Fusion Energy Sciences, and LANL Laboratory Directed Research and Development (LDRD) Programs. 

The work led by Justus Liebig University and THM University of Applied Sciences is supported by the German Aerospace Agency (DLR). We thank the research group led Markus H. Thoma and M. Kretschmer at Justus Liebig University Giessen for providing the data used in this study and 
the German Aerospace Society for providing powerful PCs in accordance with the funding decision, which facilitated the studies. The project was supported by the German Federal Ministry of Economic Affairs and Climate Action under contract No. 50WK2270B.

The work at Donghua University was supported by the National Natural Science Foundation of China under Grant No. 12035003.

The work at Soochow University was supported by the National Natural Science Foundation of China under Grant No. 12175159.

University of Michigan work is supported by the National Science Foundation ECLIPSE program award No. 2206039, the U.S. Department of Energy, Office of Fusion Energy Sciences, Grant DE-SC0018058, and Department of Energy Contract No. 89233218CNA000001.

The work at the  University of Greifswald is supported by the German Aerospace Agency (DLR) under grant Nos 50WM2161 and 50WM2561.

This research used resources of the Magnetized Plasma Research Laboratory at Auburn University. ET and SCT are supported by US Department of Energy Grant No. DE-SC-0019176 and by National Science Foundation, Plasma Physics program via grant 2308948.

The work by Mississippi State University is supported by the National Science Foundation, Plasma Physics program via grant 2308947. This work is also supported by the U.S. Department of Energy, Office of Science, Office of Fusion Energy Sciences under award number DE-SC0026203. 

Work at Iowa was supported by the U.S. Department of Energy grant DE-SC0025444 and National Science Foundation grant No. PHY-2510501.

\end{acknowledgments}

\bibliography{DustNET}% Produces the bibliography via BibTeX.

\end{document}